\documentclass{aa}
\usepackage{hyperref}%
\usepackage{graphicx}
\usepackage{txfonts} %
\usepackage{natbib}  
\usepackage{breqn} 
\usepackage{xcolor} 
\usepackage[normalem]{ulem}  

\begin{document}

   \title{Spectral index$-$mass accretion rate correlation and evaluation of  
 black hole masses in 
AGNs 3C~454.3 and M87}
   \titlerunning{Spectral index$-$mass accretion rate correlation}
    
\author{ Lev Titarchuk \inst{1,2,3} 
\and
Elena Seifina\inst{3,4}
\and
Alexandre Chekhtman\inst{5}         
\and
Indira 
Ocampo \inst{6}         
          }

\institute{ $^1$ Dipartimento di Fisica, University di Ferrara, Via Saragat 1,
I-44122, Ferrara, Italy,  \email{titarchuk@fe.infn.it};
$^2$ Astro Space Center, Lebedev Physical Institute, Russian Academy of Science, 
Profsousnaya ul., 84/32, Moscow, 117997 Russia; 
$^3$ Lomonosov Moscow State University/Sternberg Astronomical Institute,
Universitetsky Prospect 13, Moscow, 119992, Russia,  \email{seif@sai.msu.ru}; 
$^4$ All-Russian Institute of Scientific and Technical Information,
RAS, Usievich st. 20, Moscow, 125190, Russia;
$^5$ George Mason University, College of Science, 4400 University Drive, 
Fairfax, VA 22030, 
\email{achekhtm@gmu.edu};
$^6$ Lomonosov Moscow State University/Physics Department, 
Vorobievy Hills, 1, Moscow, 119234, Russia}

    %

   \date{Received 29 of March 2019 ; accepted 1 of November 2019}

\abstract
   {We present the discovery of correlations between the X-ray spectral (photon)  index 
and  mass accretion rate  observed  in  active galactic nuclei (AGNs) 
3C~454.3 and M87.  
We analyzed spectral  transition episodes observed in these AGNs 
using {\it Chandra}, {\it Swift}, {\it Suzaku}, $Beppo$SAX, ASCA and 
 {\it RXTE} data. We applied a scaling technique for a black hole (BH) mass 
evaluation which uses  a  correlation between the photon index and 
normalization of the    seed (disk)  component which is proportional to a mass accretion rate. 
We developed an analytical model that shows that the 
photon index of the BH emergent 
spectrum  undergoes an evolution from  lower to higher values depending 
on disk mass accretion rate.
To estimate a BH mass in 3C~454.3 we consider extra-galactic SMBHs NGC~4051 and NGC~7469 as well as Galactic BHs Cygnus X--1 and GRO~J1550--564   as reference sources for 
which distances, inclination  angles  are known and the BH masses are already   evaluated. 
For M87 on the other hand, we provide the BH mass scaling  using extra-galactic 
sources (IMBHs: ESO~243--49 HLX--1 and M~101  ULX--1) and Galactic sources (stellar mass BHs: XTE~J1550--564, 4U~1630--472, GRS~1915+105 and H~1743--322) as reference sources. 
Application of the scaling technique for the photon  index$-$mass 
accretion rate correlation provides  estimates of the BH masses in 3C~454.3 and M87 to 
be about $3.4\times10^9$ and $5.6\times10^7$ 
solar masses, respectively.  We also compared our scaling BH mass estimates
with  a recent BH mass estimate of $M_{87}=6.5\times 10^9 M_{\odot}$ made using the {Event Horizon Telescope} which gives an image at 1.3 mm  and is  based on the angular size of the
`BH event horizon'.    Our BH mass estimate in M87 is at least two orders of 
magnitude lower than that  made by  the EHT team.
}
      
\keywords{accretion, accretion disks --
                black hole physics --
                stars, galaxies: active -- galaxies: Individual: 3C~454.3, M87  --
                radiation mechanisms 
               }


\titlerunning{On evaluation of BH mass in 3C~454.3 and M87}  
  
  \maketitle
%

\section{Introduction}

Active galactic nuclei (AGNs) are ideal laboratories for studying  the properties of matter in 
critical conditions. Especially interesting are radio-loud AGNs such as blazars and FR-I radio 
sources, which more often enter into 
active states and are accompanied by jet ejection events. Usually they are sources of 
multiwavelength radiation. It is also assumed that their energy is controlled by a supermassive 
black hole (BH) at their center.

Blazars constitute a sub-class of radio-loud 
AGNs characterized by broadband nonthermal
emission from radio to $\gamma$-rays. It is generally proposed that 
bipolar relativistic jets closely align
to our line of sight. They exhibit rapid variability and a high
degree of polarization. Blazars with only weak or entirely absent emission lines in the optical band are 
historically classified as BL Lacertae objects, while others are
classified as flat-spectrum radio quasars  (FSRQs, see Urry \& Padovani 1995).
 
Among the FSRQ class of blazars, 3C~454.3 (PKS~2251+158) is one of the
brightest and most variable sources. This source is a well-known AGN 
and demonstrates significant variability in all energy bands 
(Unwin et al. 1997): optical (Villata et al. 2009; Raiteri et al. 1998, 2011), radio (Vol'vach et al. 2008, 2010, 
2011, 2013, 2017), X--rays (Raiteri et al. 2011; Abdo et al. 2010), and
$\gamma$-rays (Bonning et al. 2009; Abdo et al. 2009).

In the optical range, the blazar 3C~454.3 shows a change of active states. 
For example, in 2006--2007 the source remained
in a faint state (Raiteri et al. 2007). Then, from July 2007 the source began 
an active phase, which was detected several times in $\gamma$-rays by the
AGILE satellite\footnote{http://agile.iasf-roma.inaf.it/}. The source  was monitored from the optical to the
radio bands by the Whole Earth Blazar Telescope\footnote{http://www.oato.inaf.it/blazars/webt/} 
(WEBT) and its GLAST-AGILE Support Program (GASP, Villata et al. 2008).  Vercellone et al. (2008), Raiteri et al. (2008a,b), Vercellone et al. (2009), Donnarumma et al. (2009), Villata et al. (2009), and Vercellone et al. (2010) published the results of these observations.
GASP-WEBT (Global Aviation Safety Plan-Whole Earth Blazar Telescope) observations in 2009--2010 were performed in a number of observatories:
Abastumani, Calar Alto data were acquired as part of the MAPCAT project\footnote{http://www.iaa.es/~iagudo/research/MAPCAT}, Crimean, Galaxy View, Goddard
(GRT- Gamma Ray Telescope), Kitt Peak [(MDM), Michigan-Dartmouth-MIT Observatory] , Lowell (Perkins-Lowell Observatory), Lulin, New Mexico
Skies, Roque de los Muchachos (KVA-Kungliga Vetenskapsacademien), Sabadell, San Pedro
Martir, St. Petersburg, Talmassons, Teide (BRT-the Bradford Robotic Telescope ), Tijarafe, and
Valle d'Aosta. Optical observations were also carried out by the 
UVOT instrument 
onboard the $Swift$ satellite (Raiteri et al. 2011).
 This  optical object with  a magnitude of $V = 16.1^m$  was also identified as a radio source and   classified as a highly polarized quasar with a redshift of $z = 0.859$ (Jackson \& Browne 1991).
  

A multi-wavelength study of the flaring behavior of
3C~454.3 during 2005--2008 was carried out by Jorstad et al. (2010). These latter authors 
discussed the activity of this object in terms of a core-jet with a knot structure and 
suggested that the emergence of a superluminal
knot from the core relates to a series of optical and high-energy
outbursts and that the millimeter-wave core lies at the
end of the jet acceleration and collimation zone.

 $Suzaku$ observations of 3C~454.3 in November 2008 analyzed
by Abdo et al. (2010) confirmed the earlier suggestions
by Raiteri et al. (2007, 2008b) that there is a soft
X-ray excess, which becomes stronger when the $\gamma-$ radiation   in this source
gets fainter. These authors interpret this excess
as either a contribution of the high-energy
tail of the synchrotron component or bulk-Compton radiation
produced as a result of Comptonization of the disk UV photons due to
the relativistic jet plasma.

Despite the 3C~454.3  study  based on multi-wave observations, the question of the mass 
of the central engine in 3C~454.3 still remains open. On the other  hand, there is a  problem with its 
extended structures, which are clearly visible on the radio images. We  should also emphasize
its strong variability  which can be interpreted in  the framework of the binary system model 
(see Volvach et al., 2017 and references there).

The mass of the super-massive black hole  (SMBH) centered in 3C~454.3 is estimated to be within a the range of 
 $M = (0.5 - 1.5)\times 10^9$ M$_{\odot}$ (Woo \& Urry 2002; Liu et al. 2006; Sbarrato et al. 2012).
Gu et al.  (2001) concluded that the quasar contains a binary system of BHs with masses of 
$0.5 - 4 \times 10^9 M_{\odot}$, and  Eddington
luminosity that ranges from $6 \times 10^{46}$ to $5 \times 10^{47}$ erg/s (Khangulyan et al. 2013).
The flow velocity of the 
relativistic jet probably ranges between $0.97c$ and $0.999c$ (Jorstad et
al. 2005; Hovatta et al. 2009; Raiteri et al. 2011) and  the angle to
our line of sight  is between 1$^{\circ}$ and 6$^{\circ}$ (Raiteri et al. 2011; Zamaninasab
et al. 2013). Gupta et al. (2017) 
 using  optical data from the  Steward observatory 
produced a more precise estimate of the BH mass of the central source using 
the broad Mg II line width and the continuum luminosity at 3000 \AA : $M_{BH} = 2.3\pm 0.5 \times 10^9$ M$_{\odot}$ (see also Vestergaard \& Osmer 2009).

It is known that 3C~454.3 is an active source accompanied by jet ejection events that are associated with a sharp 
increase in radio emission.
The jet is also detected in X-rays. 
It is important  to note that the most prominent extragalactic X-ray jet is well observed in AGN M87 
[$E0$-type~\citep{Hubble26}, FR-I class], 
where several bright components (blobs) of this one-sided jet can be separately observed. In particular, the blob  
HST--1   is located less than 1"{\tt } from the nucleus 
of M87, and is the closest of its kind to that position. It is interesting that the HST--1 flux sometimes  becomes significantly higher  than that of a nucleus component. 
Therefore, it is also important to separate the jet emission from the radiation of the nucleus. 

In this paper we present our analysis of  available {\it Chandra}, $Swift$, $Suzaku$, {\it ASCA}, 
{\it BeppoSAX} and $RXTE$ observations of 3C~454.3 and M87  and  also reexamine previous 
conclusions on the BH nature of  these AGNs. 
Furthermore, we find further indications to supermassive BHs in 3C~454.3 and M87.  
In Sect.  2 we present the list of observations used in our data analysis, and in Sect.  3 we provide  details of the X-ray spectral analysis.  We discuss the evolution of 
the X-ray spectral properties during the spectral 
state  transitions  and present the results of the scaling analysis to estimate a BH mass for 
these AGNs 
in Sect. 4.   We  present our final conclusions on  the results  in Sect. 5.  

%
%

\begin{figure*}[ptbptbptb]
\centering
\includegraphics[scale=0.87,angle=0]{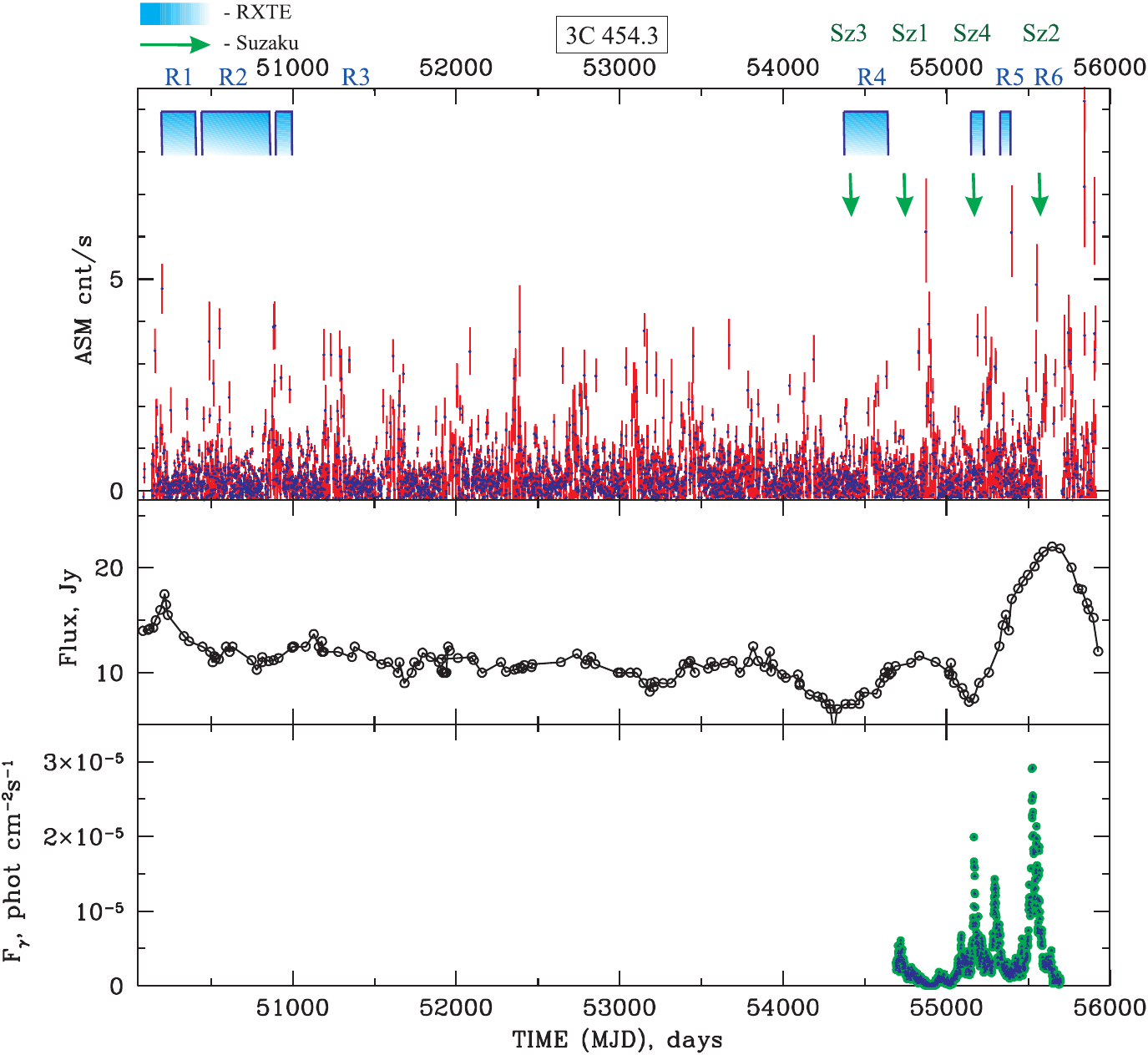}
\caption{Evolution of ASM/{\it RXTE} count rate (2-12 keV, top), radio (8 GHz, middle) and 
$\gamma$-ray radiation (0.1--300 GeV, bottom) during 1996 -- 2012 observations of 3C~454.3.
Vertical arrows (at  top  of the panels) indicate temporal distribution of the {\it RXTE} 
(blue) and $Suzaku$ (green) 
data sets listed in Tables \ref{tab:list_suzaku}$-$\ref{tab:list_RXTE}.
 }
\label{RXTE evol}
\end{figure*}

%
%
\begin{figure*}[ptbptbptb]
\centering
\includegraphics[width=12cm]{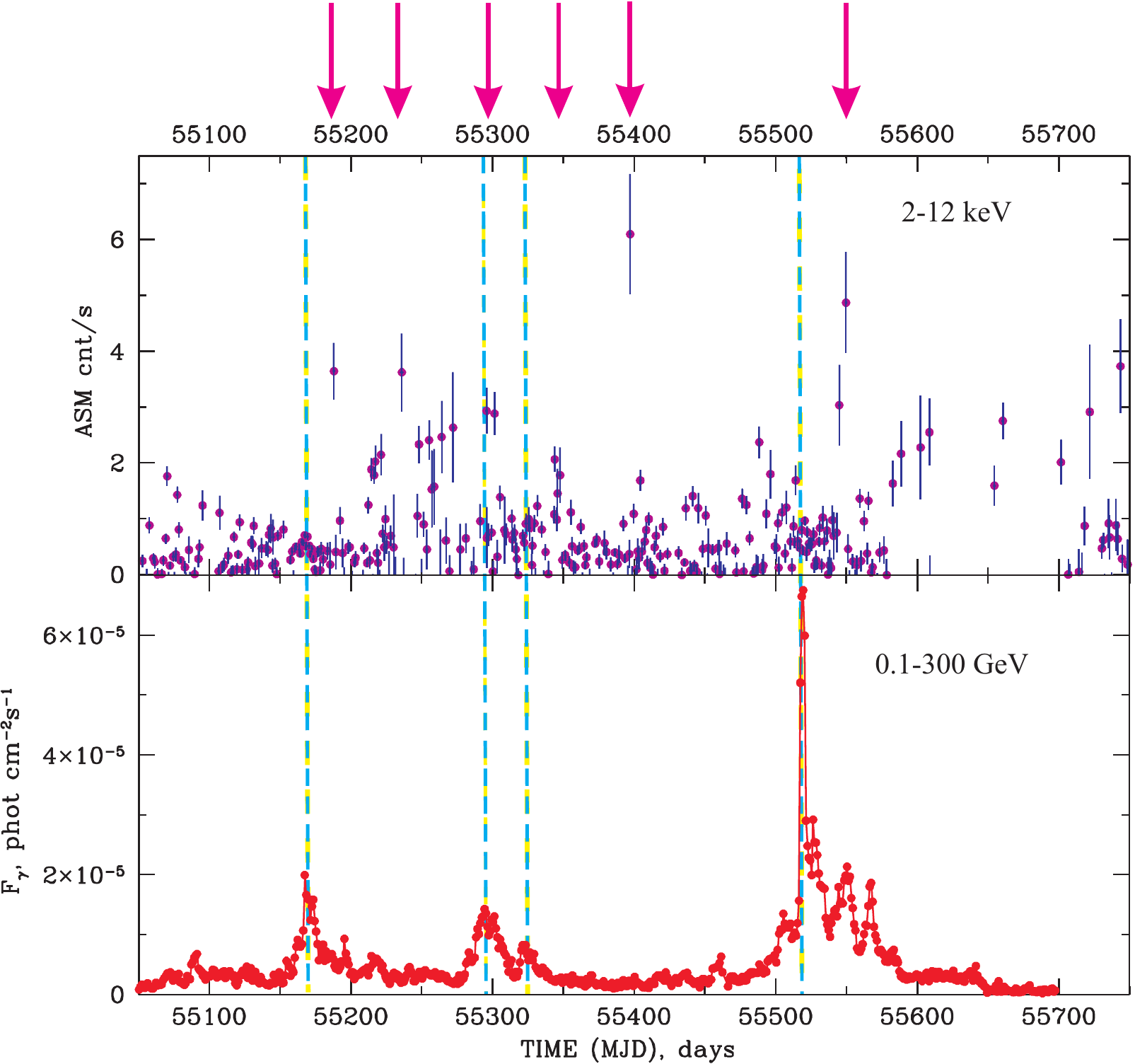}
\caption{On delays of peak-fluxes for outburst of 3C~454.3 observed in 2010--2012  for two bands: 2--12 keV (top) 
and 0.1--300 GeV (bottom). Vertical $blue$ dashed lines mark the peak of 
$\gamma$-ray flux, 
 while pink 
arrows (at  top of the panels) indicate 
the peaks of X-ray 
bursts. 
 }
\label{no_delay_3c454}
\end{figure*}

%
%
  \begin{figure*}
 \centering
\includegraphics[width=17.9cm]{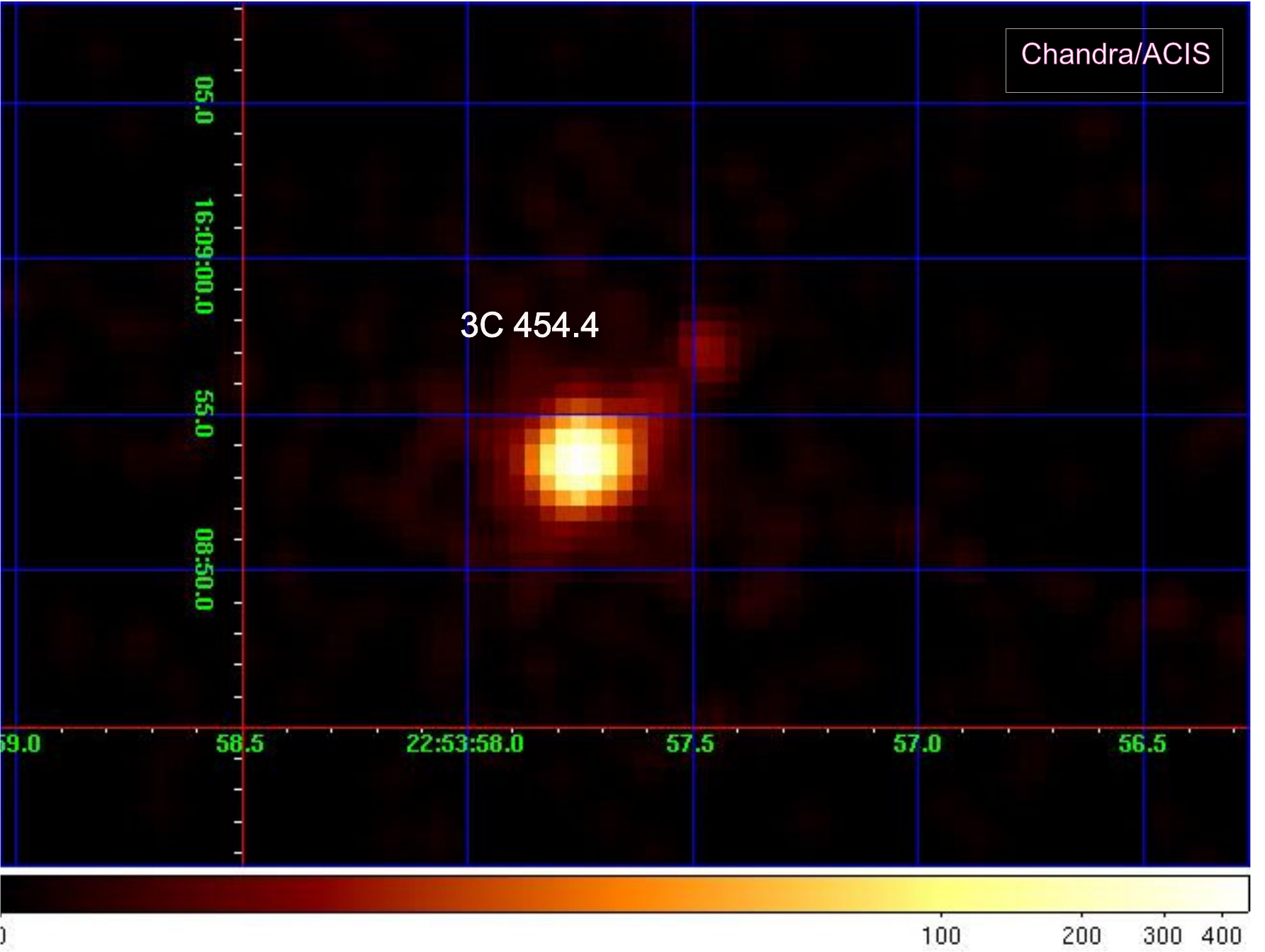}
      \caption{$Chandra$/ASIS (0.3$-$10 keV) image of 3C~454.3 field of view on November 6, 2002 
with the exposure of 4929 s (MJD=52584). 
}
      \label{imagea}
 \end{figure*}
%
%
  \begin{figure*}
 \centering
\includegraphics[width=15cm]{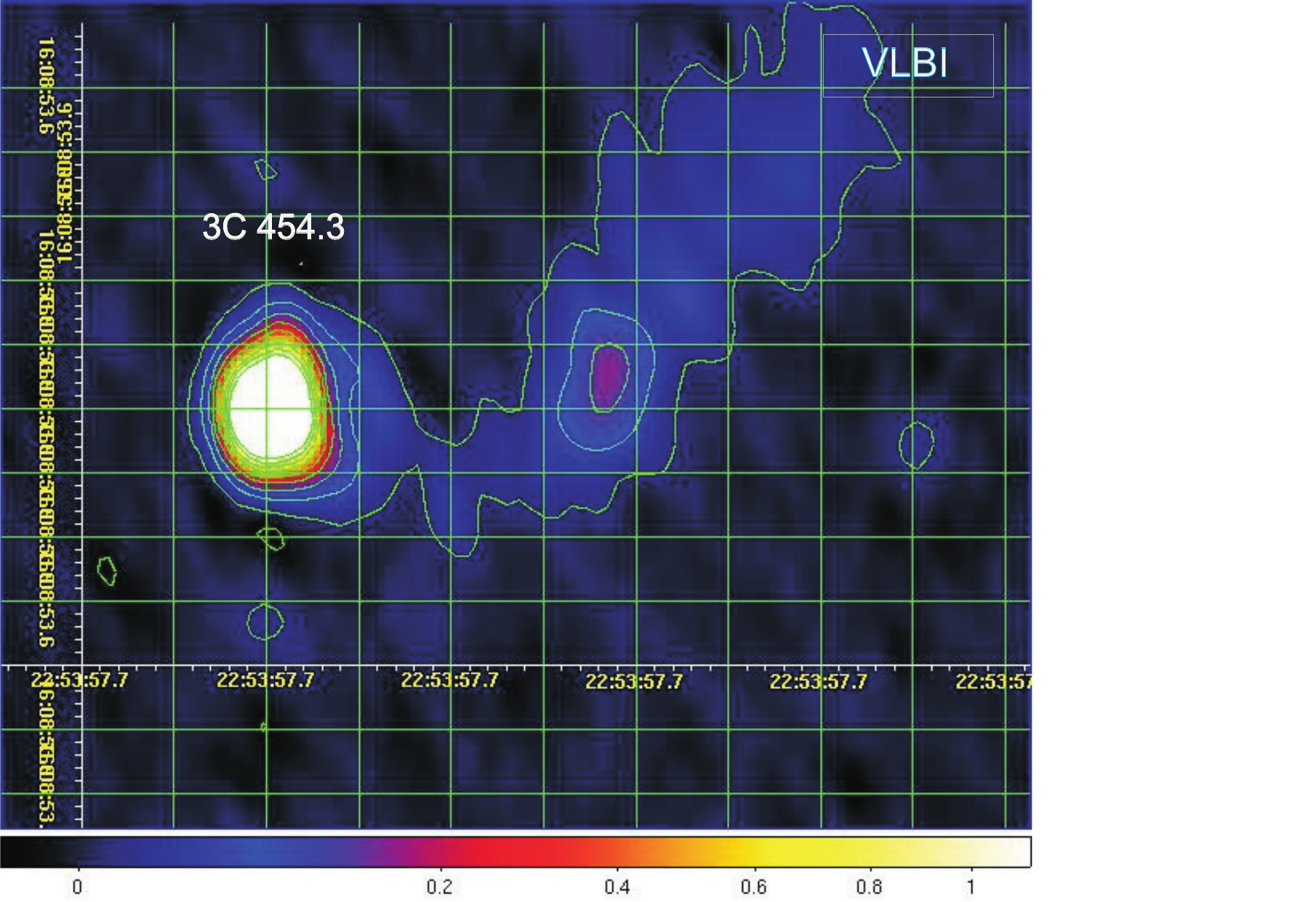}
      \caption{
Adaptively smoothed $VLBI$ image  of radio emission 
of 3C~454.3 field detected on March 10, 1997. 
Contours correspond to nine logarithmic intervals in the range of $3\times10^{-3}-$5\% with respect to the brightest pixel. 
}
\label{imageb}
\end{figure*}

%
%
\begin{figure*}[ptbptbptb]
\centering
\includegraphics[scale=0.60,angle=0]{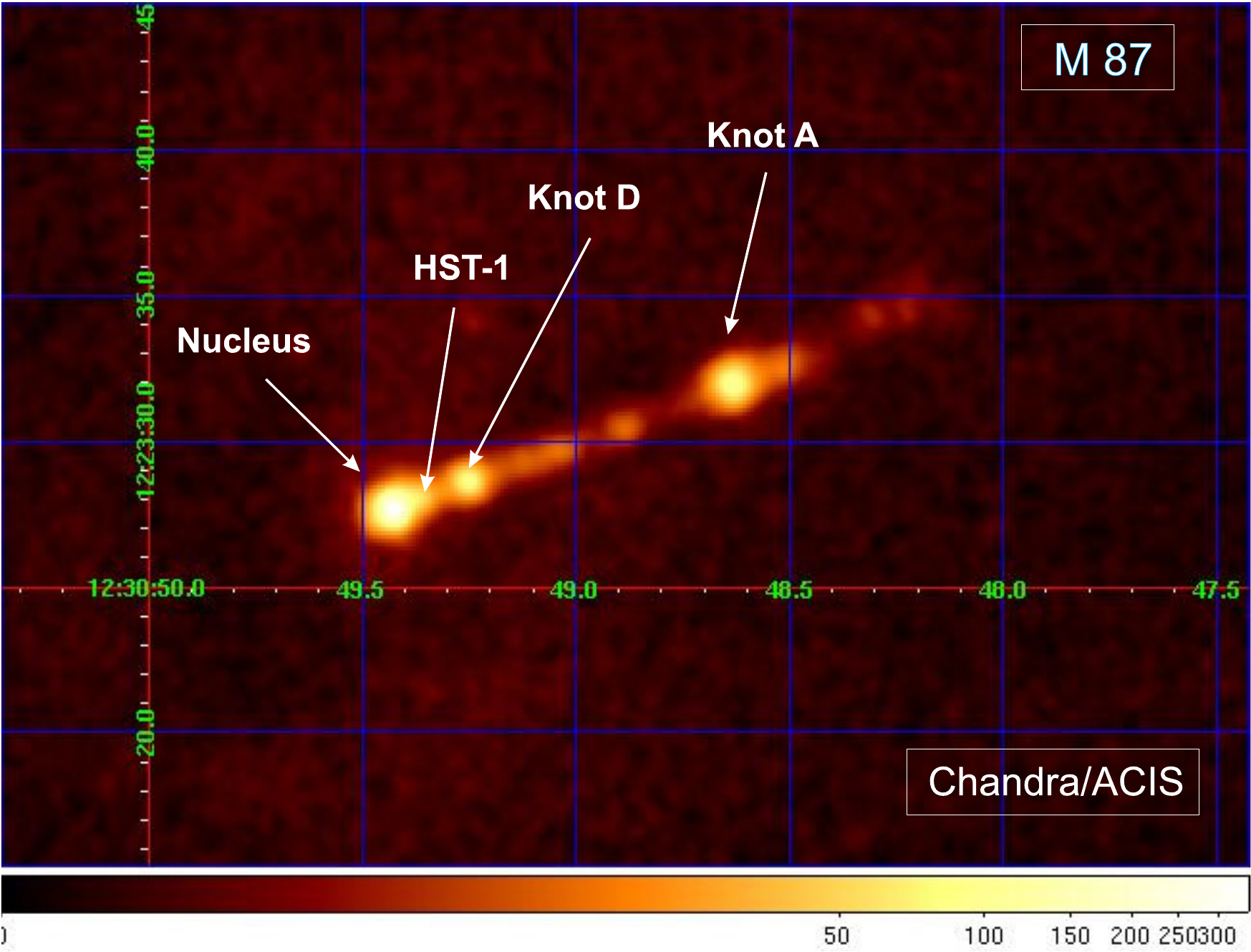}
\caption{ $Chandra$ X-ray images on April 14, 2017, with strong HST--1,
with 5 ks 
exposure. 
We indicate the nucleus, 
the flaring knot HST-1 (0.86" from the nucleus), knot D, and knot A.
 }
\label{chandra_ima_M87}
\end{figure*}

\section{Observations and data reduction \label{data}}

Along with the long-term {\it RXTE} observations (1996--2010) 
described in Sect.~\ref{rxte data}, 3C~454.3 was  also observed by 
$Suzaku$ (2007 -- 2010, see 
Sect.~\ref{suzaku data}), $Swift$ (2005--2010, see Sect.~\ref{swift data}),  
and $Chandra$ (2002, 2004, see Sect.~\ref{chandra data}).  
On the other hand, M87 was observed by {\it RXTE} (1997 -- 1998 and 2010), which is 
described in Sect.~\ref{rxte data}, $Beppo$SAX (1996; Sect.~\ref{sax data}), $Chandra$ (1996; 
Sect.~\ref{chandra data}), $Suzaku$ (2006, see Sect.~\ref{suzaku data}), and $ASCA$ (1993, 
see Sect.~\ref{asca data}). 
We extracted these data from the HEASARC archives and found that they  cover 
a wide range of X-ray luminosities for both sources.  We  should  recognize that the well-exposed {\it Suzaku} 
data are  preferable for the determination of   low-energy photoelectric absorption. Therefore,  
we start our study using  $Suzaku$ data of 3C~454.3. 
We  also apply  up-to-date SMART optical/near-infrared 
light curves that are available at {\it www.astro.yale.edu/smarts/glast/home.php}.  
%
%

\begin{figure*}[ptbptbptb]
\centering
\includegraphics[scale=0.87,angle=0]{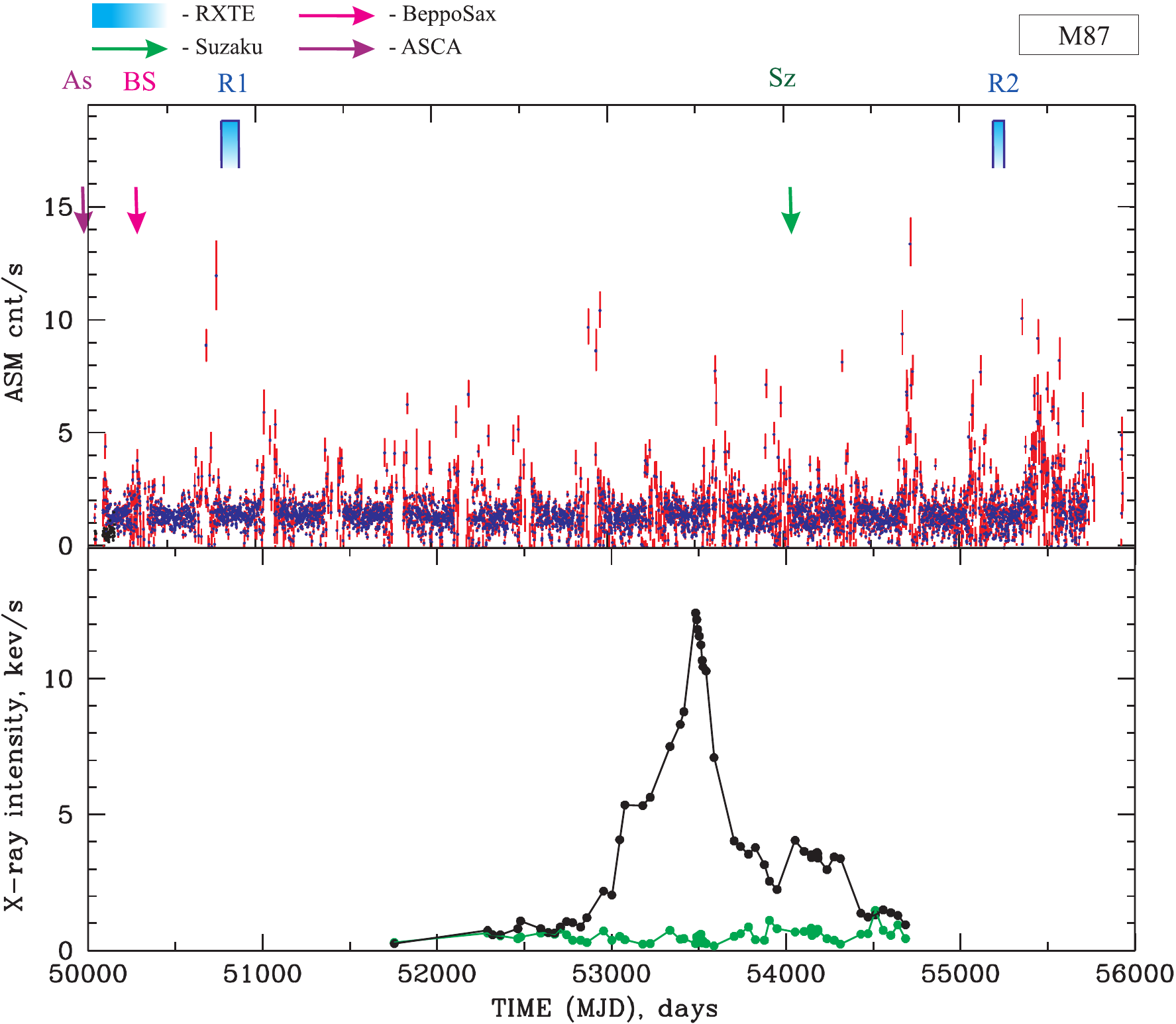}
\caption{Top: Evolution of ASM/{\it RXTE} count rate during 1996 -- 2012 observations of M87.
Vertical arrows (at the top of panel) indicate temporal distribution of the {\it RXTE} (blue) and 
$Suzaku$ (green) 
data sets listed in Tables \ref{tab:list_suzaku}$-$\ref{tab:list_RXTE}. 
Bottom: Evolution of $Chandra$ X-ray intensity of the nucleus (green) and HST-1 (black) during 
2000 -- 2008 observations of M87 according to data taken from Harris et al. (2009).
 }
\label{RXTE evol_m87}
\end{figure*}

%
%
%
\begin{figure*} 
\begin{center}
\includegraphics[scale=0.7, angle=0]{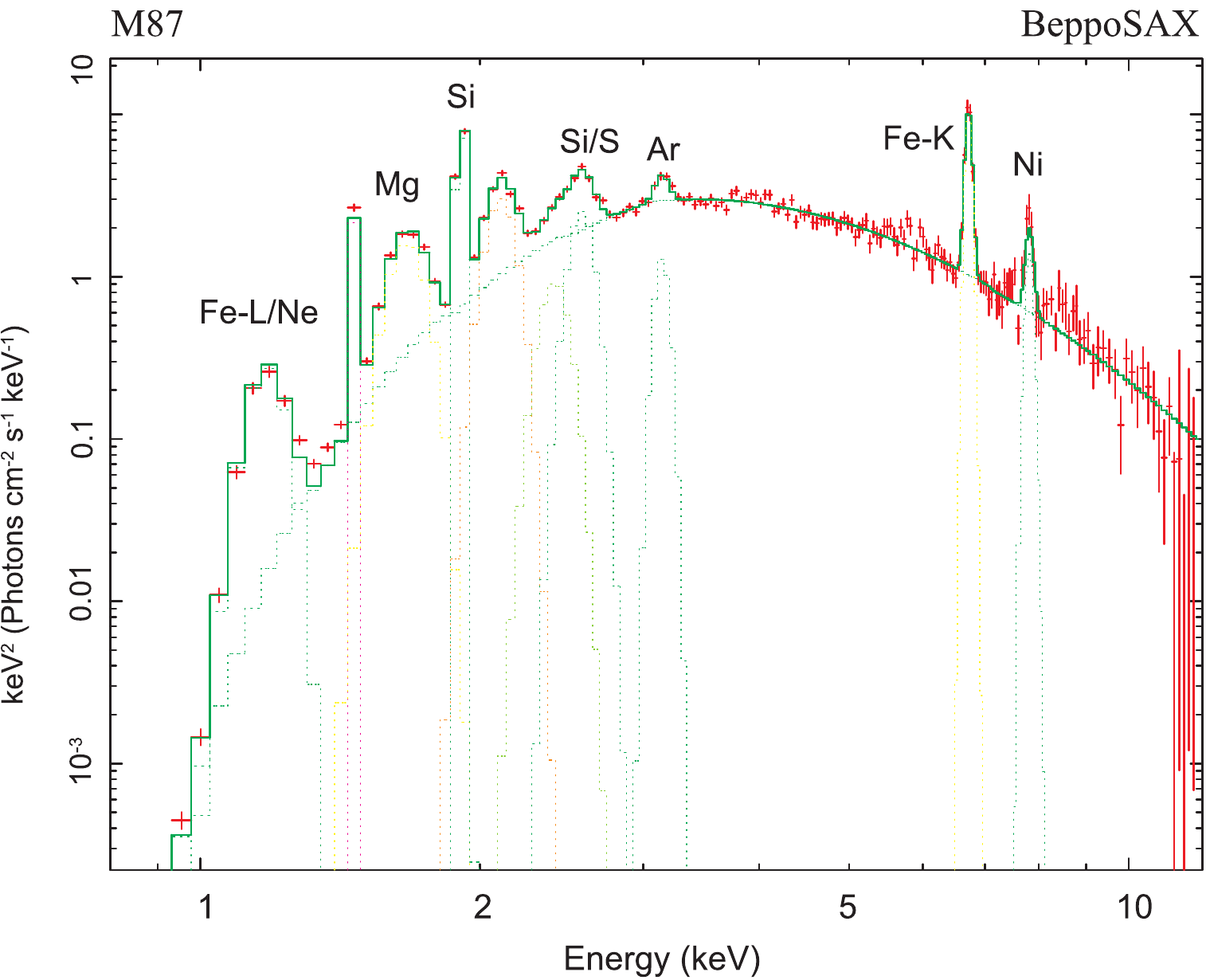}
\caption{Best-fit {\it BeppoSAX} spectrum of M87 in the IS--HSS state  (ID=60010001) using the  {\tt tbabs*bmc+N*gauss} model [$\chi^2_{red}=$ 1.05 (249 dof)]. The best-fit parameters are 
   $\Gamma=$ 2.9$\pm$0.1 and $T_s=$ 140$\pm$10 eV. 
   (see more detalis in Table~\ref{tab:par_SAX+asca_m87}). 
   The data are denoted by red {crosses}, while the spectral model is presented 
   by a green histogram. 
}
\label{sp_m87_BeppoSAX}
\end{center}
\end{figure*}

%
%

 \begin{figure*}
 \centering
\includegraphics[width=17cm]{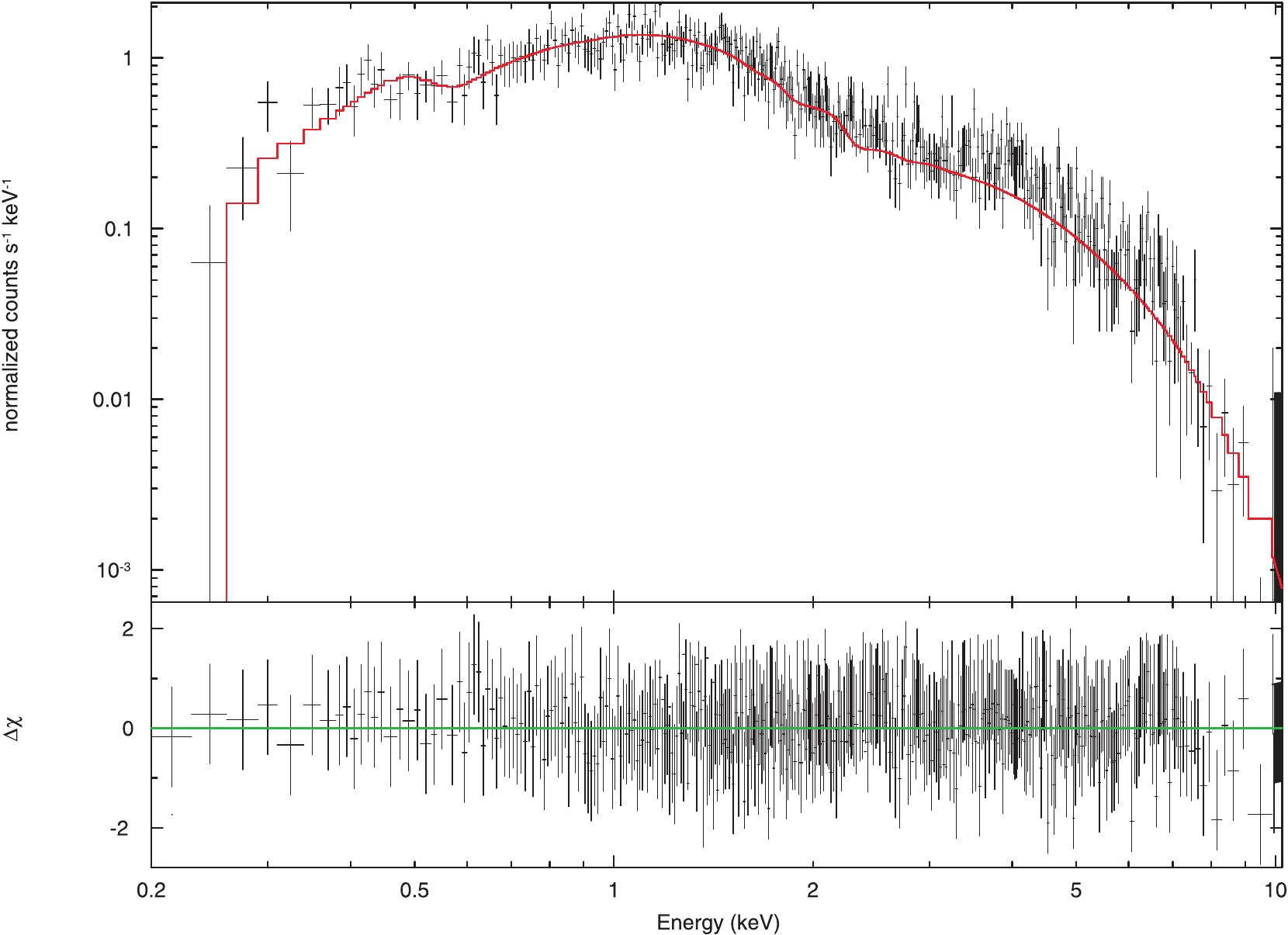} 
  \caption{Best-fit {\it Swift} spectrum of 3C~454.3 for the IS 
using the  {\tt tbabs*bmc} model    [$\chi^2_{red}=$ 1.02 (770 dof)]. The best-fit parameters are 
   $\Gamma=$1.63$\pm$0.1 and $T_s=$280$\pm$20 eV. 
   The data are denoted by black {crosses}, while the spectral model is presented 
   by a red histogram. 
}
\label{swift_interm_spectrum}
\end{figure*}

%

\begin{figure*} 
\begin{center}
\includegraphics[scale=1.4, angle=0]{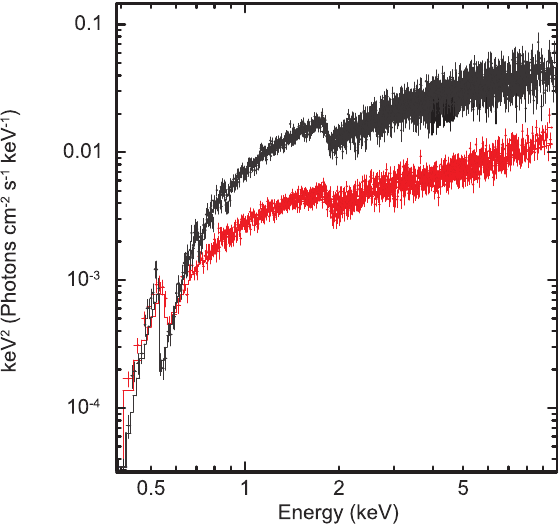}
\caption{Two  representative EF(E) spectral diagrams for the low/hard   state  (red, id=703006010) 
and  for the intermediate state   (black, id=904003010) of 3C~454.3  obtained with $Suzaku$.
}
\label{two_sp_Suzaku}
\end{center}
\end{figure*}


\begin{table*}
 \caption{List of $Suzaku$  observations of  3C~454.3 and M87 used in our analysis}              
 \label{tab:list_suzaku}      
 \centering                                      
 \begin{tabular}{l l l l l c c c}          
 \hline\hline                        
Object & Number of set  &  Obs. ID        & Start time (UT)  && End time (UT) &MJD interval \\    
 \hline                                   
3C~454.3 & Sz1  &    703006010       & 2008 Nov 22  09:18:32 & & 2008 Nov  23 16:31:19 & 54792.4 -- 54793.7$^1$ &\\
         & Sz2  &    705021010       & 2010 Nov 25  01:11:14 & & 2010 Nov  26 00:20:19 & 55525.2 -- 55526.8 &\\
         & Sz3  &    904003010       & 2009 Dec 9   02:40:43 & & 2009 Dec  9  23:31:24 & 54439.2 -- 54439.8 &\\
         & Sz4  &    902002010       & 2007 Dec 5   04:57:55 & & 2007 Dec  6  02:40:14 & 55174.1 -- 55175.4 &\\
 \hline                                   
M87      & Sz   & 801038010       & 2006 Nov 29  22:02:65 & & 2006 Dec  2 03:02:24 & 54068.9 -- 54071.1 &\\
 \hline                                             
 \end{tabular}
 \end{table*}

%
%

\begin{table*}
 \caption{Best-fit parameters  of the $Suzaku$ and $Chandra$ spectra
 of 3C~454.3 in the 0.3$-$10~keV 
 range using the following five 
models$^\dagger$: tbabs*power, tbabs*bbody,  tbabs*(bbody+power),  tbabs*bmc  and  tbabs*(bmc+power). 
}
    \label{tab:par_suzaku}
 \centering 
 \begin{tabular}{lllllllll}
 \hline\hline
     & Parameter & 703006010 & 705021010 & 904003010 & 902002010 & 3127 & 4843 \\
 \hline                                             
Model &      &        &        &       &      &    &  & \\
 \hline                                             
tbabs       & N$_H$ ($\times 10^{21}$ cm$^{-2}$) & 2.8$\pm$0.7 & 3.28$\pm$0.06 & 3.1$\pm$0.06  & 3.2$\pm$0.4 & 0.5$\pm$0.1 & 0.8$\pm$0.4\\
Power-law   & $\Gamma_{pow}$    & 1.46$\pm$0.01 & 1.48$\pm$0.01 & 1.54$\pm$0.01 & 1.56$\pm$0.01 & 0.79$\pm$0.03 & 1.29$\pm$0.02\\
            & N$_{pow}^{\dagger\dagger}$ & 32.5$\pm$0.9 & 130$\pm$2 & 139$\pm$2 & 139.6$\pm$0.1 & 6.24$\pm$0.04 & 12.72$\pm$0.03\\
      \hline
           & $\chi_{red}^2$ {\footnotesize (d.o.f.)} & 1.23 (1058)     & 1.24 (1797)     & 1.06 (1449) & 2.00 (1643)  & 1.28 (282) & 1.36 (441) \\
      \hline
tbabs      & N$_H$ ($\times 10^{21}$ cm$^{-2}$) & 0.09$\pm$0.01 & 0.08$\pm$0.03 & 0.06$\pm$0.01  & 0.08$\pm$0.2 & 0.03$\pm$0.01 & 0.01$\pm$0.01\\
Bbody      & T$_{BB}$  (keV)   & 1.00$\pm$0.01   & 1.04$\pm$0.01  & 1.10$\pm$0.01    & 0.83$\pm$0.01  & 1.09$\pm$0.02 & 0.78$\pm$0.01\\
           & N$_{BB}^{\dagger\dagger}$ & 2.1$\pm$0.3 & 8.68$\pm$0.08 & 8.62$\pm$0.09 & 7.57$\pm$0.05 & 0.93$\pm$0.05 & 0.9$\pm$0.1\\
      \hline
           & $\chi_{red}^2$ {\footnotesize (d.o.f.)} &  4.76 (1058) & 13.2 (1797) & 6.96 (1449)& 13.08 (1643) & 2.59 (282) & 6.7 (441)\\
      \hline
tbabs      & N$_H$ ($\times 10^{21}$ cm$^{-2}$) & 5.5$\pm$0.2 & 5.4$\pm$0.1 & 5.3$\pm$0.4  & 5.8$\pm$0.6 & 0.85$\pm$0.02 & 0.9$\pm$0.3\\
Bbody      & T$_{BB}$ (keV)  & 2.46$\pm$0.09 & 2.75$\pm$0.08  & 2.16$\pm$0.05  & 2.04$\pm$0.3 & 1.9$\pm$0.6 & 5.7$\pm$0.4 \\
           & N$_{BB}^{\dagger\dagger}$ & 2.8$\pm$0.2  & 10.6$\pm$0.5   & 8.3$\pm$0.2   & 8.58$\pm$0.2 & 15$\pm$4 & 170$\pm$10\\
Power-law  & $\Gamma_{pow}$    & 2.74$\pm$0.08 & 2.43$\pm$0.05 & 2.59$\pm$0.06  & 2.85$\pm$0.05 & 1.19$\pm$0.09 & 1.44$\pm$0.05 \\
           & N$_{pow}^{\dagger\dagger}$ & 69$\pm$4 & 240$\pm$8 & 251$\pm$9  & 284$\pm$8 & 7.09$\pm$0.08 & 0.13$\pm$0.05 \\
      \hline
           & $\chi_{red}^2$ {\footnotesize (d.o.f.)}& 0.95 (1056) & 0.93 (1795) & 0.86 (1447) & 0.86 (1641) & 1.18 (280) & 1.27 (439)\\
      \hline
      \hline
tbabs      & N$_H$ ($\times 10^{21}$ cm$^{-2}$)  & 5.6$\pm$0.2 & 7.4$\pm$0.5 & 3.0$\pm$0.2  & 3.9$\pm$0.2 & 5.8$\pm$0.2 & 5.0$\pm$0.1 \\
bmc        & $\Gamma_{bmc}$    & 1.51$\pm$0.03   & 2.01$\pm$0.09& 1.99$\pm$0.02   & 1.98$\pm$0.01   & 1.11$\pm$0.03 & 1.24$\pm$0.02 \\
           & T$_{s}$   (eV)    &  180$\pm$50     & 132$\pm$20   & 120$\pm$9   &  210$\pm$10    & 260$\pm$20 & 100$\pm$10 \\
           & $\log A$            & -0.14$\pm$0.01  & -0.35$\pm$0.01& -0.80$\pm$0.01   & -0.41$\pm$0.04 & 2.$^{\dagger\dagger\dagger}$ & 2.$^{\dagger\dagger\dagger}$ \\
           & N$_{bmc}^{\dagger\dagger}$ & 2.51$\pm$0.09 & 18.0$\pm$0.1 & 20.3$\pm$0.1  & 11.2$\pm$0.1  & 1.1$\pm$0.1 & 0.9$\pm$0.1 \\
      \hline
           & $\chi_{red}^2$ {\footnotesize (d.o.f.)}& 1.19 (1056) & 1.06 (1795) & 1.03 (1447)& 1.02 (1641) & 1.05 (279) & 1.14 (438)\\
      \hline
 \hline                                             
tbabs      & N$_H$ ($\times 10^{21}$ cm$^{-2}$)  & 5.6$\pm$0.2 & 7.4$\pm$0.5 & 3.0$\pm$0.2  & 3.9$\pm$0.2 & 0.2$\pm$0.1 & 1.2$\pm$0.1\\
bmc        & $\Gamma_{bmc}$    & 1.51$\pm$0.03   & 2.01$\pm$0.09& 1.99$\pm$0.02   & 1.98$\pm$0.01   & 1.12$\pm$0.02 & 1.36$\pm$0.06 \\
           & T$_{s}$   (eV)    &  170$\pm$40     & 130$\pm$10   & 120$\pm$9   &  200$\pm$10     & 112$\pm$2 & 104$\pm$2\\
           & $\log A$            & -0.18$\pm$0.02  & -0.37$\pm$0.04& -0.83$\pm$0.05   & -0.42$\pm$0.08 & 0.22$\pm$0.09 & 0.42$\pm$0.06 \\
           & N$_{bmc}^{\dagger\dagger}$ & 2.51$\pm$0.09 & 17$\pm$1 & 20.2$\pm$0.3  & 11.3$\pm$0.2 & 0.76$\pm$0.09 & 0.5$\pm$0.1 \\
Power-law  & $\Gamma_{pow}$    & 2.83$\pm$0.07 & 2.49$\pm$0.04 & 2.61$\pm$0.02  & 2.91$\pm$0.04 & -2.5$\pm$0.3 & -2.5$\pm$0.4\\
           & N$_{pow}^{\dagger\dagger}$ & 70$\pm$3 & 330$\pm$5 & 139$\pm$  & 98$\pm$5 & 0.01$\pm$0.01 & 0.01$\pm$0.01\\
      \hline
           & $\chi_{red}^2$ {\footnotesize (d.o.f.)}& 10.08 (1054) & 1.02 (1793) & 1.00 (1445)& 1.03 (1639) & 1.09 (278) & 1.2 (437)\\
      \hline
 \hline                                             

 \end{tabular}
\tablefoot{ 
$^\dagger$     Errors are given at the 90\% confidence level. 
$^{\dagger\dagger}$ Normalization parameters of blackbody and bmc components are in units of $L^{soft}_{35}/d^2_{10}$ erg s$^{-1}$ kpc$^{-2}$, 
where $L^{soft}_{35}$ is  soft photon luminosity in units of $10^{35}$ erg s$^{-1}$, $d_{10}$ is the distance to the 
source in units of 10 kpc. 
$^{\dagger\dagger\dagger}$  when parameter $\log(A)\ge 2$ we fixed this parameter at $\log(A)=2$. 
$T_{BB}$ and $T_{s}$ are the temperatures of 
the blackbody and seed photon components, respectively (in keV and eV). 
$\Gamma_{pow}$ and $\Gamma_{bmc}$ are the indices of the { power law} 
and BMC, respectively. 
}
 \end{table*}


\begin{table*}
 \caption{List of $ASCA$ and $Beppo$SAX  observations of  M87 used in our analysis}              
 \label{tab:list_asca+sax}      
 \centering                                      
 \begin{tabular}{l l l l l c c}          
 \hline\hline                        
Satellite  &  Obs. ID        & Start time (UT)  && End time (UT) &MJD interval \\    
 \hline                                   
$ASCA$     & 60033000        & 1993 Jun 7   19:15:48 & & 1993 Jun  8 07:10:36 & 49145.8 -- 49146.3 &\\
$Beppo$SAX & 60010001        & 1996 Jul 14  21:31:00 & & 1996 Jul 15 10:57:10 & 50278.8 -- 50279.4 &\\
 \hline                                             
 \end{tabular}
 \end{table*}

%
%
\begin{table*}
 \caption{Best-fit parameters  of the $Beppo$SAX and $ASCA$  spectra of M87 in the 0.3$-$10~keV 
 range$^\dagger$. 
    Errors are given at the 90\% confidence level.
}
    \label{tab:par_SAX+asca_m87}
 \centering 
 \begin{tabular}{lllllllllllllll}
 \hline\hline
      ID         & $\alpha=$       & $T_s$       &  $log(A)$           & $N_{bmc}^{\dagger\dagger}$ & $E_{cut}$ & $E_{fold}$ &  $E_{Fe}$ & $\sigma_{Fe}$ &  $N_{Fe}^{\dagger\dagger}$ & $\chi_{red}$   \\
                  & $\Gamma-1$                   &  (eV)         &                         &                                              & (keV)        & (keV)        &  (keV)       & (keV)  &     & (dof)   \\
      \hline
6001000     & 1.9$\pm$0.1 & 140$\pm$20 & 0.72$\pm$0.6 &  1.1$\pm$0.1    & 3.6$\pm$0.1 & 9.8$\pm$0.4 &  6.5$\pm$0.1 & 0.7$\pm$0.1 &  0.10$\pm$0.01 & 0.95 (40)   \\
60033000  & 1.4$\pm$0.1  & 139$\pm$30 & 2.00$^{\dagger\dagger\dagger}$ &  0.59$\pm$0.08 & 3.5$\pm$0.1 & 9.9$\pm$0.6 &  6.4$\pm$0.1 &  0.6$\pm$0.1&  0.09$\pm$0.01 &0.95 (40)    \\
      \hline
 \hline                                             
 \end{tabular}
\tablefoot{ 
$^\dagger$ The spectral model is  tbabs*(bmc+N*Gauss);  where $N_H$ is fixed at 
a value 5.0$\times 10^{21}$ cm$^{-2}$  (see Sect.~\ref{model choice}); 
$^{\dagger\dagger}$ 
for normalization parameter $N_{BMC}=L_{37}/d^2_{10}$ 
where 
$L_{37}$ is the source luminosity in units of 10$^{37}$ erg/s and  
$d_{10}$ is the distance to the source in units of 10 kpc;
$^{\dagger\dagger\dagger}$ when parameter $\log(A_2)\gg1$, this parameter is fixed at 2.0 (see comments in the text). 
}
 \end{table*}


\begin{table*}
 \caption{List of $Swift$ observations of  3C~454.3 used in our analysis}              
 \label{tab:list_Swift}      
 \centering                                      
 \begin{tabular}{l l l l l c}          
 \hline\hline                        
  Obs. ID& Start time (UT)  && End time (UT) &MJD interval \\    
 \hline                                   
00030024(001,002)   & 2005 May 11           && 2005 May 19    & 53501 -- 53509 &\\
00031018(001,003-008)& 2007 Nov 15          && 2007 Dec 15    & 54419 -- 54449    & \\
00031216(001-011,013-016,018-027,029-044,046-063) & 2008 May 27 && 2008 Oct 2& 54613 -- 54741 & \\
00031493(003,004)   & 2009 Sep 18           && 2009 Sep 19    & 55092 -- 55093 & \\
  00035030(001,003,005-010, 013-016,027-067) & 2005 Apr 24 && 2009 Sep 16 & 53484 -- 55090 &\\
  00035030(113,114,175,178,180-189,190-193,195-204) & 2010 Nov 3   && 2011 Jun 17 &   55503 -- 55729 &\\  
  00090023(001-008)  & 2008 Sep 9           && 2009 Jun 1          & 54718 -- 54983 &\\
  00090081(001,002) & 2009 Aug 13 12:56:32  && 2009 Sep 13 20:37:05& 55056 -- 55087 &\\
 \hline                                             
 \end{tabular}
 \end{table*}

%
%



\subsection{ {\it Suzaku} data 
\label{suzaku data}}

$Suzaku$ observed 3C~454.5  throughout the time period from 2008 to 2010  and investigated M87 in 2006. 
Table~\ref{tab:list_suzaku} summarizes the start time, end time, and the 
MJD interval  for each of these observations.
One can see a description of the {\it Suzaku} experiment in Mitsuda et al. (2007). 
For observations obtained using a focal X-ray CCD camera (XIS, X-ray Imaging Spectrometer, Koyama et al. 2007), 
which is sensitive over  the 0.3--12~keV range, we used  software of  the {\it Suzaku} data processing 
{\tt pipeline} (ver. 2.2.11.22).  We carried out the data reduction and analysis following the standard procedure using the latest {\tt HEASOFT software package} (version 6.25)  and following  the {\it Suzaku} 
Data Reduction Guide\footnote{http://heasarc.gsfc.nasa.gov/docs/suzaku/analysis/}. 
The spectra of the sources were extracted using spatial regions 
centered on the nominal positions of 3C~454.3 and M87
($\alpha=22^{h}53^{m}57^s.77$, $\delta=+16^{\circ} 08{\tt '} 53{\tt ''}.6$, J2000.0  for 3C~454.3 and $\alpha=12^{h}30^{m}49^s.42$, $\delta=+12^{\circ} 23{\tt '} 28{\tt ''}.04$, J2000.0 for M87),  while a background was extracted from source-free regions which have a comparable size away from the source. 
The spectrum data were re-binned to provide at least 20 counts per spectral bin to validate the use of the $\chi^2$-statistic. We carried out  spectral fitting  applying XSPEC v12.10.1. 
 The energy ranges around 1.75 and 2.23 keV were not  used for spectral fitting because of the known artificial structures in the XIS spectra around the Si and Au edges.
 Therefore, for spectral fits we chose  the 0.3 -- 10 keV  range  for the XISs 
(excluding 1.75 and 2.23 keV points). 

The source count rate was variable by 40\%   for 3C~454.3 and 60\% for M87.  We fitted the spectral data using a number of models (see Sec.~3.2.1) but the best-fits are obtained using  the  {\tt XSPEC} BMC model  (see Titarchuk et al. 1997) modified by 
 absorption by neutral gas of the solar composition  
(the  {\tt XSPEC} $tbabs$ model). 
Using this model  we found that the amplitude of X-ray flux variability  changed by up to a factor of two. 
The results of the fits are given in Table~\ref{tab:par_suzaku}. 

\subsection{\it ASCA data \label{asca data}}

{\it ASCA} observed M87 on  June 7--8, 1993. 
Table~\ref{tab:list_asca+sax} summarizes the start time, end time, and the MJD interval  for this  observation.
A description of the {\it ASCA} experiment can be found in Tanaka, Inoue, \& Holt (1994). 
The solid imaging spectrometers (SIS)  operated in Faint CCD-2 mode.
The {\it ASCA} data were screened using the ftool ascascreen and the standard screening criteria. The
spectrum for the source was extracted using spatial regions with a diameter of 3${\tt '}$ (for SISs) and 4${\tt '}$ (for GISs)
centered on the nominal position of M87, 
while background was extracted from source-free regions of comparable size away from the source. The spectrum data 
were rebinned to provide at least 20 counts per spectral bin to validate  the use of the $\chi^2$-statistic. The SIS and GIS data were fitted applying {\tt
XSPEC} in the energy ranges of 0.6 -- 10 keV and 0.8 -- 10 keV, where the spectral responses
are best known. 

\subsection{\it BeppoSAX data \label{sax data}}

We used $Beppo$SAX data 
of M87 carried out on June 14--15, 1996. 
In Table~\ref{tab:par_SAX+asca_m87} we show the log of the {\it Beppo}SAX 
observation analyzed  in this paper. 
Generally, broadband energy spectra of M87 Lac were obtained combining data from  three 
{\it Beppo}SAX narrow-field instruments (NFIs): the Low Energy Concentrator
Spectrometer [LECS; \citet{Parmar97}] for the 0.3 -- 4 keV range, the Medium Energy Concentrator Spectrometer
[MECS; \citet{boel97}] for the 1.8 -- 10 keV range,  and the Phoswich Detection
System [PDS; \citet{fron97}] for the 15 -- 200 keV range. 
The SAXDAS data analysis package is used for the data processing. 
We performed a spectral analysis for each of the instruments in a corresponding 
energy range within 
which a response matrix is well specified. 
The LECS data have been renormalized to match the MECS data. Relative normalizations of the NFIs were treated 
as free parameters in the
 model fits, except for the MECS normalization that was fixed at unity.  
While the source is bright and background is low and stable, we checked its uniform distribution 
across the detectors. Furthermore, we extracted a light curve from a source-free region far from source and 
stated that background was not varying during the whole observation. 
Additionally, 
spectra were rebinned  in accordance with 
the energy resolution of the instruments 
using
 rebinning template files
 in GRPPHA of
 XSPEC\footnote{http://heasarc.gsfc.nasa.gov/FTP/sax/cal/responses/grouping} 
to obtain a better signal-to-noise ratio. 
Systematic uncertainties  of 1\% 
were applied to these analyzed spectra. 

\subsection{{\it Swift} data\label{swift data}}

We used the $Swift$ observation of 3C~454.3 and M87  carried out from 2005 to 2011 (Table~\ref{tab:list_Swift}). 
The {\it Swift}/XRT data were taken in 
Photon-Counting (PC) and Windowed Timing (WT) modes. 
The PC mode is characterized by high sensitivity but can be affected by the photon pile-up effect when the count rate is higher than $\sim$0.5 cts/s, while  the WT mode does not suffer from a photon pile-up effect up to $\sim$200 cts/s. 
Data were processed using the HEASOFT v6.14, the tool XRTPIPELINE v0.12.84, and the
calibration files (CALDB version 4.1).
The ancillary response files were created using XRTMKARF v0.6.0 and exposure maps generated by XRTEXPOMAP v0.2.7. 
We selected source events for which  the grades are accumulated
in the range 0$-$12 and used default screening parameters 
to produce level 2  cleaned event files. To avoid photon pile-up effect during high-count-rate events, 
we extract the spectral data obtained in the PC mode  in an annular region with inner and outer radii 
of 6 and 20 pixels, respectively (Vaughan et al. 2005). The background was estimated in a nearby source-free 
circular region of 50 pixels in radius. For the WT data mode we extracted spectra in a 40$\times$20 pixel rectangular 
region centered on the nominal position of 3C~454.3. The background was estimated in a nearby 50$\times$20 
pixel source-free rectangular region. Spectra were re-binned to include at least 20 photons in each energy channel 
in order to use $\chi^2$-statistics. 
We fitted the spectrum using the response file SWXPC\-0TO12S6$\_$20010101v012.RMF.
We also used the online XRT data product generator\footnote{http://www.swift.ac.uk/user\_objects/} to obtain
the image, 
light curves, and spectra (including background and ancillary response files;  
Evans et al. 2007, 2009). 
\begin{table*}
 \caption{List of $Chandra$ observations of 3C~454.3 and M87}
 \label{tab:list_Chandra}
 \centering 
 \begin{tabular}{l l l l r l l l l}
 \hline\hline                        
Object & Obs. ID & Start time (UT) & MJD &  Exposure time (s) \\ 
 \hline                                   
3C~454.3 & 3127$^{1}$ & 2002-11-06 21:26:22 & 52584 & 4930  \\
               & 4843$^{1}$   & 2004-10-29 05:01:50 & 53221 & 18261 \\
 \hline                                             
M87 & 11514$^{2,3}$ & 2010-04-15 20:32:42 & 55301 & 4528  \\
       & 18783$^{2,3}$ & 2016-04-20 08:32:11 & 55301 & 36114  \\
       & 21075$^{2,3}$ & 2018-04-22 03:31:16 & 58230 & 9132  \\
 \hline                                             
 \end{tabular}
 \tablebib{
(1) Gupta et al. 2017; 
(2) Abeamowski et al. 2012;
(3) Harris et al. 2009.
}
 \end{table*}



\begin{table*}
 \caption{List of {\it RXTE} observations of 3C~454.3  and M87}
 \label{tab:list_RXTE}
 \centering 
 \begin{tabular}{l l c c c }
 \hline\hline                        
Object & Number of set  & Dates, MJD & RXTE Proposal ID&  Dates UT \\
 \hline    
3C 454.3 & R1  &    50181--50358      & 10360$^1$    & Apr 8 -- Oct 29, 1996       \\
 & R2  &    50389--50802      & 20346$^1$    & Nov 2, 1996 -- Dec 20, 1997   \\
 & R3  &    50816--50986      & 30264$^1$    & Jan 3 -- Jun 22, 1998  \\
 & R4  &    54309--54636      & 93150$^1$    & July 28, 2007 -- June 19, 2008   \\
 & R5  &    55170--55182      & 94150$^1$    & Dec 5  -- Dec 17, 2009    \\
 & R6  &    55293--55297      & 95149$^1$    & Apr 7  -- Apr 11, 2010    \\
 \hline                                             
M87 & R$^{m87}_1$  &    50812--50846      & 30216$^2$    & Dec 30, 1997 -- Feb 2, 1998       \\
 & R$^{m87}_2$  &    55296--55301      & 95145$^2$    & Apr 10 -- Apr 15, 2010   \\
 \hline                                             
 \end{tabular}
   \label{tab_rxte}
 \tablebib{
(1) Rivers et al. 2013. 
(2) Jorstad et al. 2010. 
}
 \end{table*}

%
%

\begin{table*}
 \caption{Best-fit parameters  of the $Chandra$  spectra of M87 in the 0.3$-$10~keV 
 range using the following four 
models$^\dagger$: tbabs*power, tbabs*bbody,  tbabs*(bbody+power) and  tbabs*bmc. 
}
    \label{tab:par_chandra}
 \centering 
 \begin{tabular}{lllllllll}
 \hline\hline
     & Parameter & 11514 & 18783 & 21075  \\
 \hline                                             
Model &      &        &        &       &      &    \\
 \hline                                             
tbabs       & N$_H$ ($\times 10^{20}$ cm$^{-2}$) & 4.03$\pm$0.01 & 5.4$\pm$0.6 & 7.2$\pm$0.1  \\
Power-law   & $\Gamma_{pow}$    & 1.98$\pm$0.06 & 2.18$\pm$0.07 & 1.96$\pm$0.07  \\
            & N$_{pow}^{\dagger\dagger}$ & 4.3$\pm$0.3 & 1.9$\pm$0.2 & 7.1$\pm$0.1 \\
      \hline
           & $\chi_{red}^2$ {\footnotesize (d.o.f.)} & 0.76 (113)     & 0.62 (174)     & 0.65 (176) \\
      \hline
tbabs      & N$_H$ ($\times 10^{20}$ cm$^{-2}$) & 0.09$\pm$0.01 & 0.08$\pm$0.03 & 0.06$\pm$0.01  \\
Bbody      & T$_{BB}$  (keV)   & 0.47$\pm$0.02   & 0.50$\pm$0.01  & 0.62$\pm$0.02    \\
           & N$_{BB}^{\dagger\dagger}$ & 0.14$\pm$0.09 & 0.07$\pm$0.01 & 0.27$\pm$0.03 \\
      \hline
           & $\chi_{red}^2$ {\footnotesize (d.o.f.)} &  2.73 (113) & 1.95 (174) & 1.49 (176) \\
      \hline
tbabs      & N$_H$ ($\times 10^{20}$ cm$^{-2}$) & 3.9$\pm$0.6 & 3.3$\pm$0.1 & 3.9$\pm$0.6  \\
Bbody      & T$_{BB}$ (keV)  & 0.45$\pm$0.09 & 0.28$\pm$0.03  & 0.44$\pm$0.09    \\
           & N$_{BB}^{\dagger\dagger}$ & 0.10$\pm$0.08  & 0.02$\pm$0.01  & 0.05$\pm$0.01    \\
Power-law  & $\Gamma_{pow}$    & 1.96$\pm$0.09 & 1.88$\pm$0.05 & 1.83$\pm$0.09    \\
           & N$_{pow}^{\dagger\dagger}$ & 4.4$\pm$0.3 & 1.26$\pm$0.6 & 5.2$\pm$0.1   \\
      \hline
           & $\chi_{red}^2$ {\footnotesize (d.o.f.)}& 0.79 (111) & 0.59 (172) & 0.86 (174) \\
      \hline
      \hline
tbabs      & N$_H$ ($\times 10^{20}$ cm$^{-2}$)  & 1.0$\pm$0.1 & 7.4$\pm$0.5 &1.0$\pm$0.2 \\
bmc        & $\Gamma_{bmc}$    & 1.94$\pm$0.07   & 2.83$\pm$0.09& 2.09$\pm$0.07  \\
           & T$_{s}$   (eV)    &  136$\pm$7     & 168$\pm$10   & 220$\pm$9  \\
           & $\log A$            & 1.08$\pm$0.02  & -2.29$\pm$0.07& 1.47$\pm$0.05 \\
           & N$_{bmc}^{\dagger\dagger}$ & 0.10$\pm$0.04 & 0.50$\pm$0.04 & 0.17$\pm$0.05 \\
      \hline
           & $\chi_{red}^2$ {\footnotesize (d.o.f.)}& 0.87 (111) & 1.01 (172) & 0.96 (174)\\
      \hline
 \hline                                             

 \end{tabular}
\tablefoot{ 
$^\dagger$     Errors are given at the 90\% confidence level. 
$^{\dagger\dagger}$ Normalization parameters of blackbody and bmc components are in units of $L^{soft}_{35}/d^2_{10}$ erg s$^{-1}$ kpc$^{-2}$, 
where $L^{soft}_{35}$ is  soft photon luminosity in units of $10^{35}$ erg s$^{-1}$, $d_{10}$ is the distance to the 
source in units of 10 kpc. 
$T_{BB}$ and $T_{s}$ are the temperatures of 
the blackbody and seed photon components, respectively (in keV and eV). 
$\Gamma_{pow}$ and $\Gamma_{bmc}$ are the indices of the { power law} 
and bmc, respectively. 
%
}
 \end{table*}

\subsection{{\it Chandra} data \label{chandra data}}

The object 3C~454.3 was also observed by {\it Chandra} in 2002 and 2004, and M87 was extensively 
investigated by  {\it Chandra}: here we apply  observations of 2010, 2016, and 2018. 
The used log of {\it Chandra} observations   is presented 
in Table~\ref{tab:list_Chandra}. 
We extracted spectra from the ACIS-S detector
via the standard pipeline CIAO v4.5 package and calibration database CALDB 2.27.
In Figure \ref{imagea} we demonstrate the Chandra/ASIS (0.3$-$10 keV) image of the 3C~454.3 field of view on November 6, 2002, 
with the exposure of 4929 s (MJD=52584).
We also identified  intervals of high background level to exclude all 
high-background events. 
 The {\it Chandra} spectra were produced and  
modeled over the 
0.3$-$7.0 keV energy range.


\subsection{\it RXTE data \label{rxte data}}
For our analysis, we also applied publicly available data of the {\it RXTE}  \citep{bradt93}   
from October 1996  to December 2009   (for 3C454.3) and  from December  1997 to February 1998 as well as April 2010 (for M87). 
These data consist  of 160 observations related to   the different spectral states of the source.
For data processing we  utilized standard tasks of the LHEASOFT/FTOOLS 6.25 software package.
Spectral analysis was implemented  using  PCA {Standard 2} mode data, collected 
in the 3$-$22~keV  range.  The standard dead-time correction procedure
has been applied to these data. 
 The data are available through the GSFC public archive\footnote{http://heasarc.gsfc.nasa.gov}. 
 We modeled the {\it RXTE}  spectra using XSPEC astrophysical fitting software and
implemented a systematic uncertainty  of 0.5\%  to all  analyzed spectra. 
In  Table~\ref{tab:list_RXTE}
we present a list of  the  groups
 of  observations which covers the complete sample of the state evolution of the source. 
In Figure \ref{RXTE evol} we show an evolution of ASM/{\it RXTE} count rate (2-12 keV, top), radio (8 GHz, middle), and 
$\gamma$-ray radiation (0.1--300 GeV, bottom) during 1996 -- 2012 observations of 3C~454.3.
Vertical arrows (at  top  of the panels) indicate temporal distribution of the {\it RXTE} 
(blue) and $Suzaku$ (green) 
data sets listed in Tables \ref{tab:list_suzaku}$-$\ref{tab:list_RXTE}. 

{ 

\subsection{{\it FERMI/LAT} data \label{fermi data}}

Gamma-ray data  observed with Fermi/LAT were obtained from MJD 54690 to  58446
\footnote{http://fermi.gsfc.nasa.gov/ssc/data/analysis/scitools/extract \_latdata.html}. The gamma-ray source 
1FGL2253.9+1608 was positionally identified with 3C~454.3~\citep{sasada10}.  Each point on the $\gamma$-ray 
light curve (Fig.~\ref{no_delay_3c454})
corresponds to a daily averaged photon flux integrated over energies from 
100 MeV to 300 GeV.  This plot  shows that there is no correlation between gamma and X-ray flashes in  3C~454.3 observed in 2010--2012  for two bands: 2--12 keV (top) 
and 0.1--300 GeV (bottom).



}

\section{Results \label{results}}

\subsection{Images and light curves of 3C~454.3  and M87 \label{lc}}

For a deep image analysis, 
we used the {\it Chandra} images with sub-arcsecond-resolution data quality provided by ACIS-S onboard {\it Chandra}.  
The {\it Chandra}/ACIS-S (0.2-8 keV) image obtained during observations of 3C~454.3 on November 6, 2002, 
(with exposure of 5 ks, ObsID=3127) is shown in Fig. \ref{imagea}.  
For each observation, we extracted the source spectrum from a
circular region of  1.25{\tt "}
radius centered on the source position of 3C~454.3 [$\alpha=22^{h}53^{m}57^s.7$, 
$\delta=16^{\circ} 08{\tt '} 53{\tt ''}.6$, J2000.0  (see details in \citep{Abdo10}) and 0.75{\tt "} radius circular region centered on the source position of M87 ($\alpha=12^{h}30^{m}49^s.4$, 
$\delta=12^{\circ} 23{\tt '} 28{\tt ''}$, J2000.0  (see details in \cite{Akiy19}). For comparison in Figure \ref{imageb} we show an image of the same area around 3C~454.3 in the radio band (15 GHz, VLBA) at the time of our observations (see $R2$  and Kellermann et al., 1998). It can be seen that even during a radio outburst with jet ejection,  we observe a point source in  X-rays. 
$Swift$ data processing is  described in our previous paper (Titarchuk \& Seifina, 2016). 

The {\it Chandra} (0.3 -- 10 keV) image of the 3C~454.3 field of view is shown in 
Fig.~\ref{imagea}. 
In the center we can see a bright point-like X-ray source, which is well matched 
with the optical one ($\alpha=23^{h}53^{m}57^s.77$, $\delta=+16^{\circ} 08{\tt '} 52{\tt ''}.7$, J2000.0).  
This image was obtained during observations of 3C~454.3 between May 11, 2005, and June 17, 2011 
(with exposure time of 277 ks). 
Figure 3 also demonstrates  a very weak signature of the X-ray jet-like (elongated) structure  
and the minimal contamination by other point sources and diffuse emission around 3C~454.3. 

 On the other hand  the jet-like structure in M87 is strong and knotted. The knot, which is close to the nucleus of M87 (so called as HST--1), 
is variable. Figure~\ref{chandra_ima_M87}
shows a $Chandra$ X-ray image on April 14, 2017, 
with the strong HST--1 for 5 ks exposure. Here, we indicate the nucleus, the flaring knot HST-1 (0.86" from the nucleus), knot D, and knot A.



Before proceeding to details of the spectral fitting we study the long-term behavior of 3C~454.3 and M87,  particularly in its active phases. We discuss a long-term one-day average X-ray 
light curve of 3C~454.3 detected by the {\it RXTE}  ASM 
over the total lifetime of the mission (1996 -- 2012, 
Fig.~\ref{RXTE evol}). 
{ The bright-blue rectangles  indicate the MJD intervals of {\it RXTE} observations used in our analysis.} 
Vertical green arrows (top of the panel) indicate temporal distribution of the {\it Suzaku} data sets listed in Table~\ref{tab:list_suzaku}.
It is clear that 3C~454.3 became brighter, 
on average, in 2009 -- 2010 ($R5$, $R6$) in a soft X-ray band (1--12 keV)  than in the 1996 -- 1997 period 
(see $R1$, $R2$). 
{Blue} points show  the source signal 
and {red} line indicates its error bars. The  ASM monitoring observations are 
distributed more densely over time (1996 -- 2012) than those of $Suzaku$ (2007 -- 2010, Abdo et al. 2010). However, it can be seen that the $Suzaku$ observations 
cover the time interval with increased X-ray count-rate events. 

 We also  wish to bring attention to  the synchronous light curves in radio (8 GHz, middle panel in Fig. \ref{RXTE evol}) and 
$\gamma$-ray bands (0.1--300 GeV, bottom panel) during 1996 -- 2012 observations of 3C~454.3.
(green) 
It can be seen here that although the radio flash of 3C~454.3 (middle panel) correlates with the amplification of the $\gamma-$ray radiation (bottom panel) from the source, the X-ray light curve of 3C~454.3 (top panel) shows a more frequent irregular variability, which does not correlate with global outburst in the radio and gamma-ray ranges.
 
In Figure \ref{RXTE evol_m87} we display the evolution of ASM/{\it RXTE} count rate during 1996 -- 2012 observations of M87 (top).
Vertical arrows (at the top) indicate temporal distribution of the {\it RXTE} (blue) and 
$Suzaku$ (green) 
data sets listed in Tables \ref{tab:list_suzaku}$-$\ref{tab:list_RXTE}. 
We also  demonstrate the evolution of the X-ray intensity of the nucleus
(green) and HST-1 (black) seen by $Chandra$  during 2000 -- 2008 observations of M87 according to data taken from Harris et al. (2009) in the lower panel.


Regarding an issue of possible correlation between flares in the X-ray and $\gamma$-ray ranges (delay/lead), we 
did not find any correlation for the data of 3C~454.3 observed in 2010--2012 (Fig.~\ref{no_delay_3c454})  
for two bands: 2--12 keV (top) 
and 0.1--300 GeV (bottom). In Fig. \ref{no_delay_3c454} we mark the peak of 
$\gamma$-ray flux by vertical blue dashed lines, 
 while pink 
arrows (at  top of the panels) indicate 
the peaks of X-ray 
bursts.
Obviously, there is no correlation between $\gamma$ and X-ray flashes. Our conclusion is somewhat different from the 
conclusions of Volvach et al. (2019), but we may have used a different set of data.

\subsection{Spectral Analysis \label{spectral analysis}}

We used different spectral models in order to test them using all available data  
of 3C~454.3 and M87. We want to establish  evolution between the low/hard  state (LHS) and the intermediate state (IS) 
using  these spectral models.
We investigate  the  {\it Chandra}, $Susaku$, $Swift$ and $RXTE$ spectra 
to test  the following XSPEC spectral models: 
powerlaw, Bbody, BMC, and their possible combinations. 

\subsubsection{Choice of spectral model\label{model choice}}



As a first step, we present $Suzaku$ and $Chandra$ 
spectral data of 3C~454.3 in the 0.3 -- 10 keV energy range.  
We establish that the thermal model (black body) fits the low-energy part well, while providing an 
excess emission for $E>$~3 keV (e.g., for $Sz2$ and $Sz4$ 
spectrum, $\chi^2_{red}$=13.2 (1797 d.o.f.) and  $\chi^2_{red}$=13.08 (1643), respectively, at 
the  top part of Table~\ref{tab:par_suzaku}). We find that the  black body  model indicates very low 
absorption (less than $0.9\times 10^{20}$ cm$^{-2}$) for all $Sz1-Sz4$ 
spectra, 
and moreover, that this model  gives unacceptable fit quality, $\chi^2,$  
 for all spectra of $Swift$ data. 
However, we should note that the {\tt tbabs*power-law} model provides better fits than 
  the thermal one. No {\it Susaku} data can be fitted well by a single-component model. 
Indeed, a simple power-law model produces a soft excess below 0.6 keV. These significant positive 
residuals at low energies, less than 1.3 keV, suggest the presence of additional emission  components 
in the spectrum. Therefore,  we also tested a model consisting of blackbody and power-law components. The model 
parameters of this combined model are  $N_H=5\times 10^{21}$ cm$^{-2}$; 
$kT_{bb}=100-260$
eV and $\Gamma=1.1 -2$. 
(see more 
in Table~\ref{tab:par_suzaku}). 
The best fits of the {\it Suzaku} spectra were found  using of  the  
{Bulk Motion Comptonization model} ({BMC XSPEC} model, \cite{tl97}).   
We should emphasize that all $Suzaku$ best-fit results were obtained using the same model for 
 all spectral  states. Therefore, further we applied the BMC model modified by the interstellar absorption 
 to all our $Swift$ and $RXTE$ observations of 3C~454.3.

 A similar result concerning the choice of spectral model was obtained for M87  using the Chandra spectra of the M87 core (see Table~\ref{tab:par_chandra}). Therefore, further we applied the BMC model modified by the interstellar absorption 
 to all our Beppo$SAX$, $Suzaku$, $ASCA$, $Swift$ and $RXTE$ observations of M87. In Figure~\ref{sp_m87_BeppoSAX} we show one  EF(E) spectral diagram for the IS using the {\it Beppo}SAX data. 
Interestingly, we found many emission features in the $Beppo$SAX spectrum and fitted the spectrum with the {\tt tbabs * (bmc + N * Gauss)} model. However, these features may be associated with a remote shell. In fact, $Beppo$SAX data come from a vast area that includes both the nucleus and the jet and from the external parts of the AGN. Moreover, the spectra obtained using $Chandra$ for a narrower 
region of the nucleus of M87 (<0.85{\tt "}) no longer contain emission features
 (see Fig.~\ref{3_chandra_sp_m87}). 

The same conclusion was drawn for the XMM observations~\citep{Kennea2000}. Therefore, we continued to work
with the Chandra spectra in order to select a model for fitting the radiation of the nucleus of M87. We also proceeded with a powerlaw model; it fits the spectrum well, but the data are poorly matched by the $\chi$-square criterion ($\chi^2_{red}=0.62$ for 174 dof, ID=18783, see top of the Table~\ref{tab:par_chandra}). The thermal black body model also fits the spectrum well, however it poorly approximates the regions of 0.3--1 keV (possibly related to the underestimation of absorption) and the energy region above 
7 keV ($\chi^2_{red}=1.5$ for 176 dof, ID=21075). In this case, taking absorption into account  unnecessary. It is worth noting  that the above models provide a soft excess below 0.3 keV. To eliminate this drawback, we used a combined model consisting of 
black body and power-law components. This latter model also  fits the spectrum of the M87 nucleus well, but this combined model is more phenomenological than the physical one. Therefore, in the subsequent step, we used the generic BMC model. This model describes the spectra of the M87 nucleus well in all states (see bottom in the Table~\ref{tab:par_chandra}).


It is important  to emphasize that the log-parabolic power-law models (Kardashev, 1962; Massaro et al., 2004a,b, 2006) were used for  analysis of the blazar spectra.
{     
Indeed, the log-parabolic model is one of the simplest ways to describe curved spectra when these show mild and nearly symmetric curvature around the maximum, instead of a sharp high-energy cut-off like that of an exponential. 
As it is described by  Massaro et al. (2006) the log-parabolic model law has only one more parameter than a simple power-law.   For example, one can see  it applying Eq. (3)  of Massaro et al. (2004) who apply this formula  for photon spectra.  This appropriate formula is 
\begin{equation}
F(E) = K(E/E_1)^{-[a + b\lg(E/E_1)]}. 
\label{1}
\end{equation}
in which the reference energy $E_1$ is fixed at 1 keV and thus the spectrum is determined only by three parameters $K$, $a$ and $b$ and the energy dependent photon index is
\begin{equation}
\Gamma(E) = a +  b \lg(E/E_1). 
\label{2}
\end{equation}



}

However, it is easy to show that this log-parabolic model is approximated by a single power-law for large E and frozen pivotE. In fact, if the index of this simplified  broken power-law and log-parabolic model is too large then it leads to  very high residuals at low-energies. This is a well-known effect of this simplification. 
{
The log-parabolic model often uses a  $N_H$  fixed at galactic values (Massaro et al., 2004), and for sources where the internal absorption is not relevant. In fact, this is one of the motivations for not using this model in the context of the present analysis.
}

There are several advantages to using the BMC model with respect to other common approaches applied to studies of X-ray spectra of accreting compact objects, including a broken power-law and the log-parabolic model. First, the BMC is by nature applicable to the general case where there is an energy gain through not only thermal Comptonization but also via dynamic (bulk) motion Comptonization (see Titarchuk et al. 1997; Laurent \& Titarchuk 1999; Shaposhnikov \& Titarchuk 2006, for details). Second, with respect to the  log-parabolic model, the BMC spectral shape has an appropriate low-energy curvature, which is essential for a correct representation of the lower-energy part of the spectrum. Long-term experience with log-parabolic components shows that the model fit with this component is often inconsistent with the $N_H$ column values and produces an unphysical component ``conspiracy'' with the highecut part. Specifically, when a multiplicative component highecut is combined with the BMC, the cutoff energies $E_{cut}$ are in the expected range of 20--30 keV, while in a combination with the log-parabolic, $E_{cut}$ often goes below 10 keV, resulting in unreasonably low values of the photon index. Furthermore,   implementation of the  log-parabolic model makes it much harder or even impossible to correctly identify the spectral state of the source, which is an imminent task for our study.  An  even more important property of the BMC model is that it consistently  calculate\subsection{Spectral Analysis \label{spectral analysis}}

We used different spectral models in order to test them using all available data  
of 3C~454.3 and M87. We want to establish  evolution between the low/hard  state (LHS) and the intermediate state (IS) 
using  these spectral models.
We investigate  the  {\it Chandra}, $Susaku$, $Swift$ and $RXTE$ spectra 
to test  the following XSPEC spectral models: 
powerlaw, Bbody, BMC, and their possible combinations. 

\subsubsection{Choice of spectral model\label{model choice}}



As a first step, we present $Suzaku$ and $Chandra$ 
spectral data of 3C~454.3 in the 0.3 -- 10 keV energy range.  
We establish that the thermal model (black body) fits the low-energy part well, while providing an 
excess emission for $E>$~3 keV (e.g., for $Sz2$ and $Sz4$ 
spectrum, $\chi^2_{red}$=13.2 (1797 d.o.f.) and  $\chi^2_{red}$=13.08 (1643), respectively, at 
the  top part of Table~\ref{tab:par_suzaku}). We find that the  black body  model indicates very low 
absorption (less than $0.9\times 10^{20}$ cm$^{-2}$) for all $Sz1-Sz4$ 
spectra, 
and moreover, that this model  gives unacceptable fit quality, $\chi^2,$  
 for all spectra of $Swift$ data. 
However, we should note that the {\tt tbabs*power-law} model provides better fits than 
  the thermal one. No {\it Susaku} data can be fitted well by a single-component model. 
Indeed, a simple power-law model produces a soft excess below 0.6 keV. These significant positive 
residuals at low energies, less than 1.3 keV, suggest the presence of additional emission  components 
in the spectrum. Therefore,  we also tested a model consisting of blackbody and power-law components. The model 
parameters of this combined model are  $N_H=5\times 10^{21}$ cm$^{-2}$; 
$kT_{bb}=100-260$
eV and $\Gamma=1.1 -2$. 
(see more 
in Table~\ref{tab:par_suzaku}). 
The best fits of the {\it Suzaku} spectra were found  using of  the  
{Bulk Motion Comptonization model} ({BMC XSPEC} model, \cite{tl97}).   
We should emphasize that all $Suzaku$ best-fit results were obtained using the same model for 
 all spectral  states. Therefore, further we applied the BMC model modified by the interstellar absorption 
 to all our $Swift$ and $RXTE$ observations of 3C~454.3.

 A similar result concerning the choice of spectral model was obtained for M87  using the Chandra spectra of the M87 core (see Table~\ref{tab:par_chandra}). Therefore, further we applied the BMC model modified by the interstellar absorption 
 to all our Beppo$SAX$, $Suzaku$, $ASCA$, $Swift$ and $RXTE$ observations of M87. In Figure~\ref{sp_m87_BeppoSAX} we show one  EF(E) spectral diagram for the IS using the {\it Beppo}SAX data. 
Interestingly, we found many emission features in the $Beppo$SAX spectrum and fitted the spectrum with the {\tt tbabs * (bmc + N * Gauss)} model. However, these features may be associated with a remote shell. In fact, $Beppo$SAX data come from a vast area that includes both the nucleus and the jet and from the external parts of the AGN. Moreover, the spectra obtained using $Chandra$ for a narrower 
region of the nucleus of M87 (<0.85{\tt "}) no longer contain emission features
 (see Fig.~\ref{3_chandra_sp_m87}). 

The same conclusion was drawn for the XMM observations~\citep{Kennea2000}. Therefore, we continued to work
with the Chandra spectra in order to select a model for fitting the radiation of the nucleus of M87. We also proceeded with a powerlaw model; it fits the spectrum well, but the data are poorly matched by the $\chi$-square criterion ($\chi^2_{red}=0.62$ for 174 dof, ID=18783, see top of the Table~\ref{tab:par_chandra}). The thermal black body model also fits the spectrum well, however it poorly approximates the regions of 0.3--1 keV (possibly related to the underestimation of absorption) and the energy region above 
7 keV ($\chi^2_{red}=1.5$ for 176 dof, ID=21075). In this case, taking absorption into account  unnecessary. It is worth noting  that the above models provide a soft excess below 0.3 keV. To eliminate this drawback, we used a combined model consisting of 
black body and power-law components. This latter model also  fits the spectrum of the M87 nucleus well, but this combined model is more phenomenological than the physical one. Therefore, in the subsequent step, we used the generic BMC model. This model describes the spectra of the M87 nucleus well in all states (see bottom in the Table~\ref{tab:par_chandra}).


It is important  to emphasize that the log-parabolic power-law models (Kardashev, 1962; Massaro et al., 2004a,b, 2006) were used for  analysis of the blazar spectra.
{     
Indeed, the log-parabolic model is one of the simplest ways to describe curved spectra when these show mild and nearly symmetric curvature around the maximum, instead of a sharp high-energy cut-off like that of an exponential. 
As it is described by  Massaro et al. (2006) the log-parabolic model law has only one more parameter than a simple power-law.   For example, one can see  it applying Eq. (3)  of Massaro et al. (2004) who apply this formula  for photon spectra.  This appropriate formula is 
\begin{equation}
F(E) = K(E/E_1)^{-[a + b\lg(E/E_1)]}. 
\label{1}
\end{equation}
in which the reference energy $E_1$ is fixed at 1 keV and thus the spectrum is determined only by three parameters $K$, $a$ and $b$ and the energy dependent photon index is
\begin{equation}
\Gamma(E) = a +  b \lg(E/E_1). 
\label{2}
\end{equation}



}

However, it is easy to show that this log-parabolic model is approximated by a single power-law for large E and frozen pivotE. In fact, if the index of this simplified  broken power-law and log-parabolic model is too large then it leads to  very high residuals at low-energies. This is a well-known effect of this simplification. 
{
The log-parabolic model often uses a  $N_H$  fixed at galactic values (Massaro et al., 2004), and for sources where the internal absorption is not relevant. In fact, this is one of the motivations for not using this model in the context of the present analysis.
}

There are several advantages to using the BMC model with respect to other common approaches applied to studies of X-ray spectra of accreting compact objects, including a broken power-law and the log-parabolic model. First, the BMC is by nature applicable to the general case where there is an energy gain through not only thermal Comptonization but also via dynamic (bulk) motion Comptonization (see Titarchuk et al. 1997; Laurent \& Titarchuk 1999; Shaposhnikov \& Titarchuk 2006, for details). Second, with respect to the  log-parabolic model, the BMC spectral shape has an appropriate low-energy curvature, which is essential for a correct representation of the lower-energy part of the spectrum. Long-term experience with log-parabolic components shows that the model fit with this component is often inconsistent with the $N_H$ column values and produces an unphysical component ``conspiracy'' with the highecut part. Specifically, when a multiplicative component highecut is combined with the BMC, the cutoff energies $E_{cut}$ are in the expected range of 20--30 keV, while in a combination with the log-parabolic, $E_{cut}$ often goes below 10 keV, resulting in unreasonably low values of the photon index. Furthermore,   implementation of the  log-parabolic model makes it much harder or even impossible to correctly identify the spectral state of the source, which is an imminent task for our study.  An  even more important property of the BMC model is that it consistently  calculate the normalization of the original ``seed'' component, which is expected to be an indicator of a correct mass accretion rate. We should  point out  that the Comptonized fraction is also properly evaluated by the BMC model.

Figure~\ref{swift_interm_spectrum} shows the best-fit model of the spectrum of 3C~454.3 (top panel). 
The data are taken from $Swift$ observations using the  {\tt tbabs*bmc} model 
   ($\chi^2_{red}=$ 1.02 for 770 degrees of freedom). The best-fit parameters are 
   $\Gamma=$ 1.63$\pm$0.02 and $T_s=$280$\pm$20 eV. 
   The data are denoted by black {crosses}, while the spectral model is presented 
   by a red histogram.
In the { bottom panel} we show  the corresponding $\Delta \chi$ versus photon energy (in keV). 
Using the 
$Swift$ data  we  find that the seed temperature $kT_s$ of the $BMC$ model  varies only slightly 
from 130 to 280 eV.

\begin{figure*}[ptbptbptb]
\centering
\includegraphics[scale=0.9,angle=0]{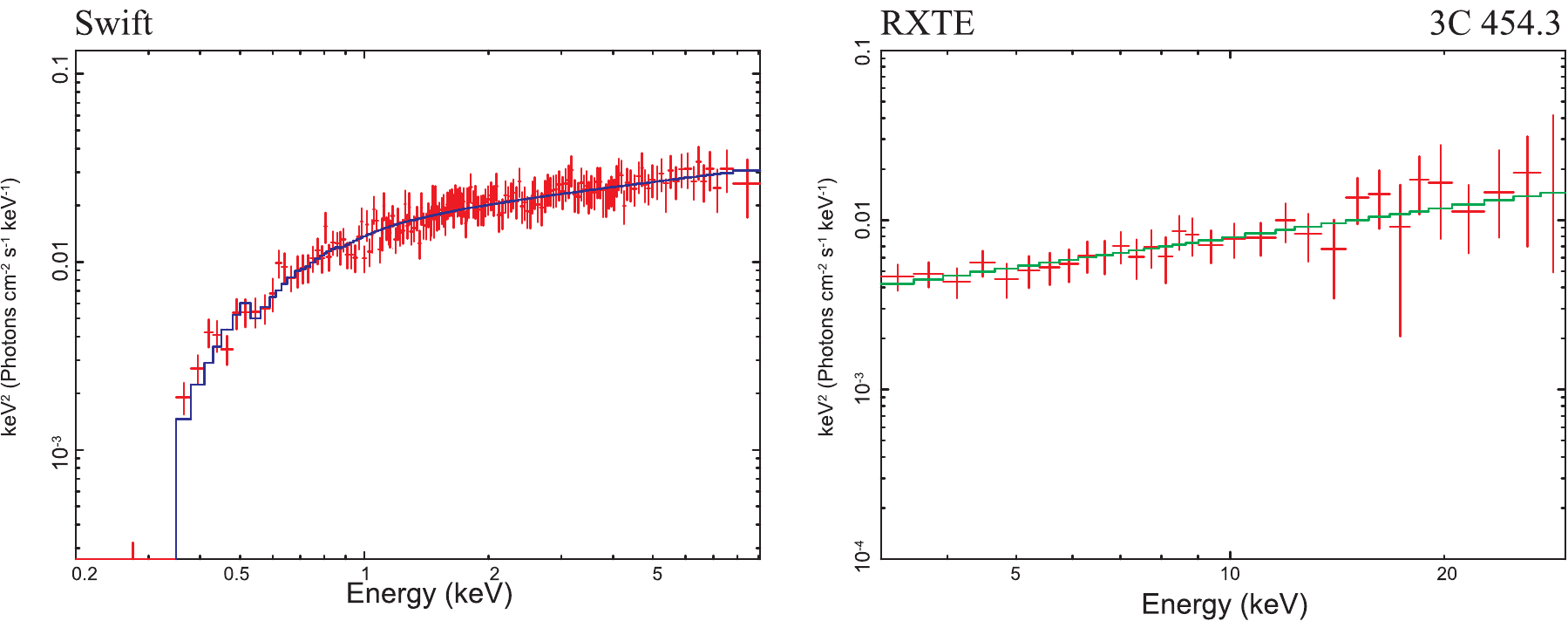}
\caption{Two $E F_E$ spectral diagrams during the LHS (2009 September, 18, blue line, left panel)  and the LHS 
(1996 November, 2  (green line, right panel)). Data taken from 
$Swift$ observation 00031493003 (low/hard 
state) and from the {\it RXTE} observation, 20346-01-01-00 ($R2$ set, low/hard state).
}
\label{swift_rxte_spectra}
\end{figure*}

The  spectral shape change  during the spectral transition can be seen in Figure~\ref{two_sp_Suzaku}, where  two  representative EF(E) spectra are shown for the low/hard   state  (red, id=703006010) 
and  for the intermediate one    (black, id=904003010) of 3C~454.3 detected by $Suzaku$.
We also  use the {\tt tbabs*bmc} model to fit  all {\it RXTE} data.
In order to  fit all of these spectra we use neutral column $N_H$  fixed at 
$5.0\times 10^{21}$ cm$^{-2}$ obtained using $Suzaku$ data  (see bottom of Table~\ref{tab:par_suzaku}). Figure~\ref{swift_rxte_spectra} shows two $E F_E$ spectral diagrams during the LHS 
(left panel)  and the LHS (right panel) events in 3C~454.3; data were taken from 
$Swift$ observation ID=00031493003 and {\it RXTE} observation ID=20346-01-01-00. Here, 
 the adopted spectral model is seen to accurately describe the source spectra obtained onboard  different 
spacecrafts
for the same  spectral state of 3C~454.3.

\begin{figure*} 
\includegraphics[width=14cm]{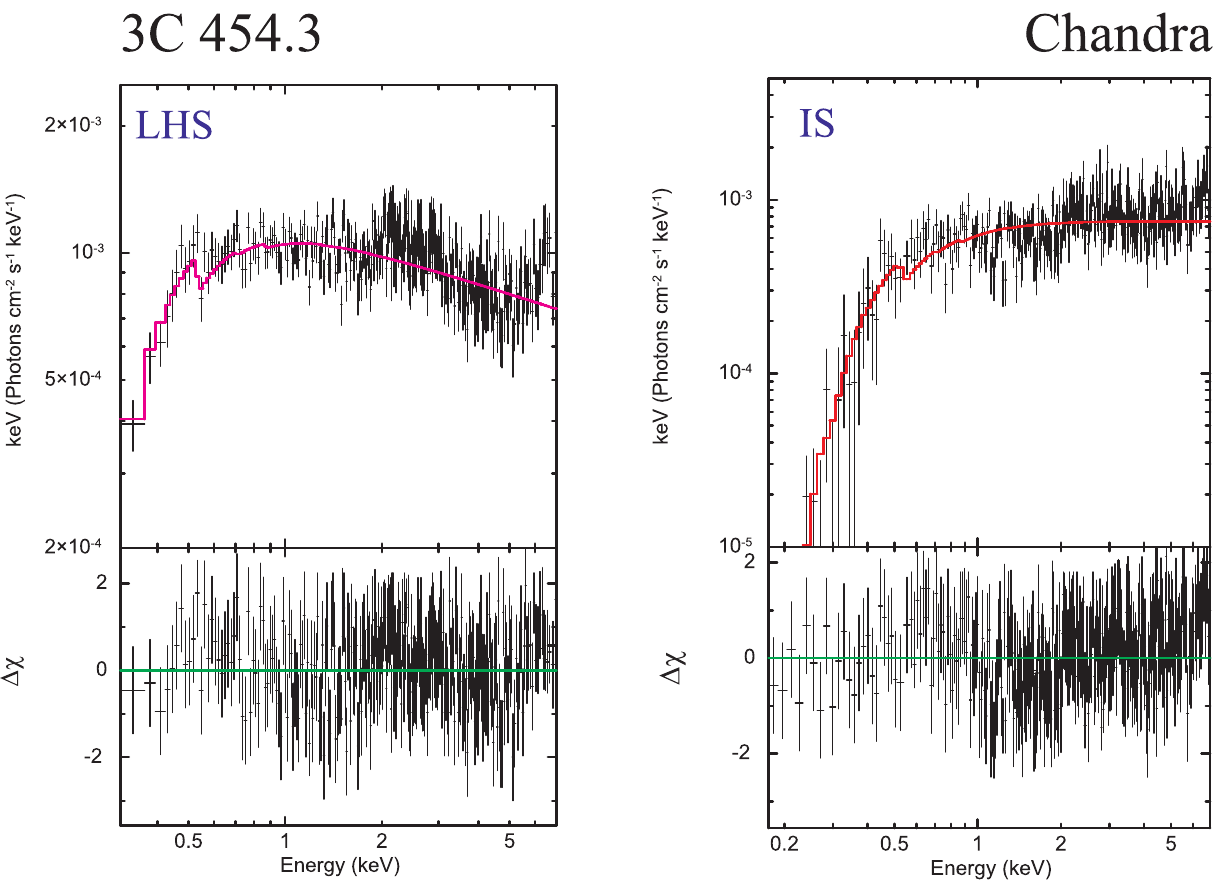}
\caption{Two representative EF(E) spectral diagrams, for different states of 3C 454.3. Data are taken from $Chandra$ observations 
ID=4843 (left panel, LHS) and ID=3127 (right panel, HSS). The data are shown 
by black crosses and the spectral model (tbabs*BMC) is displayed as a colored line.
}
\label{2_chandra_sp_454}
\end{figure*}

\begin{figure*} 
\includegraphics[width=16cm]{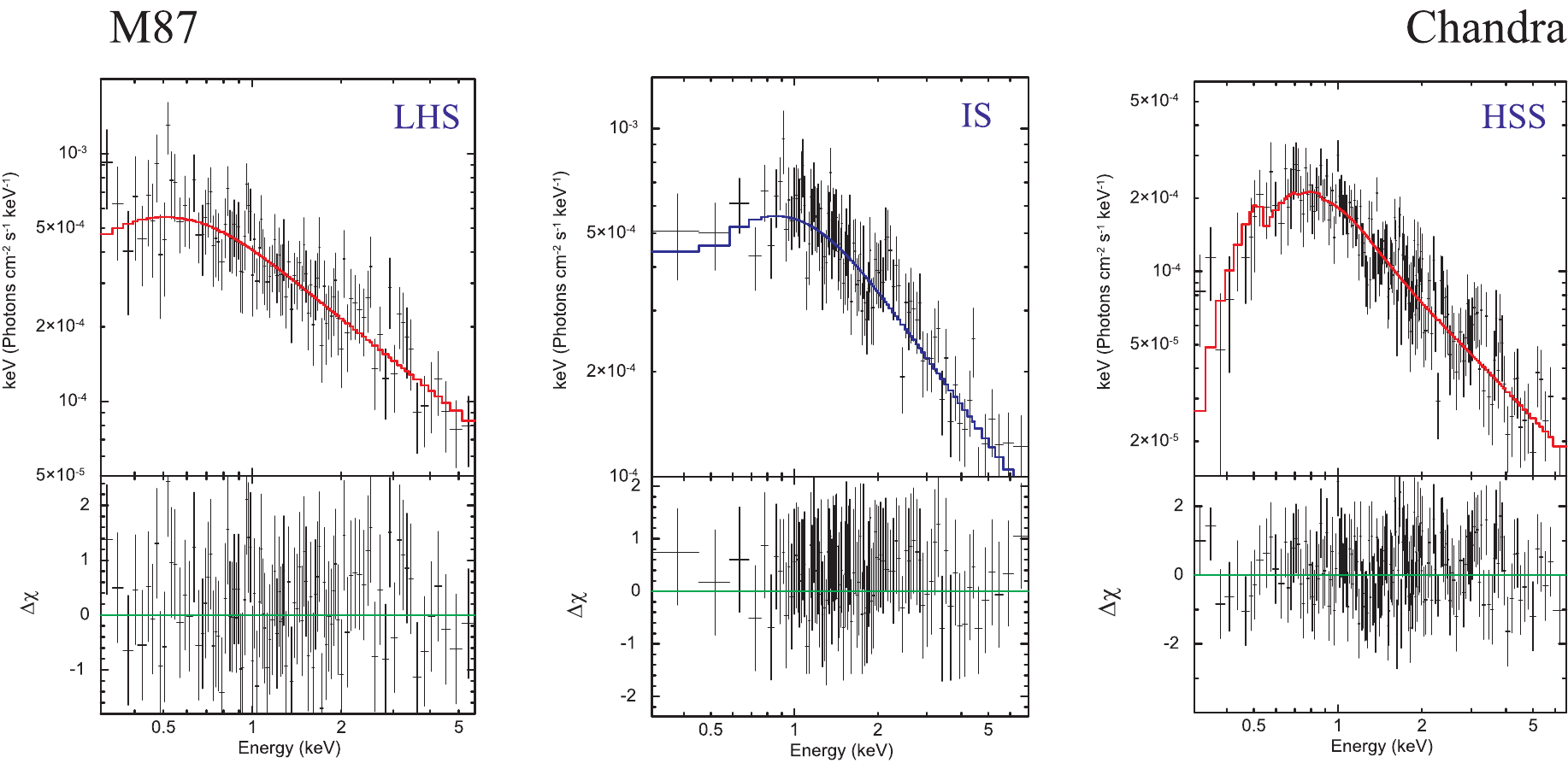}
\caption{Three representative energy spectra  for different states of M87. Data were taken from $Chandra$ observations 
ID=11514 (left panel, LHS), ID=21075 (central panel, IS), and ID=18783 (right panel, HSS). The data are shown 
by black crosses and the spectral model (tbabs*BMC) is displayed as a colored line.
}
\label{3_chandra_sp_m87}
\end{figure*}

Figure~\ref{2_chandra_sp_454} demonstrates two representative $Chandra$ spectra for different states of 3C 454.3. Data are taken using observations ID=4843 (left panel, LHS) and ID=3127 (right panel, HSS) and extracted from the nuclear region (less than 1,25 arcsec circ) around the central source. Here, the data are shown by black crosses and the spectral model (tbabs*BMC) is displayed as a colored line.  This Figure shows the spectral change is typical to that 
 observed in a BH  
(Galactic and extragalactic  ones).



%
%

%

%
%

Figure \ref{3_chandra_sp_m87} shows  three representative spectra  for different states of M87.
These {\it Chandra}  energy spectra demonstrate  an evolution of the source from the low/hard state to the softer states which was observed in  the particular dates of 2010, 2016 and 2018 (see Table \ref{tab:list_Chandra}). 

We use  the   geometry  for the X-ray spectral formation  in 3C 454 and M87 shown in   Seifina et al. (2018b); see Fig.~7 there.    
Regarding   this geometry and taking into account our  X-ray spectral analysis, we see that that the soft (disk)  photons illuminate the Compton cloud (CC) surrounding a BH
hole, and matter from the CC is accreted onto the BH  following a converging flow  (the Bulk Comptonization region).

In Figure~\ref{fragm} we present the evolution of the X-ray properties of 3C~454.3. One can see that  the photon indices  change in the range from 1.3 to about 1.8  when the corresponding normalization  slightly increases.  These spectral evolution was observed by the {\it RXTE}
during 1996 -- 1997 outburst transition events ($R1$--$R2$). 

In addition, we present  the evolution of the optical and X-ray properties of 3C~454.3 (Fig.~\ref{fragm_2009-2010}):
the optical light curve 
(in stellar magnitudes) of 3C~454.3 
(top panel), 
the {\it RXTE}/ASM count rate, and 
the BMC normalization during the outburst event of 2009 -- 2010 ($R5-R6$ sets). In the bottom panel, we see again   the slight  change of the photon index $\Gamma=\alpha+1$.
The high X-ray flux  (MJD 55550) is seen when the optical flux is low. 
At the same time, during the optical flash at MJD=55200 we see correlation of 
the optical flux in  all filters (top panel). 
It is important to conclude that the optical variability (e.g., MJD 55400 -- 55550) 
is weakly related to X-ray variability. This could indicate that the origins of   the optical 
and X-ray emissions are different. 

Applying  the {\it Suzaku} data 
(see black triangles 
in Figure~\ref{saturation}) 
we obtain  that the photon  index, $\Gamma,$
monotonically increases   from 1.5 to 2.0  
when the normalization  of the BMC component (or mass accretion rate)  changes  by a factor of 10. 
We illustrate this index versus mass accretion rate behavior  in Fig.~\ref{saturation} 
using {\it RXTE}, {\it Swift}, {\it Suzaku,}  and {\it Chandra} observations (red squares, blue squares, black triangles, and green points,  respectively).

\subsubsection{X-ray spectral modeling
\label{bmc-results}}

As a result of the model selection (see Sect.~\ref{model choice}), 
we assume one model 
to fit all spectral data (Tables \ref{tab:list_suzaku} $-$ 
\ref{tab:list_RXTE}). 
We  briefly reiterate  the physical picture described  by the Comptonization  model~(see \cite{tl97}, 
ST09, and Appendix A), and  its basic assumptions and  parameters. 
The BMC  Comptonization spectrum is  the sum of the portion of the blackbody 
emission  directly seen by the Earth observer [a fraction of $1/(1+A)$] and a fraction of  the blackbody, $f=A/(1+A)$, convolved 
with the up-scattering  Green's function, $G(E,E_0)$ which is, in  the BMC model,  a broken power-law: 
\begin{equation}
F_{\nu}=N_{BMC}[(1-f)*BB(E) + f*\int_0^{\infty}BB(E_0)G(E,E_0)dE_0].
\label{bmc_spectrum}
\end{equation}
It is worth emphasizing that this   Green's function is  characterized by the main parameter, the spectral index $\alpha=\Gamma-1$. As one can see, the BMC model 
has the parameters, $\alpha$, $A$, the seed blackbody temperature $T_s,$ and the blackbody 
normalization, which is proportional to the seed blackbody luminosity and inversely proportional 
to $D^2$ where $D$ is the distance to the source. 
{ We also apply a multiplicative Tbabs component characterized by an equivalent hydrogen column $N_H$ 
in order to take into account  absorption by neutral material.   

Figures \ref{two_sp_Suzaku}  and \ref{fragm} show the spectral evolution  from the LHS to the IS. 
The BMC model  successfully  fits  the 3C~454.3 spectra for all spectral states.
The $Swift$ spectrum obtained for  3C~454.3  in its IS 
using the BMC 
model is shown in Fig.~\ref{swift_interm_spectrum}.
In Table~\ref{tab:par_suzaku} 
(at the bottom), we present the results of spectral fitting  $Suzaku$ data of 3C~454.3 using our
 main spectral model tbabs*bmc. 

Applying the  {\it RXTE} data (see Tables \ref{tab:list_RXTE}, \ref{tab:par_rxte_2010}),  we find  the LHS$-$IS 
transition  
that is  related to 
 the photon index evolution, when  $\Gamma$ changes from 1.2 
to 2.1 with an increase of the seed photon  normalization (proportional to the mass accretion rate).
For the {\it RXTE} fits, we fix  the seed photon temperature at  200 eV.  
The BMC normalization, $N_{BMC}$, varies by  a factor of 100,  in the range of 
$0.1<N_{BMC}<10\times L_{35}/d^2_{10}$ erg s$^{-1}$ kpc$^{-2}$. 
The Comptonized (illumination) fraction changes only slightly around $\log{A}\sim$ 1.5 
[$f=A/(1+A)$]. 

%
%
 \begin{figure*}
 \centering
\includegraphics[width=12cm]{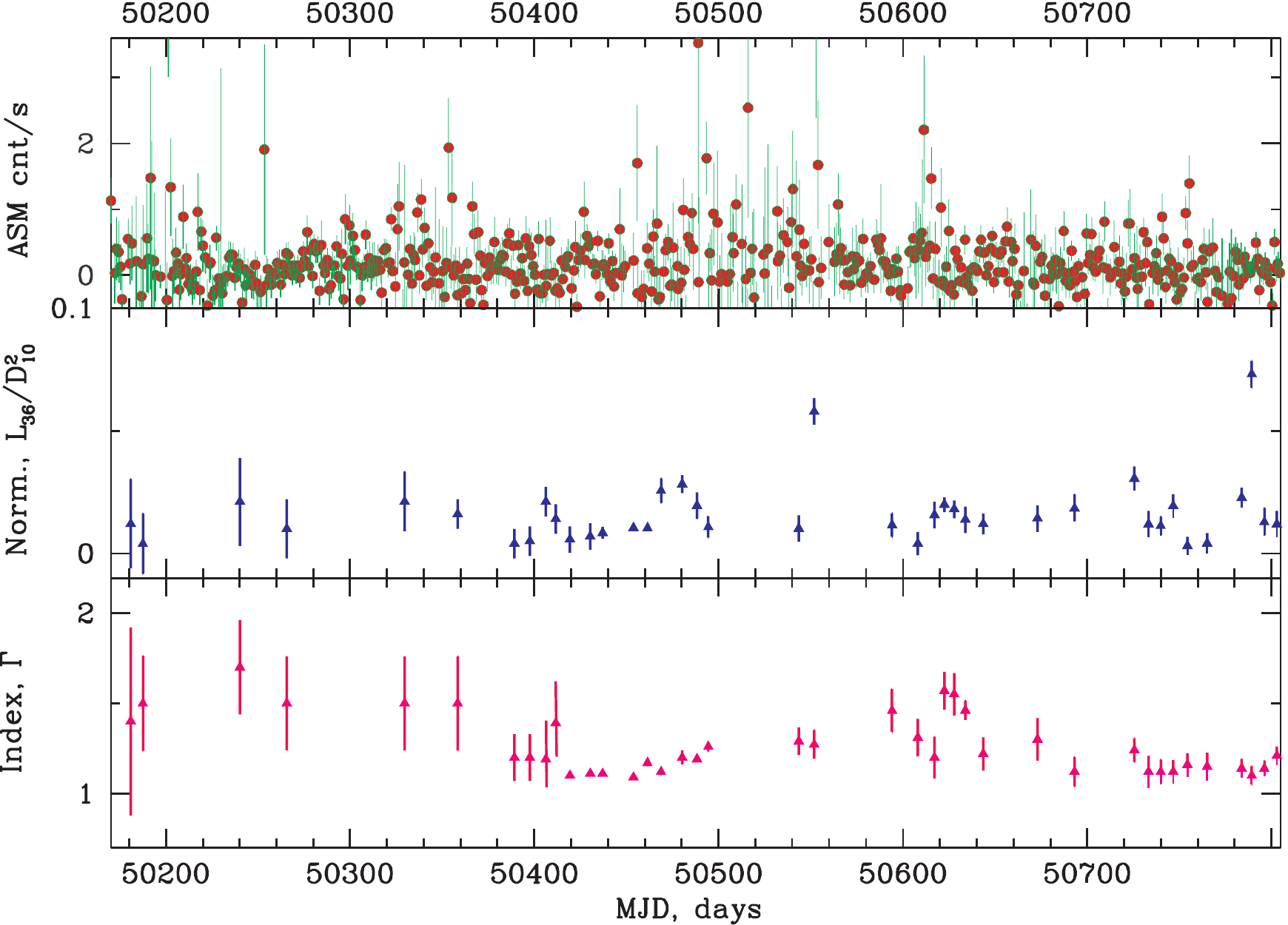} 
  \caption{
Top:  
Evolution of the {\it RXTE}/ASM count rate for 3C 454. Middle: BMC normalization for 3C 454 during 1996 -- 1997 outburst events ($R1-R2$ sets). Bottom: Evolution 
of the photon index $\Gamma=\alpha+1$.
}
\label{fragm}
\end{figure*}

%
 \begin{figure*}
 \centering
 \includegraphics[width=8cm]{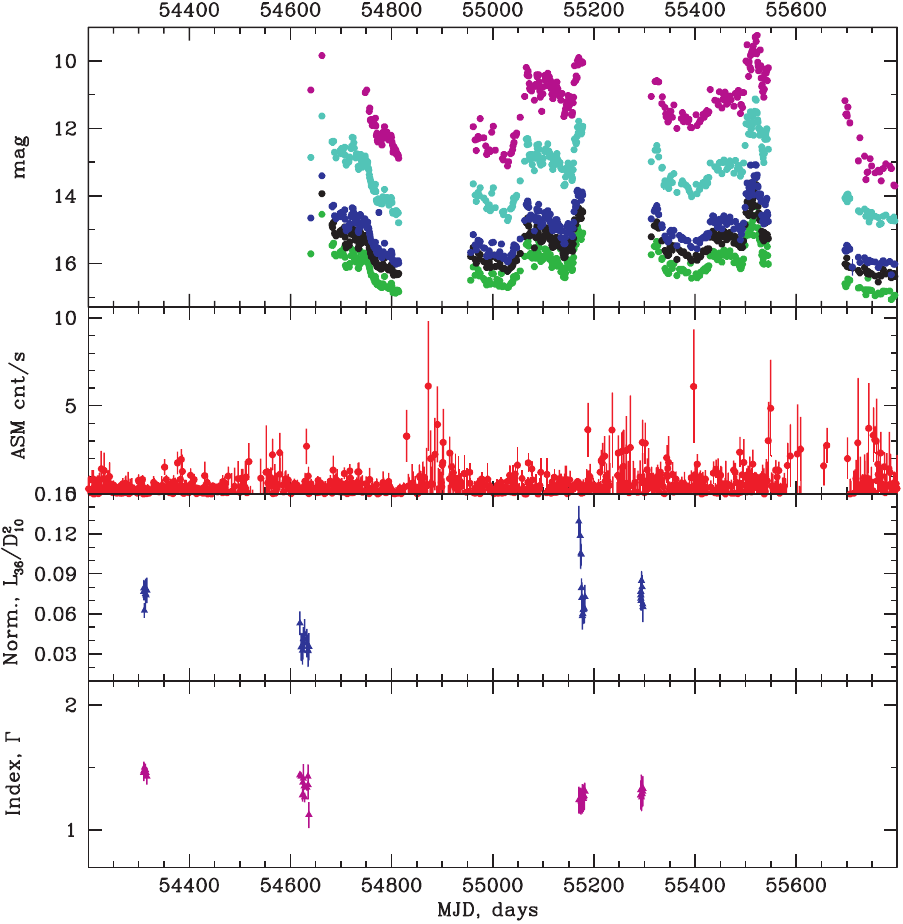}
   \caption{
{Top:}
Evolution of the optical flux (in stellar magnitudes) of 3C~454.3 in B-filter (green), 
V-filter (black), R-filter (blue), J-filter (bright blue), and K-filter (pink). Second row:  
{\it RXTE}/ASM count rate. Third row: 
 BMC normalization during 2009 -- 2010 outburst events ($R5-R6$ sets). Bottom: Evolution 
of the photon index $\Gamma=\alpha+1$.
}
\label{fragm_2009-2010}
\end{figure*}
As discussed  above, the spectral evolution of 3C~454.3 was previously 
investigated  using X-ray data by many  authors. In particular, Abdo et al. (2010) 
and Raiteri et al. (2011) studied the 2008 ($Suzaku$), 2008--2009 ($Swift$), and 2008 -- 2010 ($Swift$) data sets, respectively 
(see also Tables \ref{tab:list_suzaku} and \ref{tab:list_Swift}), using single power law  and absorbed power-law models, 
respectively. 
It is worth noting that 
Abdo et al. (2010)  and Raiteri et al. (2011) fixed  the hydrogen column to $N_H = 1.67\times 10^{21}$ cm$^{-2}$ and $N_H = 1.34\times 10^{21}$ cm$^{-2}$,  respectively  
(applying the $Chandra$ observations in 2005, see Villata et al. 2006)
These  qualitative models   were used to establish  the evolution of the spectral model parameters throughout 
state transitions during the outbursts. 


 We  also found similar 
spectral behavior using our model and the full set of the $RXTE$ observations.  
In particular, like in the aforementioned Guainazzi's and Rivers's et al works, we also found that 
3C~454.3 demonstrates a change of the photon index $\Gamma$ between $\sim$1.2 and 2.1
during the LHS -- IS transition. Furthermore, we revealed that  $\Gamma$ tends to  saturate  at 2.1 
at high values of $N_{bmc}$. In other words  we  find  that  $\Gamma$  saturates at 2.1 when  
 the mass accretion rate increases. 

Our spectral model shows  very good performance throughout
all data sets. 
{In Tables~\ref{tab:par_suzaku}, \ref{tab:par_rxte}$-$\ref{tab:par_rxte_2010}, and Figures~\ref{swift_interm_spectrum}$-$\ref{2_chandra_sp_454} and \ref{fragm}$-$\ref{fragm_2009-2010}
 we demonstrate the good performance of the BMC model in application to the $Chandra$,  $Swift$, $Suzaku,$ and $RXTE$ data}
for which  the reduced  $\chi^2_{red}=\chi^2/N_{dof}$ 
($N_{dof}$ is the number of degree of freedom) is  less than or around  1 
for  all 
observations ($0.75<\chi^2_{red}<1.19$).

We performed a similar simulation of the M87 spectrum using the available observations. 
The  {\it RXTE} data reveal a weak activity of the M87 nucleus in X-rays.
However,  for 13 years there is a set of all states from the LHS to the HSS. 
In particular, M87 spends most of its time in the HSS (1997--1998, $\Gamma \sim 3$), 
although it can slowly reach LHS (2010, $\Gamma \sim 1.7-1.9$). The data of  
two other satellites fill the gaps between these states.   $ASCA$ (1993) and $Suzaku$ (2006) 
find the M87 core in the IS ($\Gamma\sim 2.5$). Finally, $BeppoSAX$ (1996) 
points to the LHS ($\Gamma\sim 1.6$).

In Table~\ref{tab:par_chandra} 
 we present the results of spectral fitting  $Chandra$ data of M87 using our
 main spectral model tbabs*bmc. 
Using the  {\it RXTE} data (see Table \ref{tab:par_rxte_m87})  the LHS$-$IS 
transition  is related when  $\Gamma$ changes from 1.8 
to 3 and  the normalization of the seed photon increases.
For the {\it RXTE} fits we fix  the seed photon temperature at  200 eV.  The BMC normalization, $N_{BMC}$, varies by  a factor of five  in the range of 
$0.33<N_{BMC}<1.5\times L_{37}/d^2_{10}$ erg s$^{-1}$ kpc$^{-2}$. 

As we have already discussed  above, the spectral evolution of M87 was previously 
investigated  using X-ray data by many  authors. In particular, Abdo et al. (2010) 
and Raiteri et al. (2011) studied data sets from 2008 ($Suzaku$) and 2008--2009 ($Swift$), 
and  2008 -- 2010 ($Swift$), respectively 
(see also Tables \ref{tab:list_suzaku} and \ref{tab:list_Swift}), using a single power law  and 
an absorbed power-law model, 
respectively. For their spectral analysis, 
Abdo et al. (2010)  and Raiteri et al. (2011) used   the hydrogen column
fixed to $N_H = 1.67\times 10^{21}$ cm$^{-2}$ and $N_H = 1.34\times 10^{21}$ cm$^{-2}$, 
 respectively, 
as determined by the $Chandra$ observations in 2005 and  Villata et al. 2006.
These  qualitative models   describe the evolution of these spectral model parameters throughout 
the state transition during the outbursts. 

%
%

 \begin{figure*}
 \centering
\includegraphics[width=10cm]{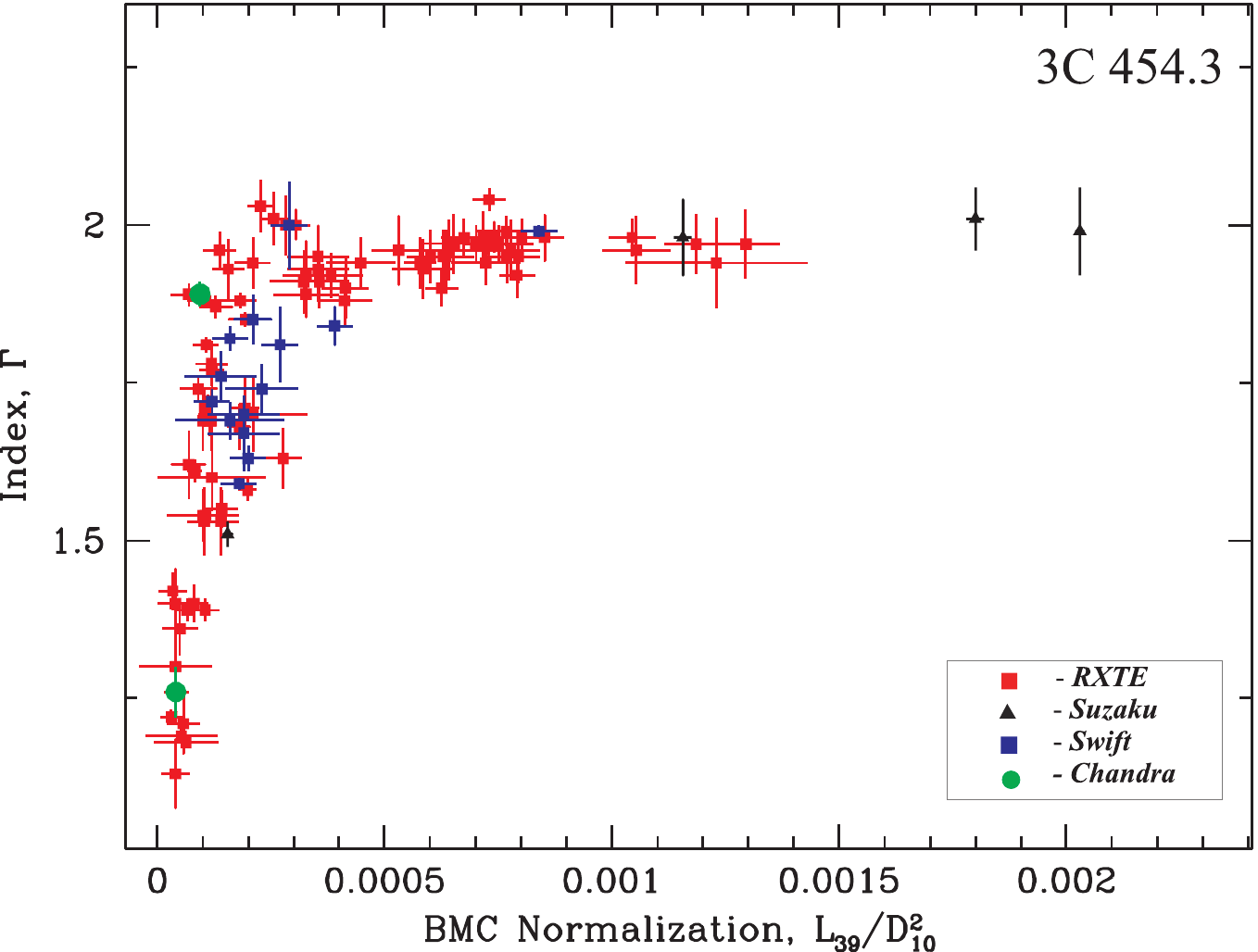} 
  \caption{
3C 454: Correlations of photon index $\Gamma$ ($=\alpha+1$) 
vs. BMC normalization, $N_{BMC}$ (proportional to mass accretion rate) in units of $L_{39}/D^2_{10}$. 
Red 
and blue squares 
show  $RXTE$ and $Swift$ observations, respectively, while blue triangles and green points correspond to {\it Suzaku} and {\it Chandra} observations, respectively.
%
%
}
\label{saturation}
\end{figure*}
%
%

 \begin{figure*}
 \centering
\includegraphics[width=10cm]{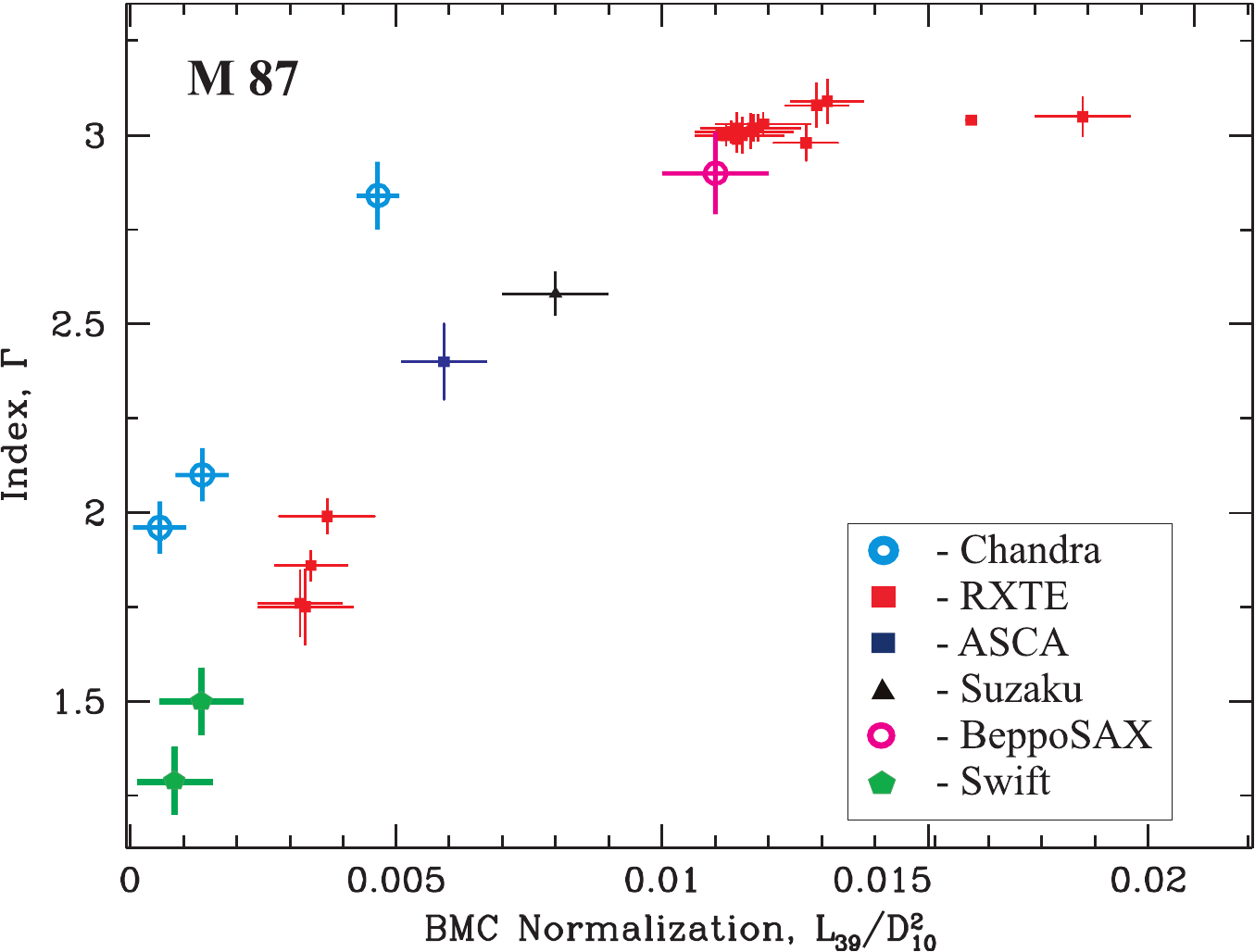}
   \caption{
M87: Correlations of photon index $\Gamma$ ($=\alpha+1$) 
vs. BMC normalization, $N_{BMC}$ (proportional to mass accretion rate) in units of $L_{39}/D^2_{10}$. 
Red and  
 blue squares 
show  $RXTE$ and $ASCA$ observations, respectively, green points correspond to the 
{\it Swift} observations, black triangles correspond to $Suzaku$ data, a pink circle denotes a point related to  $BeppoSAX$ observations, and blue circles show those related to $Chandra$ observations.
}
\label{saturation_m87}
\end{figure*}

%
%

\begin{figure*}
 \centering
\includegraphics[width=8cm]{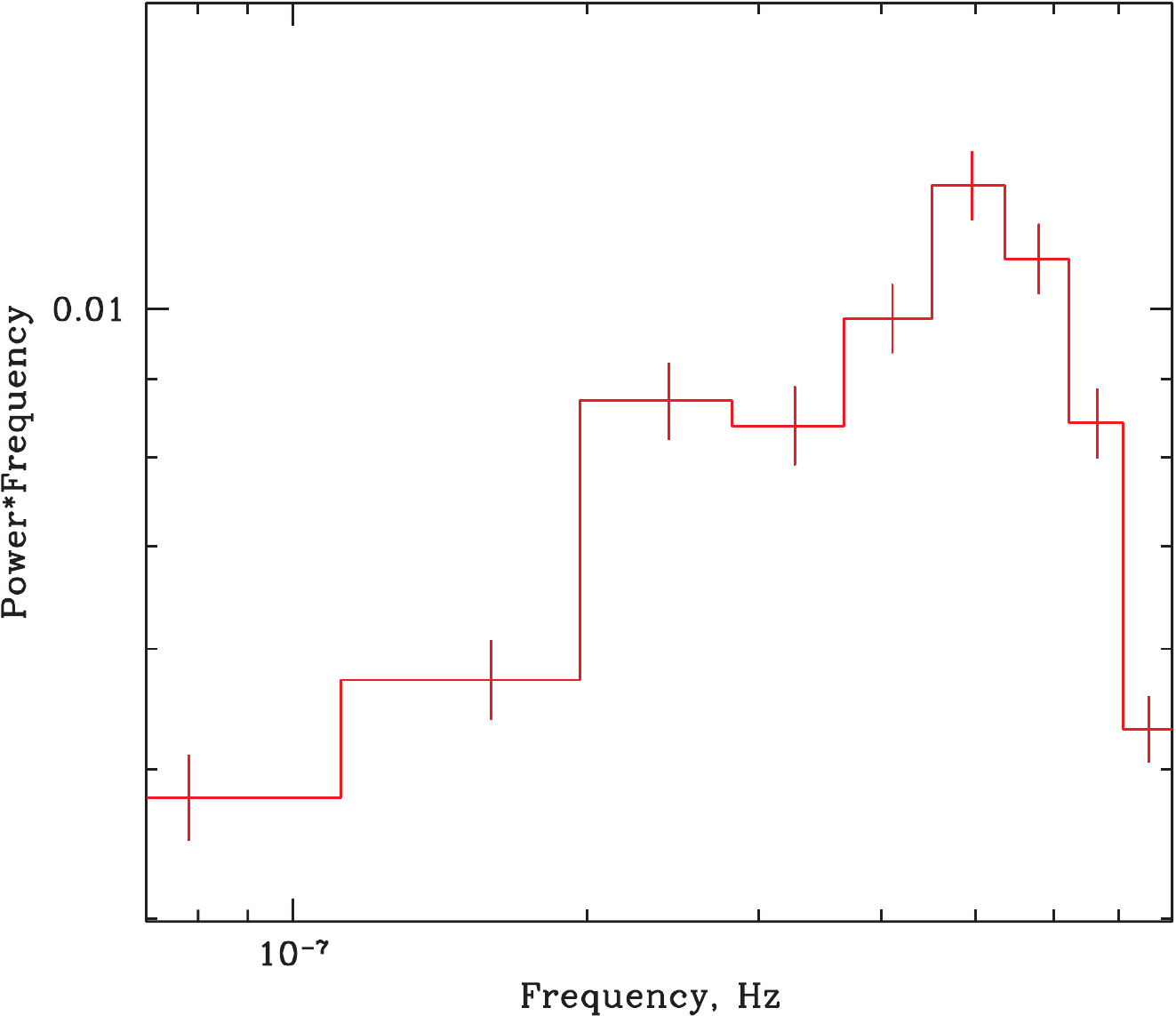}
   \caption{
Representative power spectrum of M87. Data are taken from $RXTE$/ASM observations (1996--2011)  with 
a time bin of 100 ks.
}
\label{pds_m87}
\end{figure*}

We estimated a radius of the blackbody emission region, $R_{BB}$,  using a relation 
$L_{BB} = 4\pi R^2_{BB}\sigma T^4_{BB}$, where $L_{BB}$ is the seed blackbody  luminosity  and 
$\sigma$ is  the Stefan-Boltzmann constant. With  a distance $D$ to the source  of $5\times 10^3$ Mpc, we found 
$R_{BB}\sim 1.6\times 10^{12}$ cm.  
Such a large  black body  region would only be expected  around  a supermassive black hole (SMBH)  and therefore 3C~454.3 is probably   a SMBH source.  
}
 
We also observed the photon index, $\Gamma,$ versus the normalization of the X-ray spectrum (proportional to the mass accretion rate)  in the case of M87. In this case,  $\Gamma$ increases from a value of 1.3 and saturates at 3 (see Fig. \ref{saturation_m87}). 
{ The inclusion of high-quality observations using the $Chandra$ data allows us to take into account  the radiation of the nucleus only (without outer AGN parts and jet contaminations)}.  
In Figure \ref{saturation}  we can see that $Chandra$ observations for 3C 454.3 (green points) agree well with the general 
$\Gamma$--Norm trend, while for M87, {\it Chandra} observations (blue points) show a slightly lower normalization for the 
same photon indices (see Fig. \ref{saturation_m87}) along with that for {\it RXTE}, $Suzaku,$ and $Swift$ points.

\section{Discussion \label{disc}}

 The spectral data of 3C~454.3 and { M87} are accurately fitted by the BMC model for all 
analyzed LHS and IS spectra (see Figures~\ref{swift_interm_spectrum}--\ref{swift_rxte_spectra} 
and  Tables~\ref{tab:par_suzaku}, \ref{tab:par_rxte} and \ref{tab:par_rxte_2010}). 
 Our results of spectral analysis 
are consistent with previous results of the spectral fitting  
by other authors using various X-ray observations of 3C~454.3 and {M87}.

 In particular, Rivers et al. (2011)
found a value for $\Gamma$ of about 1.5 on average using the {\it RXTE} data of 3C~454.3, while Giommi et al. (2006) 
obtained   $\Gamma \sim 1.7$ using $Swift$ observations of 3C~454.3. 
 Chitnis et al. (2009) showed that $Swift$ spectra of 3C~454.3 can be fitted by an absorbed power-law model with $\Gamma$ of 1.3 using  the $Swift$ data. Rivers et al. (2013) 
demonstrated that soft X-ray emission can be described by an absorbed power law  
with  $\Gamma\sim 1.6$ using the {\it RXTE} data. Raiteri et al. (2008) showed that $\Gamma$ 
changes  from 1.5 to 1.6 using the appropriate XMM-$Newton$ observations. Raiteri et al.  also found
$\Gamma$ to vary from 1.38 to 1.85, with an average value of $\Gamma \sim 1.6$ based on 
  $Swift$ data obtained in 2008--2010 (Raiteri et al. 2011). $Susaku$ observations also demonstrated that  $\Gamma=1.58\pm 0.01$ for 2008 observations (our $Sz1$ state, see 
Table~\ref{tab:par_suzaku}). The combined $IBIS$ and $Jem-X$ data  of 3C~454.3 obtained onboard $INTEGRAL$  gave   $\Gamma=1.8$ (Pian et al. 2006). Finally, Pacciani et al. (2010) found that $\Gamma$  was very close to 1.7 using  the  {\it RXTE} data and $\Gamma=1.5-1.7$ using $Swift$/XRT observations of 3C~454.3.

For M87, Wilson and Yang (2002)  described the spectrum of the nucleus obtained using Chandra by a simple power-law model and found that the values of the photon index vary from 2 to 3, which is in good agreement with our results.

\subsection{Timing analysis \label{bh_mass}}

We examined archival {\it RXTE}/ASM observations of M87 obtained from January 6, 1996, to December 28, 2011. The ASM light curve is presented in the  top panel of Figure \ref{RXTE evol_m87}. This light curve was analyzed using the {\it powspec} task from FTOOLS 6.26.1. We generated power density spectra (PDSs) in the $10^{-7} - 10^{-3}$ Hz frequency range and subtracted the contribution due to Poissonian statistics for all PDSs. To model PDSs we used the QDP/PLT plotting package. The PDS continuum shape is usually characterized by band-limited noise shape, which is well presented by an empirical model $P_X\sim (1.0+(x)^2)^{-in}$ ($KING$ model in QDP/PLT plotter). The parameter $in$ is a slope of PDS continuum. 
As a  result, we find a statistically significant ($3\sigma$)  feature  at the frequency $\nu_{max} \sim 5\times 10^{-7}$ Hz (see Fig.~\ref{pds_m87}) with a time bin of $\sim$100 ks. One can estimate an  upper limit  of  a BH mass magnitude using this PDS feature as follows  
\begin{equation}
\nu_{max}\sim V_{pl}/L_{tr}=10^8 \frac{[V_{pl}/(10^8~\rm {cm/s})]}{L~(\rm cm)} ~{\rm cm/s}
,\end{equation} 
where $L_{tr}$ is the size of the Compton cloud (CC) in centimeters (or the transition layer)
and $V_{pl}$ is a plasma velocity in the CC of the order of $10^8$ cm/s related to a typical  plasma temperature of the order of 10 keV (see Shaposhnikov \& Titarchuk 2009).
We use a value of the maximum of $\nu*$ PDS frequency which is $\sim 5 \times 10^{-7}$ Hz in order to estimate a BH mass in M87:
\begin{equation}
L_{tr}\sim \frac{V_{pl}}{\nu_{max}}\sim 2 \times 10^{14}~
\frac{V_{pl}/10^8~{\rm cm/s}}{\nu_{max}/5\times 10^{-7}~{\rm Hz} }~  {\rm cm} 
.\end{equation} 
One can easily compare this size $L_{tr}$ with the Schwarzschild radius $R_{\rm S}$ and find that 
\begin{equation}
M_{87}<L/ R_{{\rm S}, \odot} 
\sim  0.6\times 10^9~ {\rm solar~masses}.  
\end{equation} 
This BH estimate in M87  is at least one order of a magnitude lower than that obtained by other methods for M87 [see \cite{Akiyama2019_VI}]. 
In Sect.  4.3 we show  that  $m_{87}=M_{87}/M_{\odot}$ can be more precisely  obtained using 
 the scaling method.


\subsection{Saturation of the  index as a  signature of a BH  \label{constancy}}

We applied 
our analysis in order to find the evolution of  $\Gamma$  in 3C~454.3 and M87, and found 
that   $\Gamma$ saturates  with  the mass accretion rate, $\dot M$.
ST09
demonstrates this index saturation is an indication of the converging flow (CF) into a BH. 
In fact, the spectral index $\alpha\sim Y^{-1}$is inverselly proportional to the Comptonization parameter, $Y$   which is a product of an average  number of scatterings N  in the CF and an efficiency of  the energy gain at any scattering in average $\eta$
[see  \cite{lt07}]. But for the converging flow the efficient number of scattering, $N$ is proportional to the dimensionless mass accretion rate $\dot m$ [or  the optical depth $\tau$, not like $\tau^2$ in the thermal plasma, see e.g. Rybicki \& Lightman (1979)]  and the energy gain $\eta \propto {\dot m^{-1}}$. Thus one obtains that  $\alpha$  saturates when $\dot m$ increases. This is precisely  what we see in Figs. (\ref{three_scal}-\ref{three_scal_1}). Another words, when the normalization parameter, proportional the mass accretion rate, $\dot m$ increases the photon index $\Gamma=\alpha+1$ saturates. 
\cite{tz98}, hereafter TZ98,  semi-analytically discovered   this  
saturation effect. Later  \cite{LT99}, (2011) (hereafter LT99 and LT11, respectively) confirmed it using Monte Carlo simulations.

It should be also  noted that  \cite{tlm98}, hereafter TLM98,  demonstrated 
that the innermost part  of the accretion flow (the so-called transition layer)  shrank  
when 
$\dot M$ increased.

Observations of many  Galactic BHs (GBHs) and their X-ray spectral analysis 
(see  ST09, \cite{tsei09}, \cite{ST10} and Seifina et al. (2014); hereafter STS14) confirm  this TZ98  prediction.
For  3C~454.3,  we also reveal  that   $\Gamma$ 
monotonically increases  from 1.2  and then   finally saturates at a value of 2.1  (see  Figure~\ref{saturation} for 3C~454.3).  For M87, the photon index increases from 1.3 and then saturates at 3.  
The  index-$\dot M$ correlations found in 3C~454.3   and M87 allow us to estimate  BH masses  in these
sources by scaling these correlations with those detected in a  number of GBHs and extragalactic sources  
(see details below in Sect.  4.3 and in Appendix B).

%
%

\begin{table*}
 \caption{Parameterizations for reference and target sources}
 \label{tab:parametrization_scal}
 \centering 
 \begin{tabular}{lcccccc}
 \hline\hline                        
  Reference source  &       $\cal A$ &     $\cal B$     &  $\cal D$  &    $x_{tr}$      & $\beta$  &  \\
      \hline
GRO~J1655--40 & 2.03$\pm$0.02 &  0.45$\pm$0.03    &  1.0 & 0.07$\pm$0.02  &   1.9$\pm$0.2  \\
Cyg~X--1      & 2.09$\pm$0.01 &  0.52$\pm$0.02    &  1.0 & 0.4$\pm$0.1    &   3.5$\pm$0.1  \\
NGC~4051      & 2.05$\pm$0.07 &  0.61$\pm$0.08    &  1.0 &   [9.54$\pm$0.2]$\times 10^{-4}$ &   0.52$\pm$0.09  \\
NGC~7469      & 2.04$\pm$0.06 & 0.62$\pm$0.03    & 1.0  &   1.25$\pm$0.04 & 0.62$\pm$0.04  \\
XTE~J1550-564 RISE 1998 & 2.84$\pm$0.08 &  1.8$\pm$0.3    &  1.0 & 0.132$\pm$0.004   &   0.61$\pm$0.02  \\
H~1743-322    RISE 2003 & 2.97$\pm$0.07 &  1.27$\pm$0.08  &  1.0 & 0.053$\pm$0.001   &   0.62$\pm$0.04  \\
4U~1630-472   & 2.88$\pm$0.06 &  1.29$\pm$0.07  &  1.0 & 0.045$\pm$0.002   &   0.64$\pm$0.03  \\
M101 ULX-1   & 2.88$\pm$0.06 &  1.29$\pm$0.07   & 1.0  &   [4.2$\pm$0.2]$\times 10^{-4}$ &   0.61$\pm$0.03  \\
ESO 243-49 HLX-1 & 3.00$\pm$0.04 & 1.27$\pm$0.05   & 1.0  &   4.25$\pm$0.03 & 0.62$\pm$0.05  \\
 \hline\hline                        
  Target source     &      $\cal A$     &    $\cal B$   &  $\cal  D$  &   $x_{tr}$ & $\beta$ \\
      \hline
3C~454.3      & 2.01$\pm$0.06 & 0.60$\pm$0.03    & 1.0  &    [1.2$\pm$0.2]$\times 10^{-4}$  & 0.60$\pm$0.04  \\
M87      & 3.02$\pm$0.07 & 0.52$\pm$0.02    & 1.0  &   [0.8$\pm$0.1]$\times 10^{-4}$ & 0.50$\pm$0.03  \\
 \hline                                             
 \end{tabular}
 \end{table*}

\subsection{An estimate of BH mass in 3C~454.3 and M87 \label{bh_mass}}


A previous mass estimate of a central engine in 3C~454.3 was made using a spectroscopic method 
 applied to multi-wavelength data  (see Gupta et al., 2017; Woo \& Urry 2002; Liu et
al. 2006; Sbarrato et al. 2012; Gu et al. 2001). Particularly, Gupta et al.  estimated 
a BH mass of 3C~454.3 (see also Vestergaard  \& Osmer 2009) using optical emission (the broad 
Mg~II line width and the continuum luminosity at 3000 
 \AA). 
We   estimated a BH mass, $M_{BH}$, in 3C~454.3 applying X-ray data. 
In order to estimate $M_{BH}$  we chose two Galactic sources 
[GRO~J1655--40, Cygnus~X--1 (see ST09)] and an extragalactic source (NGC~4051; Seifina et al. 2018, 
hereafter SCT18) as the reference sources, whose BH masses and  distances  are well established now.
 
\begin{figure*}
 \centering
\includegraphics[width=18cm]{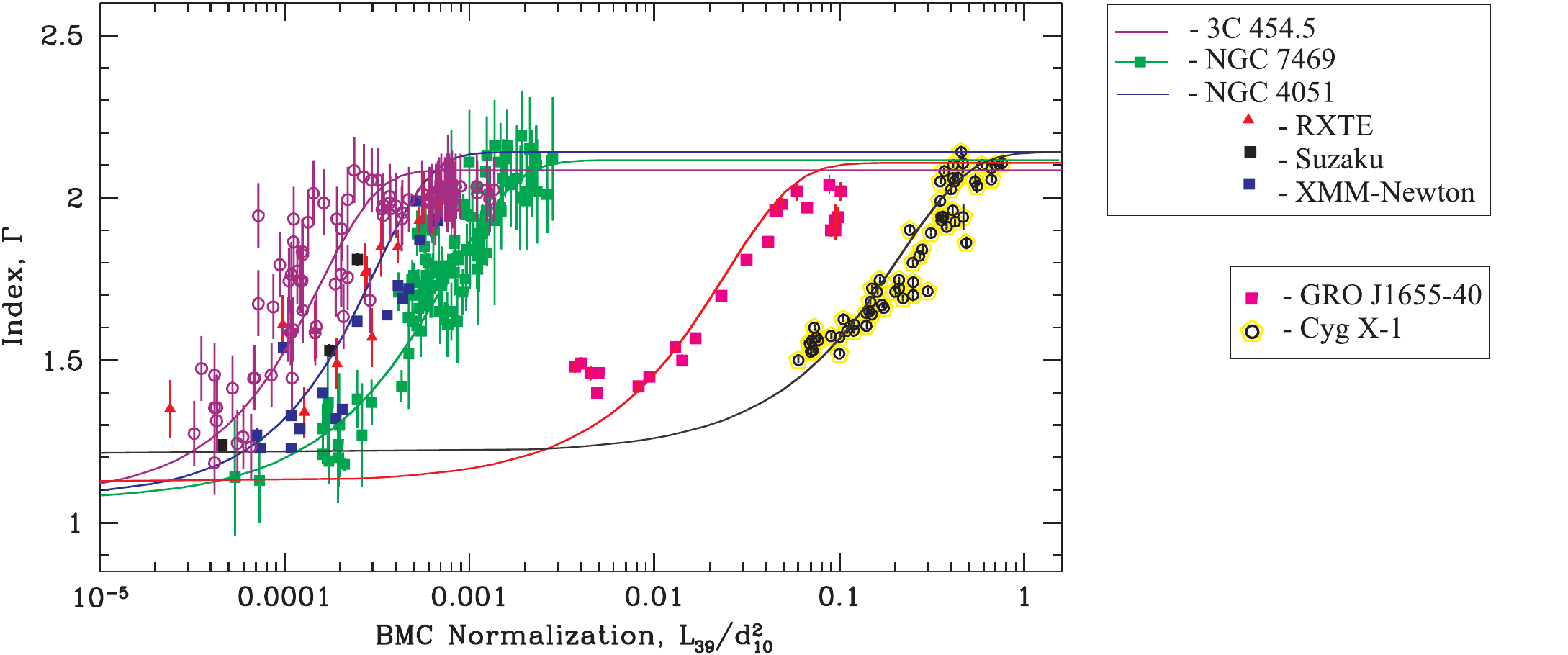}
      \caption{
Scaling of the photon index $\Gamma$ vs. the normalization $N_{BMC}$ for 3C~454.3  
(violet line -- target source) as well as NGC~4051, NGC~7469, GRO~J1655--40, and Cygnus~X--1 
(reference sources).  
Red triangles stand for {\it RXTE}, 
black squares -- $Suzaku$ and blue squares for  XMM-$Newton$ data for NGC~4051, while green squares mark 
{\it RXTE} data for NGC~7469.  Pink squares and yellow-black circles correspond to {\it RXTE} data for 
GRO~J1655--40 and Cygnus~X--1, respectively.
}
\label{three_scal}
\end{figure*}

A BH mass for GRO~J1655--40 was estimated using dynamical methods. 
For a BH mass estimate 
of   3C~454.3  
we  also used  the BMC normalizations, $N_{BMC}$ of these reference sources.  
As a result, we scale  the index versus  $N_{BMC}$  correlations for the target and reference sources, NGC~4051 and  NGC~7469 
 (see Figure~\ref{three_scal}). 
The value of the  index saturation  is   almost the same, $\Gamma\sim2$  for all these target and 
reference sources.   We apply the correlations found in  these four reference sources to make  a  
 comprehensive cross-check of  a BH mass estimate for 3C~454.3.

For M87, a BH mass estimate turns out to be highly dependent on the accuracy of the distance to it.  The distance to M87 has been estimated using several independent techniques which 
include measurement of the luminosity of planetary nebulae, comparison with nearby 
galaxies for which a distance is estimated using standard candles such as cepheid variables, and
the linear size distribution of globular clusters (SBF method). This yields a distance of 
16.6$\pm$2.3 Mpc (Blakeslee et al. 2009). Furthermore, the tip of the red-giant branch method \citep{Bird2010}  using individually resolved red giant stars (TRGB method) 
yields a distance of 16.7$\pm$0.9 Mpc~\citep{Bird2010}]. It should be noted that the SBF measurements by Blakeslee et al. (2009) used the HST ACS-VCS data, while the TRGB measurements were based on data from the Next Generation Virgo Cluster Survey (NGVCS) obtained using ground-based adaptive-optics with the 
Canada French Hawaii Telescope (Cantiello et al. 2018).

These SBF and TRGB measurements are consistent with each 
other, and their weighted averages yield a distance  of 16.4$\pm$0.5 
Mpc~\citep{Bird2010}. 
Using this distance and the modeling of surface brightness and stellar velocity
dispersion at optical wavelengths (Gebhardt \& Thomas 2009;
Gebhardt et al. 2011), a BH mass for M87  of $6.2_{-0.6}^{+1.1}\times 10^9$ 
M$_{\odot}$~was inferred \citep{Akiyama2019_VI}. On the other
hand, mass measurements modeling the kinematic structure of the
gas disk (Harms et al. 1994; Macchetto et al. 1997) give us a BH mass of
$3.5_{-0.3}^{+0.9}\times 10^9$ M$_{\odot}$ (Walsh et al. 2013).
Therefore,  a wide range of SMBH mass estimates exist from $(3.5 \pm 0.8)\times 10^9$ 
M$_{\odot}$~\citep{Walsh2013} to $(6.6 \pm 0.4)\times 10^9$ M$_{\odot}$~\citep{Walsh2013}. 
This agrees with the BH mass estimate 
 of $7.22^{+0.34}_{-0.40} \times 10^9$ M$_{\odot}$ obtained by \cite{Oldham+Auger2016}.
In April 2019, the 
the Event Horizon Telescope (EHT) observations 
released measurements  and estimates of a BH  mass as 
(6.5 $\pm$ 0.2$_{stat} \pm$ 0.7$_{sys}$) $\times 10^9$ M$_{\odot}$~\citep{Akiyama2019_VI}, 
which is also consistent with the above estimates. 

However, it is desirable to use an identification 
for  the core of M87 that is as independent as possible,
as well as a BH mass estimate that uses an alternative 
to the above-mentioned  methods, such as the scaling technique (ST09).

%
%

%
%
%
%

 \begin{figure*}
 \centering
 \includegraphics[width=12cm]{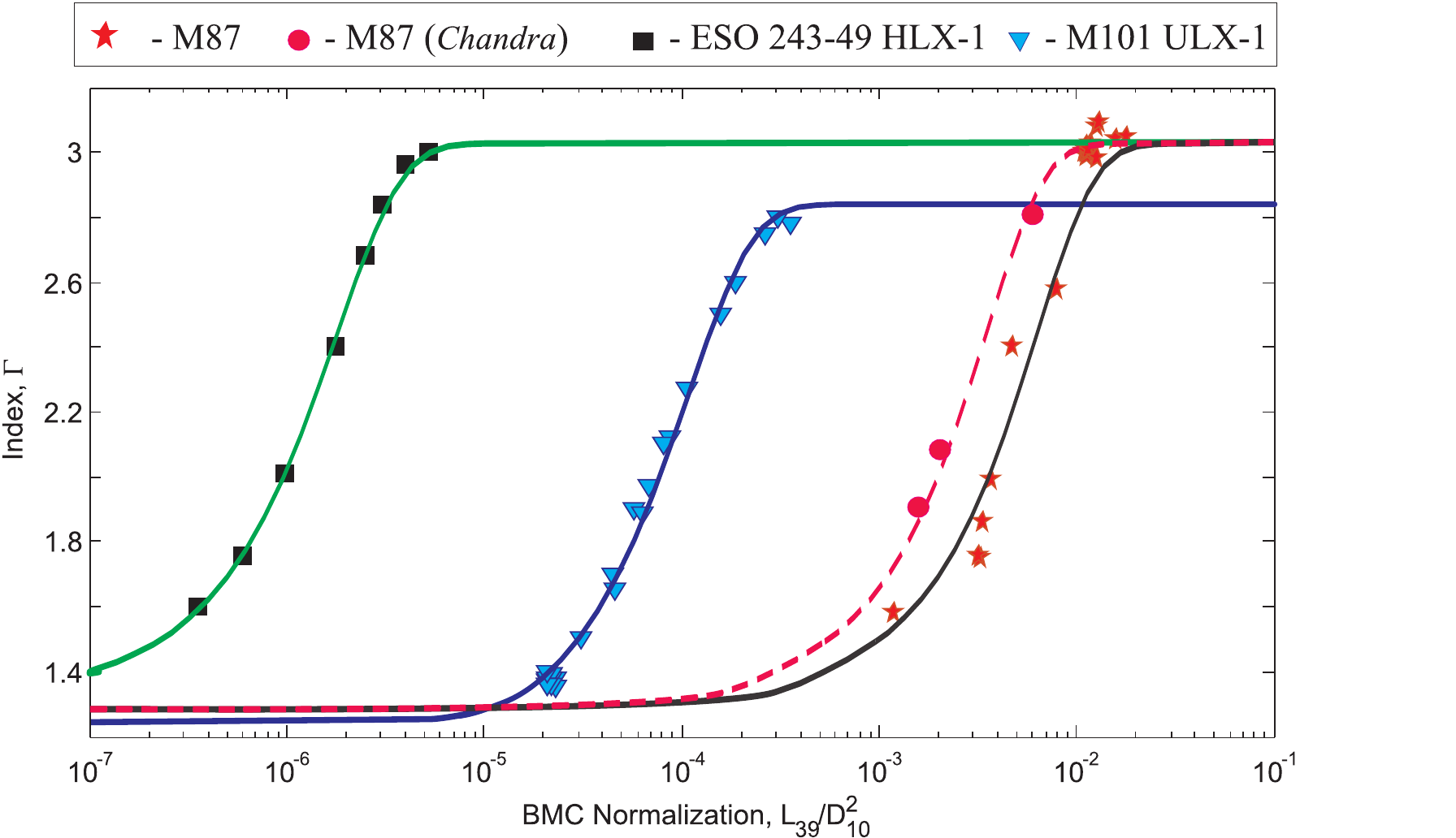}
      \caption{
Scaling of the photon index $\Gamma$ vs. the normalization $N_{BMC}$ for M87  
 (black line -- target source) using Galactic sources: XTE~J1550--564, 4U~1630--472  
and H~1743--322 (reference sources).  
Pink crosses stand for {\it RXTE} data of 4U~1630--472, green circles mark H~1743--322 data,
 and blue squares indicate XTE~J1550--564 data. Pink dashed line indicates 
the $\Gamma$-Norm correlation corrected for a nucleus emission from M87 using $Chandra$ data ($pink$ points). For 
the scaling of a BH mass of M87 we used this corrected $\Gamma$-Norm track (see details in the text).
}
\label{three_scal_1}
\end{figure*}

%
%

 \begin{figure*}
 \centering
\includegraphics[width=12cm]{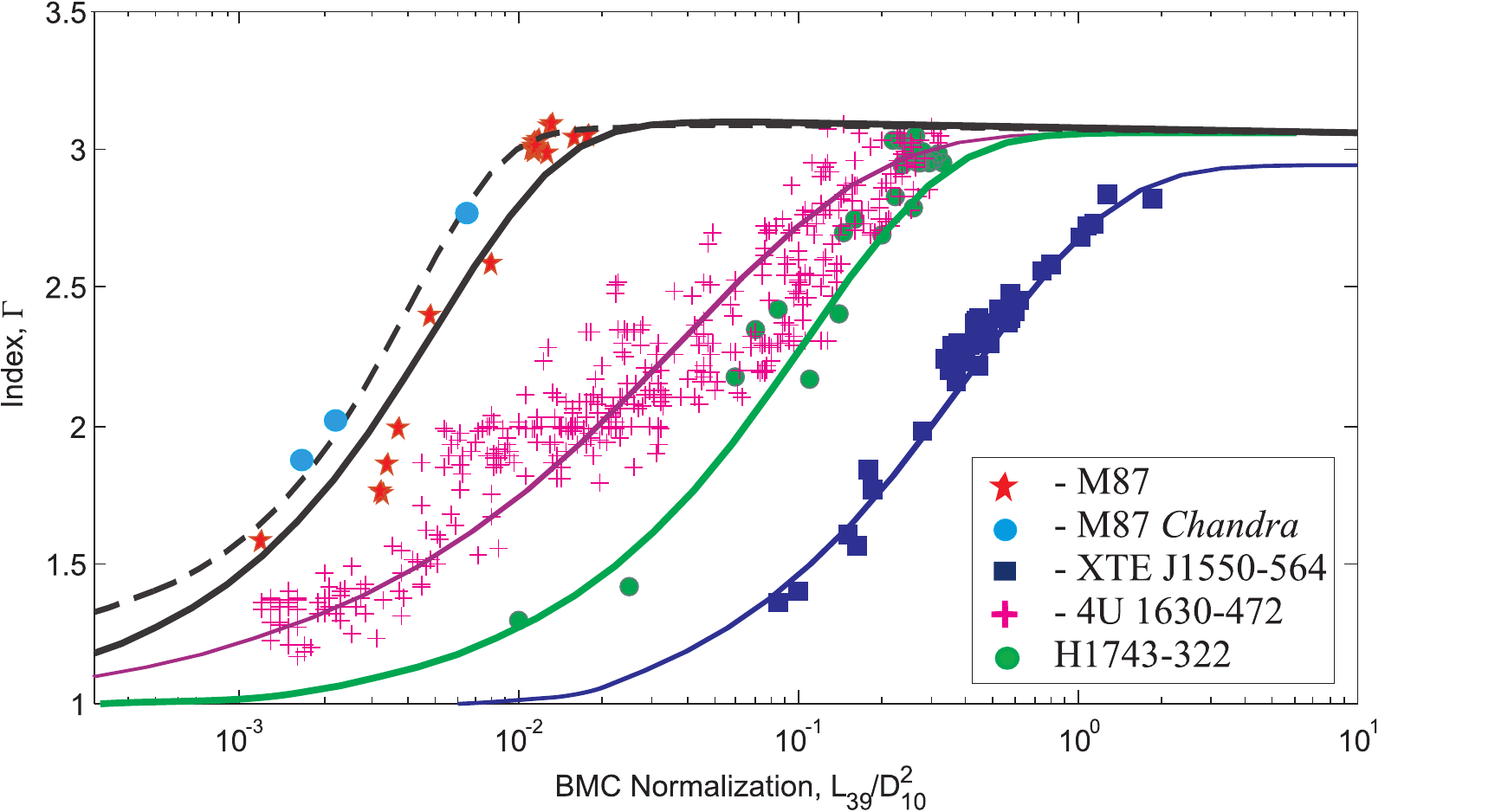}
      \caption{
Scaling of the photon index $\Gamma$ vs. the normalization $N_{BMC}$ for M87  
 (black line -- target source) using extragalactic sources: ESO~243--49 HLX--1 and M~101  ULX--1
 (reference sources).  
Black squares stand for {\it RXTE} data of ESO~243--49 HLX--1 and blue triangles indicate 
M~101  ULX--1 data. The black dashed line indicates 
the $\Gamma$-Norm correlation corrected for a nucleus emission from M87 using $Chandra$ data (bright~blue points). For 
the scaling of BH mass of M87 we used this corrected $\Gamma$-normalization track.
}
\label{three_scal_2}
\end{figure*}

The BH mass scaling process is described in detail in Appendix B. This method can be used to (i) search 
for such pairs of black holes for which the   $\Gamma$ increases with $\dot M$ (the BMC normalization, $N_{BMC}$ ) along 
the $\Gamma-N_{BMC}$ track and those for which the saturation level $\Gamma_{sat}$
is the same, and to {ii) calculate two scaling coefficients which allows determination of  the BH  mass of 
a target. 

As can be seen from Figs.~\ref{three_scal}, \ref{three_scal_1}, and \ref{three_scal_2},  
the correlations of the target sources (3C~454.3 and M87) 
and the reference sources  are characterized by similar shapes and index saturation levels.
Consequently, 
this method gives us 
a reliable scaling of these correlations with that of 3C~454.3 and M87.
In order to implement the   scaling technique we introduce an analytical 
approximation  
of the $\Gamma-N_{bmc}$ correlation, 
fitted by Eq.~(\ref{scaling function}).
 

As a result of fitting the  observed correlation  by  this function $F(x)$ [see Eq.~(\ref{scaling function})]
we obtained a set of the best-fit parameters $\cal A$, $\cal B$, $\cal D$, $N_{tr}$, and $\beta$
(see Table \ref{tab:parametrization_scal}).  
The meaning of these parameters is  described in detail in our previous papers (Titarchuk 
\& Seifina (2016), hereafter TS16; STU18; SCT18).
This function $F(x)$ is widely used for a description 
of the correlation of $\Gamma$ versus $N_{bmc}$ 
(\cite{sp09}, ST09, \cite{ST10}, 
STS14, \cite{ggt14}, Titarchuk \& Seifina 2016, 2017;  
Seifina et al. 2017, 2018a,b). 

In order to implement this BH mass determination for the target source one should rely on 
the same shape of the $\Gamma-N_{BMC}$ correlations for this target source and those for the reference sources.  
Accordingly, the only difference in values of $N_{bmc}$  for these three sources is   in 
   the ratio of the BH mass to the squared distance, 
$M_{BH}/d^2$. 
 As one can  see from Fig. ~\ref{three_scal} 
  the index  saturation value, $\cal A,$  is approximately the same for the target and reference sources (see also  the second column in Table \ref{tab:parametrization_scal}). 
The shape of the correlations for  3C~454.3 (violet line) and Cyg~X--1 
(black line)  
 are similar and the only difference between these correlations  
   is in the BMC normalization values (proportional to 
  $M_{BH}/d^2$ ratio).
  
To estimate a BH mass,  $M_t$,  of 3C~454.3 and M87 (target sources) one should slide 
the reference source correlation along the$N_{BMC}-$axis  to that of the target source 
(see Figs. \ref{three_scal}, \ref{three_scal_1} and \ref{three_scal_2}):

\begin{equation}
m_t=m_r \frac{N_t}{N_r}
\left(\frac{d_t}{d_r}
\right)^2 f_G=
m_r \frac{N_t}{N_r}
\left(\frac{z_t}{z_r}
\right)^2 f_G,
\label{scaling coefficient}
\end{equation}
\noindent where 
$m_t=M_t/M_{\odot}$,  $m_r=M_r/M_{\odot}$ are the dimensionless  BH masses with respect that of the sun, and ~$z_r$, $z_t$ are redshifts of the reference and target sources, 
correspondingly.  

In Figure~\ref{three_scal} we demonstrate   the $\Gamma-N_{bmc}$ correlation  for 3C~454.3 
(violet points) obtained  using  the $RXTE$ spectra  along with the correlations  for  the  
two Galactic reference sources  (GRO~J1655--40 (pink), Cygnus~X--1 (black)) and 
two extragalactic reference source (NGC~4051 (green line) and NGC~7469 (blue line)), which  are similar  to the correlation found  for the target source.

{
For M87, we performed scaling with both galactic and extragalactic sources. Specifically, 
Fig.~\ref{three_scal_1} presents  $\Gamma$ versus  normalization, $N_{BMC}$ for M87  
 (black line -- target source) along with $\Gamma$-Norm correlations for extragalactic sources, ESO~243--49 
HLX--1 and M~101  ULX--1 (reference sources). 
It can be seen that the $Chandra$ points go to the left with respect of other points, perhaps due to the fact that we applied a more compact angular region around the nucleus of M87  for the $Chandra$ spectra. 
Figure~\ref{three_scal_2} 
shows the photon index $\Gamma$ versus the normalization $N_{BMC}$ for M87  
along with $\Gamma-N_{BMC}$ correlations for four Galactic reference sources: XTE~J1550--564, 4U~1630--472  
and H~1743--322.  
The black  line indicates 
the $\Gamma-N_{BMC}$ correlation corrected for a nucleus emission from M87 using the $Chandra$ data. 
For an estimate  of a BH mass in M87 we use this corrected $\Gamma-N_{BMC}$ track.
The BH masses and distances  for each of these target-reference pairs are shown in 
Table~\ref{tab:par_scal}. 

We apply values of $m_r$, $m_t$, $d_r$, $d_t$, and $\cos (i)$ (see Table~\ref{tab:par_scal}) 
and then  we calculate the 
the BH mass for 3C~454.3 using the best-fit value of  
$N^{3c}_t= (1.2\pm 0.1)\times 10^{-4}$ taken at the beginning of the index saturation  
(see Fig. \ref{three_scal}) and measured
in units of $L_{39}/d^2_{10}$ erg s$^{-1}$ kpc$^{-2}$ (see Table \ref{tab:parametrization_scal}
 for values of the parameters of function $F(N_t)$ (Eq. 1)). 

For M87, the best-fit value is 
$N^{m87}_t= (0.8\pm 0.1)\times 10^{-4}$.  
Using $d_r$, $m_r$, $N_r$ (see ST09) we found that  $C_0\sim 2.0, 2.0, 1.9,$ and  $1.83$ 
for NGC~7469, NGC~4051, GRO~J1555--40, and Cyg~X--1, respectively (for 3C~454.3 case). 
Similarly, we obtained that  $C_0\sim 1.9, 2.0, 2.1,$ and  $2.03$ 
for XTE~J1550--564, H~1743--322, GRS~1915+105 and 4U~1630--472, respectively.

We find that   $M_{3C}\ge 3.4\times 10^9~M_{\odot}$ 
($M_{3C}=M_t$) 
assuming $d_{3C}\sim$~3$\times 10^3$ Mpc~\citep{Gupta17,Wright06}. 
To determine the distance to 3C~454.3 we used the formula
\begin{equation}
d_{3c}=z_{3c}c/H_0
\label{bllac_distance}
,\end{equation} 
where  the redshift $z_{3c}=0.859$  for 3C~454.3 (see { Wright 2006}), 
$c$ is the speed of light, and $H_0=70.8\pm 1.6$ km s$^{-1}$ Mpc$^{-1}$  is the Hubble constant.  
For M87 case, we obtain  $M_{87}\ge 5.6\times 10^7~M_{\odot}$ 
($M_{87}=M_t$) 
assuming $d_{m87}\sim$~16.8 Mpc~\citep{Akiyama2019_VI}. 

{
It is worth noting that for M87, we 
take into account the correction (dashed line) of the $\Gamma$-Normalization of the track. This correction for the jet 
contamination 
is important in the case of M87. In fact, for a reliable BH mass estimate we need to use the nucleus 
emission to scale the $\Gamma$-$N_{BMC}$ correlations. In addition, we selected reference sources with a 
well-known geometric factor (i.e., the angles at which we see the source). This allowed us to 
obtain a more accurate estimate of the BH  mass and not just its lower limit. It should also 
be emphasized that the BH mass value  in M87 turned out to be two orders of magnitude smaller 
than that estimated by standard methods [see e.g., \cite{Akiyama2019_VI}].
} We summarize all these results  in  Table~\ref{tab:par_scal}.

{
In addition, we tested the possible contribution from the broad-line region (BLR). To disentangle this  contribution  we investigated the low-energy part of the $Chandra$ spectrum of 3C 454.3 (which was calculated for an angular region that captures only the core) and compared it with the $Suzaku$ spectrum of 3C 454.3 (which captures both the core region and the jet region and the BLR zone). As a result, we find that few photons come from the jet and BLR zone in this range. Therefore, the contribution of BLR is small.

Celotti et al. (2007) argued that blazar spectra are contributed via the bulk 
Compton process by a cold, relativistic shell of plasma moving (and accelerating) 
along the jet of a blazar, scattering external photons emitted by the 
accretion disk and reprocessed in the BLR zone. However, the bulk 
Comptonization of disk photons provide a spectral component contributing in the 
far-ultraviolet band, and are therefore currently unobservable. On the other hand, 
the bulk Comptonization of broad-line photons may yield a significant feature in 
the soft X-ray band. Such a feature is rather transient however, and  only dominates over 
the nonthermal continuum when: (i) the dissipation occurs close to, 
or within, 
the BLR; and (ii) other competing processes, like the synchrotron self-Compton emission, 
yield a negligible flux in the X-ray band. The presence of a bulk Compton component 
may account for the X-ray properties of high-redshift blazars that show a flattening 
(and possibly a hump) in the soft X-rays, previously interpreted as due to intrinsic 
absorption. 

We also tested $Suzaku$ and $Chandra$ data and concluded that the spectra of 3C454.3 are 
even less influenced by jet. In fact, the $Chandra$ spectra focused on a compact area 
around the nucleus of 3C~454.3 (without the jet contamination), while $Suzaku$ spectra 
come from some wider zone (nucleus plus jet and outer parts of the AGN). We analyzed and 
compared these $Suzaku$ and $Chandra$ spectra. As a result, we concluded that the jet 
contribution (SSC+CE) is too low for the observations used in the paper.
}

The obtained  BH mass estimate is in agreement with a high bolometric luminosity for 3C~454.3 
and the $kT_s$  value which is in the range of 100--260 
eV using the {\it Suzaku} and $Chandra$ spectra. 
For example, Shakura \& Sunyaev, (1973) and 
 Novikov \& Thorne, (1973) provide an effective temperature of the accretion material,
$T_{eff}\sim T_s\propto M_{BH}^{-1/4}$.

It is also important to emphasize  
that our  
 BH mass estimate in 3C~454.3 is consistent with  a SMBH mass of 
$(0.5 - 1.5) \times 10^9$ M$_{\odot}$ (Gupta et al., 2017; Woo \& Urry 2002; Liu et
al. 2006; Sbarrato et al. 2012; Gu et al. 2001). 
{The derived BH mass is the lower limit estimate only because the photon index 
versus the normalization has the uncertainty with geometrical factor $f_G=(\cos i_d)_r/(\cos i_d)_t$. Generally, the photon index 
versus the QPO frequency correlation enables us to obtain the precise BH mass (see ST09). 
Although our BH mass estimate is only  the lower limit of that, it significantly 
constrains the BH mass for 3C~454.3 (see Table \ref{tab:par_scal}). 


%
%
\begin{table*}
 \caption{Black hole  masses and distances}
 \label{tab:par_scal}
 \centering 
 \begin{tabular}{lllllc}
 \hline\hline                        
      Source   & M$_{dyn}$ (M$_{\odot})$ & i$_{orb}$ (deg) & d (kpc) & $M_{lum}$ (M$_{\odot}$)  & M$_{scal}$ (M$_{\odot}$) \\
      \hline
GRO~J1655--40  &   6.3$\pm$0.3$^{(1, 2)}$ &  70$\pm$1$^{(1, 2)}$ &  3.2$\pm$0.2$^{(3)}$    &   ... \\
Cyg~X--1       &   6.8 -- 13.3$^{(4, 5)}$ &  35$\pm$5$^{(4, 5)}$ &  2.5$\pm$0.3$^{(4, 5)}$  & ... & 7.9$\pm$1.0 \\
NGC~4051$^{(6, 7, 8, 9, 10)}$  & ... &     ...      & $\sim$10$\times 10^3$ & ... & $\ge 6\times 10^{5}$ \\
NGC~7469$^{(11, 12, 13)}$  & ... &     ...      & $\sim$70$\times 10^3$ & ... & $\ge 3\times 10^{6}$ \\
XTE~J1550-564$^{(14,15,16)}$  &   9.5$\pm$1.1 &  72$\pm$5    &   $\sim$6           &...&   10.7$\pm$1.5$^c$ \\
H~1743-322$^{(17)}$     &   ...    &  75$\pm$3    &   8.5$\pm$0.8          &...&   13.3$\pm$3.2$^c$ \\
GRS~1915+105$^{(18)}$     &   ...    &  60-70    &   8.5$\pm$0.8          &...&   13.3$\pm$3.2$^c$ \\
4U~1630--47$^{(19)}$    &       ...     &   $\leq$70   &   $\sim$10 -- 11    &...&   9.5$\pm$1.1     \\
M101~ULX-1$^{(20,21,22,23,24)}$     & 3 -- 1000     &   60$^{(25)}$   & (6.4$\pm$0.5)$\times 10^3$,  &...& $\ge 3.2\times10^{4}$, $\ge 4.3\times10^{4}$ \\
                                                     &                    &              &  (7.4$\pm$0.6)$\times 10^3$ & &  \\
ESO 243-49 HLX-1$^{(26,27)}$  & ... &   70$^{(28)}$  & $\sim$95$\times 10^3$ &8$\pm 4\times 10^4$& $\ge 7.2\times10^{4}$ \\
3C~454.3$^{(29)}$  & ... &     $\le$4$^{(29)}$      & $\sim$30$\times 10^5$ & ... &$\sim 3.4\times 10^{9}$ \\
M87$^{(30,31)}$  & ... &    $\le$20$^{(31)}$      & (16.8$\pm$ 0.8) $\times 10^3$ & 6.5$\pm$ 0.7 $\times 10^9$ & $\sim$ 5.6 $\times 10^7$ \\
 \hline                                             
 \end{tabular}
 \tablebib{
(1) Green et al. 2001; (2) Hjellming \& Rupen 1995; (3) Jonker \& Nelemans G. 2004 
(4) \cite{Herrero95}; (5) \cite{Ninkov87}; 
(6) M$^c$Hardy et al. (2004); 
(7) \cite{haba08}; 
(8) Pounds \& King (2013); 
(9) \cite{Lobban11}; 
(10) \cite{Terashima09}; 
(11) \cite{Peterson04}; 
(12) \cite{Peterson14}; 
(13) \cite{Shapovalova17}; 
(14) Orosz et al. 2002; 
(15) S$\grave a$nchez-Fern$\grave a$ndez et al. 1999; 
(16) Sobczak et al. 1999;  
(17) Petri 2008; 
(18) Mirabel \& Rodrigues 2007 
(19) King et al. 2014 
(20) STS14;
(21) Shappee \& Stanek 2011;
(22) Mukai et al. 2005; 
(23) Kelson et al. 1996;
(24) TS15;
(25) Liu et al. 2013; 
(26) Copperwheat et al. 2007; 
(27) Farrell et al, 2009; 
(28) Soria et al. 2013 
(29) Gupta et al. (2017);
(30) Akiyama et al. 2019a 
(31) Davis et al. 2011
}
 \end{table*}

We  derived the bolometric luminosity using  the normalization of the BMC model   and found it to be between $2\times 10^{46}$ erg/s and $4\times 10^{47}$ erg/s  (assuming isotropic radiation). 
Observations of many  Galactic BHs (GBHs) and their X-ray spectral analysis 
(see  ST09, \cite{tsei09}, \cite{ST10} and STS14, SCT18, STU18)
confirm    the TZ98  prediction that the spectral (photon) index saturates with the mass accretion rate, the latter being related to the Eddington ratio.  This value is not so far from the Eddington limit for the obtained BH mass and assumed source distance (see Table~\ref{tab:par_scal}). 

The resulting  mass of the central BH in 3C~454.3 is in agreement with previous estimates of a BH mass (Gupta et al., 2017; Woo \& Urry 2002; Liu et al. 2006; Sbarrato et al. 2012; Gu et al. 2001). At the same time its high value raises the question of the origin of such SMBHs and their high variability.
The answer to this question can be given by a timing analysis using the longest interval of observations. There are indications of a duality of the central object (Volvach et al., 2017), which may partly manifest itself in the form of a high variability of its radiation and; updated scenarios and justification for feeding such a system of two BHs are however needed. Further observations of 3C~454.3 in a different range of frequencies could shed light on these and other problems.

\section{Conclusions \label{summary}} 

We found the spectral  
state transitions observed in  3C~454.3 {and M87} using the full 
set of $Suzaku$ (2007 -- 2010), $Swift$ (2005 -- 2009), {\it RXTE}  (1996 -- 2010), $Chandra$, $ASCA,$ and $Beppo$SAX data and demonstrated a validity of the spectral  BMC model for fitting   the observed spectra  
for all  these sets of observations, independently of the spectral state of the sources. 

We investigated the X-ray flaring properties of 3C~454.3 { and M87} and revealed spectral state transitions during the outbursts using the index$-$normalization (proportional to  $\dot M$) correlation observed in 3C~454.3 { and M87}, which 
were similar to those in Galactic BHs as well as in a number of extragalactic BH sources. 
In particular, we find that 3C~454.3 follows the $\Gamma-\dot M$ correlation previously obtained for extragalactic SMBH sources, NGC~4051 (SCT18), NGC~7469 (STU18), Galactic BHs GRO~J1655--40, and Cyg~X--1 (ST09) for which  we take  into account the 
particular values of the $M_{BH}/d^2$ ratio (see Fig.~\ref{three_scal}).
{
We find that M87 follows the $\Gamma-\dot M$ correlation previously obtained for extragalactic IMBHs, ESO~243--49 HLX--1 (ST16a) and M~101 ULX-1 (TS16b), and Galactic BHs, XTE~J1550--564, 4U~1630--472, GRS~1915+105, and H~1743--322 (ST09) 
which we use as reference sources, for which  we take into  account the 
particular values of the $M_{BH}/d^2$ ratio (see Figs.~\ref{three_scal_1} and \ref{three_scal_2}).

}
The photon index $\Gamma$ of the 3C~454.3 spectra is  in the range of $\Gamma = 1.2 - 2.1$, 
while the photon index $\Gamma$ of the M87 spectra varies in the range of $\Gamma = 1.3 - 3.1$. 
We tested a possible effect  of the presence of   the  jet 
for 3C~454.3 and M87 on our results 
using $Chandra$ observations. We conclude that the jet 
contribution in the spectrum is relatively small  in the case of 3C~454.3, 
but can contribute to the M87 spectra. Therefore, to estimate the BH mass for M87, it was crucial to separate the 
radiation of the nucleus and the jet.

We  also find that the peak bolometric luminosity for 3C~454.3 is about $2\times 10^{46}$ erg s$^{-1}$.  
 We use the observed index--mass accretion rate correlation to estimate $M_{BH}$  in 3C~454.3.
This scaling method was  successfully applied to find  BH masses of Galactic (e.g. ST09, STF13) 
and extragalactic BHs (TS16; \citet{sp09}; \cite{ggt14}; Titarchuk \& Seifina (2017); Seifina et al. 2017, 2018a,b).  
Applying  the scaling  technique  to the X-ray data  for {\it Suzaku}, {\it Swift}, $Chandra$, $ASCA$, $Beppo$SAX, and {\it RXTE} observations of 3C~454.3 and M87 allows us to estimate  $M_{BH}$ for these particular sources.  We find  values of $M_{BH}\sim 3.4\times 10^9 M_{\odot}$ (for 3C~454.3) and $M_{BH}\sim 5.6\times 10^7 M_{\odot}$ (for M87).   

{
It must  also be emphasized that the mass of the BH in M87 estimated in this paper 
($M^{m87}_{BH} \sim 5.6\times 10^7 M_{\odot}$) turned out to be two orders of magnitude 
smaller than that estimated by standard methods 
(Gebhardt \& Thomas 2009; Gebhardt et al. 2011; \cite{Akiyama2019_VI}) that indicate a BH mass for M87 of 
$\ge 6\times 10^9$ M$_{\odot}$.
This discrepancy in  BH mass estimates for M87 poses a problem to observers and theorists. However, the power spectrum  lead us to lower the BH mass in M87 (see Sect.  4.1).
}

Our  BH mass estimate for 3C 454.3 on the other hand is 
in agreement with the previous BH mass evaluations of $0.5 - 4.5 \times 10^9$ M$_{\odot}$ 
derived using different BH mass estimate methods and multiwavelength observations  (Gupta et al., 2017; Woo \& Urry 2002; Liu et al. 2006; Sbarrato et al. 2012; Gu et al. 2001). 
Combining all these estimates with  the inferred  low temperatures of the seed disk photons $kT_s$ 
we  establish  that the compact object of 3C~454.3 is likely to 
be a SMBH with at least $M^{3c}_{BH}> 3.4\times10^9 M_{\odot}$. 


\begin{acknowledgements}

This research was performed using  data supplied by the UK $Swift$ Science Data 
Centre at the University of Leicester. We also used the data from the GLAST-AGILE 
Support program at {\it www.oato.inaf.it/blazars/webt/}. This paper has made use of up-to-date SMART optical/near-infrared 
light curves that are available at {\it www.astro.yale.edu/smarts/glast/home.php}.   
We also acknowledge the deep analysis of  the  paper  by the referee.
We thank Lubov Ugolkova for useful discussion on the subject of the presented paper. 
\end{acknowledgements}

\bibliographystyle{aa}


\appendix

\section{Comptonized spectrum using the BMC model}

The {\it BMC} model describes the outgoing spectrum as a convolution 
of the input ``seed''  blackbody-like spectrum, whose normalization is
$N_{BMC}$ and color temperature is $kT$,  with the  Comptonization Green's function. 
It was realized a long time ago (see e.g., Titarchuk et al. (1997); Shrader \& Titarchuk (1998); Laurent \& Titarchuk (1999), hereafter LT99) that one should  compute the expected spectral energy distribution for an accreting BH  and neutron star binaries in terms of only four  model parameters: the disk color temperature, a geometric factor related to the illumination of the BH site by the disk, a spectral index related to the efficiency of the Comptonization (thermal or bulk-motion up-scattering), and the normalization (the so-called BMC model). In this case the  shape of the emergent spectrum is described  by Eq. (\ref{bmc_spectrum}). 
 If we want take into account the recoil effect we should  introduce one more parameter $E_{cf}$ which characterizes the high-energy cutoff energy (see the COMPTT or COMPTB models).  The BMC analytical model was  successfully checked by LT99 and Laurent \& Titarchuk (2011) using the Monte Carlo  simulations.
These latter authors demonstrated that the core of the spectrum,  until the recoil high-energy cut-off,  is exactly described by the BMC model.     
 

\section{The $\Gamma$-normalization correlation and details of the BH mass estimates}

The evaluation method  of a 
black holes mass   is based on a comparative 
analysis of similar $\Gamma-N_{BMC}$ tracks by scaling their characteristics.
It was shown in Titarchuk \& Zannias (1998) and LT99 that the photon index $\Gamma$
monotonically increases with an increase in the accretion rate (which is 
proportional to the normalization parameter in the Compotonization model; e.g.. 
COMPTT, COMPTB, BMC) and saturates, i.e., reaches a constant level, with 
large values of the mass accretion rate $\dot M$. The steepness for the $\Gamma-N_{BMC}$ 
track and the saturation level of $\Gamma_{sat}$ may be different for 
different black holes. The comparative analysis in the scaling problem 
is therefore simplified as searching for such pairs of black holes for which the rate of $\Gamma$ increases
with $\dot M$ ($N_{BMC}$) along the $\Gamma-N_{BMC}$ track and the saturation level $\Gamma_{sat}$
is the same. It is assumed that for each BH from such a pair, distances are known. 
In addition, the mass of one of the black holes of this pair is known. The scaling 
process itself consists in calculating two scaling coefficients and then with their 
help the mass of the desired BH is determined.


We now briefly reiterate the main points of the scaling method. 
Shaposhnikov \& Titarchuk (2007), hereafter ST07, apply an inverse proportionality  of a  frequency of 
quasi-periodic oscillations (QPO)  on BH mass in order  to estimate the latter. 
 ST07  
also present theoretical arguments in terms of the transition layer model
for  the  inverse proportionality  of QPO frequencies
on BH mass. Therefore, a first scaling law claims that
\begin{equation}
s_\nu=\frac{\nu_r}{\nu_t}=\frac{M_t}{M_r},
\end{equation}
where subscript $r$ and $t$ subscripts denote reference and target sources.

The second scaling law, which we use as a basis for our scaling technique,
is 
the detected   intensity dependence of
the source luminosity and distance:
\begin{equation}
\frac{F_r}{F_t}=\frac{L_r}{L_t}\, \frac{d^2_t}{d^2_r} .
\label{ratio_fluxes}
\end{equation}
Here, $F$ stands for the source flux detected by an observer on Earth,
$L$ is the source luminosity, and $d$ is a source distance. The
luminosity $L$ can be represented as
\begin{equation}
L=\frac{GM_{BH}}{R_*}\dot{M} \eta \sim \frac{GM_{BH}}{R_S} \dot{M} \eta \sim
\dot{M} \eta = M_{BH} \dot{m} \eta
,\end{equation}
where  $R_*$ is the effective radius where energy release occurs, $\eta$ is the
efficiency of gravitation energy conversion into radiation power, $\dot{M}$ is the
accretion rate, and $\dot{m}$ is its dimensionless analog normalized by the Eddington
luminosity. Both $\dot m$ and $\eta$ are considered to be the same for two
different sources and for the same spectral state, which lead us  to $L_r/L_t=M_r/M_t=1/s_\nu$. 
In our analysis of energy spectra from BHs  we determine  normalization of seed photon radiation supplied by an accretion flow (disk) prior to Comptonization. The ratio of this
normalization for
the two sources in the same spectral state can be written as
\begin{equation}
s_N=\frac{N_r}{N_t}=\frac{L_r}{L_t}\frac{d^2_t}{d^2_r}f_G.
\end{equation}
Here, $f_G$ is a  geometry factor that comes due to the fact that the accretion
disk which produces thermal input  for Comptonization has a plane geometry.
Therefore, in the case of radiation coming directly from the disk it would have
the value $f_G=(\cos i_d)_r/(\cos i_d)_t$, where $ i_d$  is the inclination
angle of the disk. 
When information on a (target and reference) source  inclination  is  available we
can use the values  $i_{d,r}$  and $i_{d,t}$.

We  are now ready 
to write down the final equations of our scaling
analysis.
When $s_N$ are measured directly using observations, 
 the mass 
of the target source can be calculated as
\begin{equation}
m_t= m_r \frac{d_t^2}{d_r^2}\frac{f_G}{s_N}
\label{mass}
.\end{equation}
This equation allows estimation of a ratio $m_t$ using values of $m_r$, $s_N$,  $f_G$, $d_t$ and $d_r$.  

Using Eq.~(\ref{mass}), 
the problem of
BH mass and distance measurements for a target source is reduced to
the determination of scaling coefficients  $s_N$ with
respect to the data for a reference source. This can be obtained by a technique
similar to that adopted by ST07. Specifically, after scalable state transition
episodes are identified for two sources, the correlation pattern for
a reference transition is parameterized in terms of the analytical function (see also ST09),
\begin{equation}
F(x)= {\cal A} - ({\cal D}\cdot {\cal B})\ln\{\exp[(1.0 - (x/x_{tr})^{\beta})/{\cal D}] + 1\},
\label{scaling function}
\end{equation}
where $x=N_{bmc}$.


After rearrangement,  the BH mass $M_t$  of a target source 
can be evaluated using  the formula (see TS16)
\begin{equation}
m_t=C_0 {N_t} {d_t}^2 f_G 
\label{C0 coefficient}
,\end{equation}
\noindent where 
$C_0=(1/d_r^2)(m_r/N_r)$ is the scaling coefficient for the reference source, 
BH masses $m_t$ and $m_r$ are in solar units, and $d_r$ is the distance to a particular reference source  measured in kiloparsecs.

%
%

\begin{table*}
\caption{Best-fit parameters of spectral analysis of 
{\it RXTE} 
observations of 3C~454.3 ($R1$ -- $R2$ sets) in the 3 -- 50~keV energy range$^{\dagger}$. 
Parameter errors correspond to 90\% confidence level.}\label{tab:fit_table_RXTE_1}
\centering
\begin{tabular}{lcllllccccccccc}
\hline\hline
Observational ID & MJD    & $\alpha=\Gamma-1$ & $kT_s$ & $N_{BMC}^{\dagger\dagger\dagger}$ &  $\chi^2_{red}$ (dof)\\ 
                 & (day)  &                   & (keV)  & ($L_{35}/d^2_{10}$)               &            \\%
\hline                                                           
10360-01-23-00 & 50358.56 & 0.63$\pm$0.09 & 0.5$\pm$0.2   &  1.6$\pm$0.1  &   0.75 (49)   \\
10360-01-36-00 & 50265.65 & 0.54$\pm$0.09 & 0.5$\pm$0.2   &  1.0$\pm$0.2  &   0.87 (47)   \\
10360-01-27-00 & 50329.59 & 0.7$\pm$0.1   & 0.5$\pm$0.2   &  2.1$\pm$0.2  &   0.95 (47)   \\
10360-01-40-00 & 50240.21 & 0.76$\pm$0.09 & 0.7$\pm$0.2   &  2.1$\pm$0.3  &   0.87 (49)   \\
10360-01-47-00 & 50187.57 & 0.3$\pm$0.2   & 0.5$\pm$0.2   &  0.4$\pm$0.1  &   0.85 (54)   \\
10360-01-48-00 & 50181.11 & 0.6$\pm$0.2   & 0.4$\pm$0.4   &  1.2$\pm$0.3  &   0.94 (54)   \\
20346-01-01-00 & 50389.18 & 0.4$\pm$0.2   & 0.2$\pm$0.1   &  0.4$\pm$0.1  &   0.85 (54)   \\
20346-01-02-00 & 50397.65 & 0.36$\pm$0.07 & 0.2$\pm$0.1   &  0.5$\pm$0.1  &   0.97 (54)   \\
20346-01-03-00 & 50406.32 & 0.94$\pm$0.07 & 0.6$\pm$0.4   &  2.1$\pm$0.1  &   0.89 (54)   \\
20346-01-04-00 & 50411.66 & 0.53$\pm$0.09 & 0.8$\pm$0.6   &  1.4$\pm$0.1  &   0.76 (54)   \\
20346-01-05-00 & 50419.32 & 0.21$\pm$0.10 & 0.10$\pm$0.01 &  0.57$\pm$0.09&   1.05 (54)   \\
20346-01-07-00 & 50430.41 & 0.62$\pm$0.09 & 0.11$\pm$0.01 &  0.69$\pm$0.09&   1.12 (54)   \\
20346-01-08-00 & 50437.14 & 0.61$\pm$0.03 & 0.11$\pm$0.01 &  0.83$\pm$0.04&   0.83 (54)   \\
20346-01-10-00 & 50453.87 & 0.71$\pm$0.05 & 0.09$\pm$0.01 &  1.04$\pm$0.02&   0.82 (54)   \\
20346-01-11-00 & 50461.42 & 0.71$\pm$0.05 & 0.17$\pm$0.01 &  1.04$\pm$0.02&   0.82 (54)   \\
20346-01-12-00 & 50468.77 & 1.01$\pm$0.07 & 0.12$\pm$0.01 &  2.56$\pm$0.08&   0.95 (54)   \\
20346-01-14-00 & 50480.36 & 1.00$\pm$0.08 & 0.20$\pm$0.03 &  2.82$\pm$0.06&   0.93 (54)   \\
20346-01-15-00 & 50488.36 & 0.85$\pm$0.02 & 0.19$\pm$0.01 &  1.94$\pm$0.09&   0.93 (54)   \\
20346-01-16-00 & 50494.49 & 0.81$\pm$0.02 & 0.26$\pm$0.02 &  1.07$\pm$0.07&   1.12 (54)   \\
20346-01-17-00 & 50543.51 & 0.69$\pm$0.08 & 0.29$\pm$0.06 &  1.01$\pm$0.09&   0.73 (54)   \\
20346-01-18-00 & 50551.81 & 0.94$\pm$0.07 & 0.27$\pm$0.06 &  5.78$\pm$0.09&   0.98 (54)   \\
20346-01-24-00 & 50594.23 & 0.69$\pm$0.04 & 0.46$\pm$0.09 &  1.16$\pm$0.08&   0.95 (54)   \\
20346-01-26-00 & 50608.24 & 0.13$\pm$0.09 & 0.31$\pm$0.08 &  0.40$\pm$0.08&   1.00 (54)   \\
20346-01-27-00 & 50617.22 & 0.93$\pm$0.08 & 0.20$\pm$0.09 &  1.56$\pm$0.09&   0.73 (54)   \\
20346-01-28-00 & 50622.50 & 0.58$\pm$0.03 & 0.57$\pm$0.08 &  1.99$\pm$0.05&   0.92 (54)   \\
20346-01-29-00 & 50627.83 & 0.68$\pm$0.06 & 0.55$\pm$0.09 &  1.80$\pm$0.06&   0.87 (54)   \\
20346-01-30-00 & 50633.97 & 0.96$\pm$0.05 & 0.46$\pm$0.04 &  1.37$\pm$0.09&   0.92 (54)   \\
20346-01-31-00 & 50643.58 & 0.77$\pm$0.06 & 0.22$\pm$0.07 &  1.20$\pm$0.07&   0.87 (54)   \\
20346-01-35-00 & 50673.14 & 0.55$\pm$0.05 & 0.30$\pm$0.09 &  1.42$\pm$0.09&   0.87 (54)   \\
20346-01-38-00 & 50693.06 & 0.88$\pm$0.02 & 0.12$\pm$0.06 &  1.83$\pm$0.09&   0.79 (54)   \\
20346-01-43-00 & 50725.75 & 1.00$\pm$0.04 & 0.24$\pm$0.05 &  3.05$\pm$0.08&   0.90 (54)   \\
20346-01-44-00 & 50733.43 & 0.78$\pm$0.06 & 0.12$\pm$0.07 &  1.19$\pm$0.09&   0.99 (54)   \\
20346-01-45-00 & 50739.89 & 0.72$\pm$0.02 & 0.12$\pm$0.05 &  1.12$\pm$0.06&   0.95 (54)   \\
20346-01-46-00 & 50746.76 & 0.71$\pm$0.08 & 0.12$\pm$0.05 &  1.93$\pm$0.08&   0.91 (54)   \\
20346-01-47-00 & 50754.55 & 0.22$\pm$0.02 & 0.16$\pm$0.05 &  0.31$\pm$0.06&   1.03 (54)   \\
20346-01-48-00 & 50765.24 & 0.26$\pm$0.04 & 0.15$\pm$0.06 &  0.41$\pm$0.07&   1.05 (54)   \\
20346-01-51-00 & 50783.86 & 1.03$\pm$0.07 & 0.14$\pm$0.04 &  2.27$\pm$0.07&   0.96 (54)   \\
20346-01-52-00 & 50789.39 & 1.04$\pm$0.03 & 0.10$\pm$0.04 &  7.30$\pm$0.09&   0.97 (54)   \\
20346-01-53-00 & 50796.13 & 0.87$\pm$0.03 & 0.14$\pm$0.03 &  1.28$\pm$0.09&   0.86 (54)   \\
20346-01-54-00 & 50802.97 & 0.69$\pm$0.08 & 0.21$\pm$0.04 &  1.19$\pm$0.09&   0.99 (54)   \\
\hline
\label{tab:par_rxte}
\end{tabular}
\tablefoot{ 
$^\dagger$ The spectral model is  tbabs*bmc;  where $N_H$ is fixed at 
a value 5.0$\times 10^{22}$ cm$^{-2}$ (see Sect.~\ref{model choice}); 
parameter $\log(A)$ is low variable around 1.5; 
$^{\dagger\dagger\dagger}$ 
For normalization parameter $N_{BMC}=L_{35}/d^2_{10}$ (see Eqs. \ref{bmc_spectrum}, \ref{ratio_fluxes})    
where 
$L_{35}$ is the source luminosity in units of 10$^{35}$ erg/s and  
$d_{10}$ is the distance to the source in units of 10 kpc. 
}
\end{table*}


\begin{table*}
\caption{Best-fit parameters of spectral analysis of 
{\it RXTE} 
observations of 3C~454.3 ($R3$ -- $R5$ sets) in 3 -- 50~keV energy range$^{\dagger}$. 
Parameter errors correspond to 90\% confidence level.}\label{tab:fit_table_RXTE_1}
\centering
\begin{tabular}{lcllllcrccccccccc}
\hline\hline
Observational ID & MJD    & $\alpha=\Gamma-1$ & $kT_s$ & log(A)$^{\dagger\dagger}$ & $N_{BMC}^{\dagger\dagger\dagger}$ &  $\chi^2_{red}$ (dof) & F$_1$/F$_2^{\dagger\dagger\dagger\dagger}$ \\
                 & (day)  &                   & (keV)  &        & ($L_{35}/d^2_{10}$)               &            \\%
\hline                                                           
30264-01-02-00 & 50816.05 & 0.94$\pm$0.12 & 0.38$\pm$0.05 &  -0.20$\pm$0.04 &  12.3  0.5 &  1.14 (46) &    35.87/23.97  \\
30264-01-03-00 & 50825.97 & 0.53$\pm$0.09 & 0.42$\pm$0.08 &   2.00$^{\dagger\dagger}$   &  1.04 0.07 &  1.18 (48) &    9.26/8.28   \\
30264-01-04-00 & 50831.97 & 0.53$\pm$0.03 & 0.42$\pm$0.09 &   2.00$^{\dagger\dagger}$   &  1.03 0.09 &  1.16 (48) &    9.26/8.28   \\
30264-01-05-00 & 50839.97 & 0.54$\pm$0.02 & 0.25$\pm$0.04 &   2.00$^{\dagger\dagger}$   &  1.06 0.09 &  0.95 (48) &    11.8/10.4   \\
30264-01-06-00 & 50846.97 & 0.39$\pm$0.03 & 0.19$\pm$0.08 &   2.00$^{\dagger\dagger}$   &  1.05 0.08 &  0.95 (48) &    11.8/10.4   \\
30264-01-08-00 & 50860.97 & 0.39$\pm$0.03 & 0.18$\pm$0.05 &   2.00$^{\dagger\dagger}$   &  0.66 0.05 &  0.98 (48) &     9.7/9.8   \\
30264-01-09-00 & 50911.97 & 0.40$\pm$0.05 & 0.19$\pm$0.07 &   2.00$^{\dagger\dagger}$   &  0.81 0.08 &  0.96 (41) &    11.87/12.0   \\
30264-01-10-00 & 50916.97 & 0.39$\pm$0.01 & 0.18$\pm$0.03 &   2.00$^{\dagger\dagger}$   &  0.65 0.06 &  0.96 (41) &    11.86/12.1   \\
30264-01-12-00 & 50932.97 & 0.74$\pm$0.03 & 0.13$\pm$0.02 &   2.00$^{\dagger\dagger}$   &  0.90 0.10 &  0.74 (49) &    10.05/8.45  \\
30264-01-13-00 & 50939.97 & 0.19$\pm$0.02 & 0.12$\pm$0.02 &   2.00$^{\dagger\dagger}$   &  0.53 0.20 &  1.16 (49) &    10.9/13.2  \\
30264-01-14-00 & 50945.97 & 0.18$\pm$0.01 & 0.12$\pm$0.01 &   2.00$^{\dagger\dagger}$   &  0.63 0.18 &  0.93 (49) &    13.1/15.8  \\
30264-01-15-00 & 50952.97 & 0.88$\pm$0.02 & 0.20$\pm$0.01 &   2.00$^{\dagger\dagger}$   &  1.07 0.10 &  0.95 (48) &    12.68/8.13 \\
30264-01-17-00 & 50966.97 & 0.89$\pm$0.03 & 0.20$\pm$0.01 &   2.00$^{\dagger\dagger}$   &  0.69 0.10 &  0.95 (48) &    12.68/8.13 \\
30264-01-19-00 & 50980.97 & 0.30$\pm$0.03 & 0.22$\pm$0.01 &   2.00$^{\dagger\dagger}$   &  0.41 0.09 &  1.01 (47) &    10.58/11.62 \\
30264-01-20-00 & 50986.51 & 0.42$\pm$0.05 & 0.27$\pm$0.01 &   2.00$^{\dagger\dagger}$   &  0.34 0.08 &  0.95 (47) &     8.4/8.29  \\
93150-03-01-00 & 54309.72 & 0.92$\pm$0.06 & 0.46$\pm$0.04 &   2.00$^{\dagger\dagger}$   &  7.92 0.10 &  1.06 (45) &     7.05/5.21 \\
93150-03-01-01 & 54309.93 & 0.95$\pm$0.07 & 0.47$\pm$0.06 &   1.99$\pm$0.09  &   7.69 0.11&   0.97 (45) &      6.85/5.06\\
93150-03-01-02 & 54311.27 & 0.90$\pm$0.05 & 0.50$\pm$0.03 &   1.66$\pm$0.05 &   6.26 0.09 &  1.19 (45) &      5.72/4.02\\
93150-03-01-04 & 54314.60 & 0.98$\pm$0.02 & 0.46$\pm$0.03 &   1.95$\pm$0.08 &    7.41 0.10 &  1.13 (45) &     6.58/4.83 \\
93150-03-01-06 & 54313.30 & 0.97$\pm$0.05 & 0.48$\pm$0.04 &   1.51$\pm$0.06 &    7.51 0.19 &  1.09 (45) &     6.77/4.80 \\
93150-03-02-00 & 54315.78 & 0.96$\pm$0.08 & 0.43$\pm$0.05 &   1.60$\pm$0.07 &    7.78 0.16 &  1.11 (45) &     6.86/5.03 \\
93150-03-03-03 & 54619.15 & 0.96$\pm$0.09 & 0.44$\pm$0.02 &   1.59$\pm$0.06 &    5.31 0.15 &  0.72 (45) &     4.67/3.42 \\
93150-03-03-05 & 54621.02 & 0.95$\pm$0.08 & 0.43$\pm$0.01 &   1.63$\pm$0.04 &    3.53 0.17 &  0.95 (45) &     3.10/2.29 \\
93150-03-04-00 & 54624.36 & 0.91$\pm$0.07 & 0.38$\pm$0.02 &   1.76$\pm$0.05 &    3.57 0.16 &  0.97 (45) &     3.05/2.34 \\
93150-03-04-01 & 54623.85 & 0.89$\pm$0.06 & 0.28$\pm$0.04 &   1.61$\pm$0.09 &    3.27 0.18 &  0.79 (45) &     2.64/2.18 \\
93150-03-04-02 & 54625.88 & 0.88$\pm$0.07 & 0.41$\pm$0.09 &   0.93$\pm$0.07  &    4.12 0.15 &  1.06 (45) &     3.65/2.48 \\
93150-03-04-04 & 54627.44 & 0.90$\pm$0.08 & 0.35$\pm$0.02 &   0.99$\pm$0.08  &    4.14 0.13 &  0.98 (45) &     3.54/2.54 \\
93150-03-04-05 & 54628.12 & 0.94$\pm$0.07 & 0.26$\pm$0.03 &   1.05$\pm$0.09  &    4.48 0.19 &  0.94 (45) &     3.65/2.86 \\
93150-03-05-02 & 54632.34 & 0.92$\pm$0.06 & 0.34$\pm$0.02 &   0.89$\pm$0.08  &    3.82 0.18 &  0.96 (45) &     3.29/2.31 \\
93150-03-05-04 & 54634.58 & 0.91$\pm$0.08 & 0.43$\pm$0.07 &   0.81$\pm$0.06  &    3.22 0.19 &  0.98 (45) &     2.93/1.88 \\
93150-03-05-05 & 54635.18 & 0.92$\pm$0.09 & 0.36$\pm$0.09 &   0.85$\pm$0.02  &    3.27 0.13 &  0.96 (45) &     2.87/1.95 \\
93150-03-05-06 & 54636.81 & 0.93$\pm$0.08 & 0.12$\pm$0.08 &   0.70$\pm$0.05  &    3.54 0.17 &  0.95 (45) &     2.85/2.25 \\
94150-03-01-00 & 55170.34 & 0.97$\pm$0.09 & 0.24$\pm$0.08 &   0.73$\pm$0.02  &    12.95 0.19&   1.06 45 &     11.14/7.74\\
94150-03-01-01 & 55173.36 & 0.97$\pm$0.08 & 0.24$\pm$0.07 &   0.73$\pm$0.04  &    11.86 0.18&   1.00 45 &     10.18/7.07\\
94150-03-01-02 & 55174.39 & 0.96$\pm$0.09 & 0.23$\pm$0.08 &   0.74$\pm$0.03  &    10.54 0.19 &  0.97 45 &     8.99/6.33\\
94150-03-01-03 & 55175.96 & 0.95$\pm$0.07 & 0.25$\pm$0.06 &   0.73$\pm$0.03  &    7.95 0.12  & 1.12 (45) &    6.85/4.73 \\
94150-03-01-04 & 55175.19 & 0.98$\pm$0.05 & 0.25$\pm$0.05 &   0.74$\pm$0.02  &   10.45 0.13  & 1.13 (45) &    8.97/6.25 \\
94150-03-02-00 & 55176.44 & 0.94$\pm$0.06 & 0.28$\pm$0.04 &   0.75$\pm$0.03  &    7.23 0.19  & 1.07 (45) &    6.31/4.28 \\
94150-03-02-01 & 55177.27 & 0.93$\pm$0.08 & 0.25$\pm$0.09 &   0.81$\pm$0.03  &    5.85 0.17  & 0.94 (45) &    5.01/3.56 \\
94150-03-02-02 & 55178.19 & 0.95$\pm$0.07 & 0.25$\pm$0.08 &   0.82$\pm$0.04  &    6.01 0.13  & 1.09 (45) &    5.12/3.67 \\
94150-03-02-03 & 55179.17 & 0.95$\pm$0.07 & 0.25$\pm$0.08 &   0.82$\pm$0.04  &    6.29 0.17  & 1.02 (45) &    5.38/3.83 \\
94150-03-02-04 & 55180.15 & 0.95$\pm$0.07 & 0.26$\pm$0.08 &   0.83$\pm$0.04  &    6.35 0.14  & 1.04 (45) &    5.51/3.86 \\
94150-03-02-05 & 55181.19 & 0.96$\pm$0.08 & 0.27$\pm$0.08 &   0.83$\pm$0.02  &    6.42 0.18  & 1.06 (45) &    5.57/3.91 \\
94150-03-02-06 & 55182.76 & 0.97$\pm$0.03 & 0.31$\pm$0.05 &   0.79$\pm$0.04  &    7.28 0.15  & 0.97 (45) &    6.47/4.35 \\
95149-18-01-02 & 55293.02 & 0.98$\pm$0.04 & 0.32$\pm$0.05 &   0.78$\pm$0.04  &    7.41 0.13  & 0.91 (45) &    6.61/4.40 \\
95149-18-01-03 & 55293.38 & 0.97$\pm$0.05 & 0.32$\pm$0.09 &   0.78$\pm$0.05  &    7.01 0.12  & 0.97 (45) &    6.26/4.16 \\
95149-18-01-05 & 55294.28 & 0.98$\pm$0.06 & 0.26$\pm$0.08 &   0.80$\pm$0.06  &    8.52 0.11  & 0.98 (45) &    7.43/5.15 \\
95149-18-01-08 & 55293.15 & 0.98$\pm$0.07 & 0.28$\pm$0.09 &   0.80$\pm$0.05  &    7.17 0.15  & 0.96 (45) &    6.30/4.31 \\
95149-18-01-09 & 55292.89 & 0.99$\pm$0.04 & 0.28$\pm$0.05 &   0.80$\pm$0.06  &    7.68 0.17  & 0.98 (45) &    6.74/4.62 \\
95149-18-02-00 & 55295.77 & 0.98$\pm$0.05 & 0.29$\pm$0.08 &   0.78$\pm$0.02  &    8.02 0.15  & 0.94 (45) &    7.13/4.78 \\
95149-18-02-01 & 55296.83 & 0.98$\pm$0.05 & 0.31$\pm$0.09 &   0.77$\pm$0.06  &    6.75 0.13  & 0.96 (45) &    6.06/4.00 \\
95149-18-02-02 & 55297.29 & 0.97$\pm$0.08 & 0.33$\pm$0.07 &   0.76$\pm$0.04  &    6.53 0.19  & 0.97 (45) &    5.92/3.84 \\
\hline 
\label{tab:par_rxte_2010}
\end{tabular}
\tablefoot{ 
$^\dagger$ The spectral model is  {\tt tbabs*bmc};  where $N_H$ is fixed at 
a value 5.0$\times 10^{22}$ cm$^{-2}$ (see Sect.~\ref{model choice}); 
$^{\dagger\dagger}$ when parameter $\log(A)\gg1$, this parameter is fixed at 2.0 (see comments in the text), 
$^{\dagger\dagger\dagger}$ for normalization parameter  $N_{BMC}=L_{35}/d^2_{10}$ (see Eqs. \ref{bmc_spectrum}, \ref{ratio_fluxes})
 where $L_{35}$ is the source luminosity in units of 10$^{35}$ erg/s and  
$d_{10}$ is the distance to the source in units of 10 kpc 
$^{\dagger\dagger\dagger\dagger}$spectral fluxes (F$_1$/F$_2$) in units of 
$\times 10^{-9}$ ergs/s/cm$^2$ for  (3 -- 10) and (10 -- 20) keV 
energy ranges, respectively.  
}
\end{table*}


\begin{table*}
\caption{Best-fit parameters of spectral analysis of 
{\it RXTE} 
observations of M87 ($R1$ -- $R2$ sets) in 3 -- 50~keV energy range$^{\dagger}$. 
Parameter errors correspond to 90\% confidence level.}\label{tab:fit_table_RXTE_1}
\centering
\begin{tabular}{lcllllccccccccc}
\hline\hline
Observational ID & MJD    & $\alpha=\Gamma-1$ & $N_{BMC}^{\dagger\dagger\dagger}$ & $E_{cut}$ & $E_{fold}$ & $E_{line}$ & $\sigma_{line}$ & $N_{line}^{\dagger\dagger\dagger}$ &$\chi^2_{red}$ (dof)\\ 
                 & (day)  &                   & ($L_{37}/d^2_{10}$)               & (keV)     &  (keV)     &   (keV)    &    (keV)        &            &\\%

\hline                                                           
%
30216-01-01-000 & 50812.94 & 2.02$\pm$0.06  &  1.18$\pm$0.08 &  3.6$\pm$0.1 & 9.9$\pm$0.8 &  6.4$\pm$0.1 & 0.7$\pm$0.2 & 0.10$\pm$0.01 & 0.95(40) \\
30216-01-01-010 & 50813.74 & 2.04$\pm$0.02  &  1.58$\pm$0.01 &  7.3$\pm$0.9 & 4.7$\pm$0.7 &  6.6$\pm$0.1 & 0.2$\pm$0.1 & 0.09$\pm$0.02 & 0.95(40) \\
30216-01-01-02  & 50814.74 & 2.00$\pm$0.01  &  1.11$\pm$0.01 &  6.7$\pm$0.3 & 4.2$\pm$0.3 &  6.5$\pm$0.2 & 0.5$\pm$0.2 & 0.08$\pm$0.01 & 0.93(40) \\
30216-01-02-00  & 50822.61 & 2.01$\pm$0.05  &  1.13$\pm$0.02 &  7.0$\pm$0.3 & 3.6$\pm$0.2 &  6.4$\pm$0.3 & 0.4$\pm$0.1 & 0.07$\pm$0.01 & 0.98(40) \\
30216-01-02-01  & 50823.56 & 1.99$\pm$0.06  &  1.14$\pm$0.02 &  7.2$\pm$0.4 & 3.4$\pm$0.3 &  6.5$\pm$0.1 & 0.3$\pm$0.1 & 0.09$\pm$0.01 & 0.86(40) \\
30216-01-02-020 & 50824.62 & 2.00$\pm$0.05  &  1.12$\pm$0.06 &  6.8$\pm$0.3 & 3.8$\pm$0.3 &  6.3$\pm$0.2 & 0.4$\pm$0.1 & 0.10$\pm$0.03 & 0.93(40) \\
30216-01-03-00  & 50832.95 & 2.02$\pm$0.07  &  1.14$\pm$0.07 &  7.0$\pm$0.4 & 3.6$\pm$0.3 &  6.4$\pm$0.2 & 0.5$\pm$0.2 & 0.10$\pm$0.02 & 0.86(40) \\
30216-01-03-01  & 50833.68 & 2.01$\pm$0.05  &  1.13$\pm$0.07 &  6.3$\pm$0.5 & 5.1$\pm$0.6 &  6.5$\pm$0.1 & 0.4$\pm$0.1 & 0.08$\pm$0.01 & 0.99(40) \\
30216-01-03-02  & 50833.97 & 2.00$\pm$0.08  &  1.15$\pm$0.08 &  6.7$\pm$0.4 & 4.2$\pm$0.4 &  6.5$\pm$0.2 & 0.4$\pm$0.1 & 0.09$\pm$0.01 & 0.96(40) \\
30216-01-03-03  & 50833.29 & 2.00$\pm$0.05  &  1.14$\pm$0.08 &  6.2$\pm$0.6 & 5.2$\pm$0.8 &  6.3$\pm$0.3 & 0.7$\pm$0.2 & 0.10$\pm$0.04 & 0.95(40) \\
30216-01-03-04  & 50835.56 & 2.03$\pm$0.05  &  1.19$\pm$0.09 &  4.6$\pm$0.9 & 7.2$\pm$0.9 &  6.5$\pm$0.1 & 0.4$\pm$0.1 & 0.10$\pm$0.03 & 0.98(40) \\
30216-01-04-00  & 50843.61 & 2.02$\pm$0.06  &  1.17$\pm$0.09 &  6.5$\pm$0.8 & 4.5$\pm$0.5 &  6.5$\pm$0.2 & 0.4$\pm$0.1 & 0.08$\pm$0.02 & 0.97(40) \\
30216-01-04-01  & 50844.70 & 2.01$\pm$0.08  &  1.16$\pm$0.08 &  6.8$\pm$0.4 & 4.2$\pm$0.3 &  6.4$\pm$0.2 & 0.4$\pm$0.1 & 0.10$\pm$0.01 & 0.99(40) \\
30216-01-04-02  & 50845.77 & 2.08$\pm$0.09  &  1.29$\pm$0.06 &  7.1$\pm$0.9 & 4.7$\pm$0.9 &  6.3$\pm$0.2 & 0.4$\pm$0.1 & 0.10$\pm$0.01 & 0.96(40) \\
30216-01-04-03  & 50843.89 & 2.09$\pm$0.08  &  1.31$\pm$0.07 &  7.3$\pm$1.2 & 4.5$\pm$0.9 &  6.5$\pm$0.1 & 0.6$\pm$0.2 & 0.08$\pm$0.01 & 0.98(40) \\
30216-01-04-04  & 50846.55 & 2.05$\pm$0.09  &  1.79$\pm$0.09 &  7.1$\pm$1.8 & 7.8$\pm$1.6 &  6.4$\pm$0.3 & 0.4$\pm$0.1 & 0.09$\pm$0.03 & 0.97(40) \\
30216-01-04-05  & 50846.88 & 1.98$\pm$0.08  &  1.27$\pm$0.06 &  6.9$\pm$1.0 & 5.0$\pm$1.8 &  6.5$\pm$0.2 & 0.7$\pm$0.2 & 0.10$\pm$0.01 & 0.94(40) \\
95145-01-01-00  & 55296.79 & 0.99$\pm$0.08  &  0.37$\pm$0.09 &  2.8$\pm$1.0 & 3.7$\pm$0.3 &  6.3$\pm$0.2 & 0.4$\pm$0.1 & 0.10$\pm$0.08 & 0.93(40) \\
95145-01-01-01  & 55298.94 & 0.86$\pm$0.07  &  0.34$\pm$0.07 &  3.3$\pm$1.4 & 3.4$\pm$0.2 &  6.5$\pm$0.1 & 0.5$\pm$0.1 & 0.09$\pm$0.02 & 0.97(40) \\
95145-01-01-02  & 55299.99 & 0.8$\pm$0.1    &  0.33$\pm$0.09 &  2.9$\pm$1.5 & 3.1$\pm$1.3 &  6.4$\pm$0.2 & 0.6$\pm$0.2 & 0.09$\pm$0.02 & 1.03(40) \\
95145-01-01-03  & 55301.00 & 0.76$\pm$0.09  &  0.32$\pm$0.08 &  3.1$\pm$1.8 & 3.0$\pm$1.8 &  6.5$\pm$0.2 & 0.4$\pm$0.1 & 0.10$\pm$0.06 & 0.95(40) \\
\hline
\label{tab:par_rxte_m87}
\end{tabular}
\tablefoot{ 
$^\dagger$ The spectral model is {\tt tbabs*bmc*highecut};  where $N_H$ is fixed at 
a value 1.94$\times 10^{20}$ cm$^{-2}$ (see Sect.~\ref{model choice}) 
 and the parameter $kT_s$ of  the BMC model was fixed at 0.4 keV.
parameter $\log(A)$ is low variable around 2.0; 
$^{\dagger\dagger\dagger}$ for normalization parameter  
$L_{37}$ is the source luminosity in units of 10$^{37}$ erg/s and  
$d_{10}$ is the distance to the source in units of 10 kpc; 
and the Gaussian component is in units of $10^{-2}\times total~~photons$ $cm^{-2}s^{-1}$ in line.   
}
\end{table*}

\end{document}